\documentclass[11pt, a4paper, notitlepage]{article}

\pdfoutput=1

\usepackage{setspace}
\renewcommand{\baselinestretch}{1.75}

\usepackage{algorithm}
\usepackage{algorithmic}
\usepackage{graphicx,url}
\usepackage{floatflt}
\usepackage[english]{babel}
\usepackage[utf8]{inputenc}
\usepackage{latexsym}
\usepackage{cite}
\usepackage{fancyhdr}
\usepackage{longtable}
\usepackage{array}
\usepackage{epsfig}
\usepackage{graphicx}
\usepackage{alltt}
\usepackage{verbatim}
\usepackage{amssymb,amsmath}
\usepackage{pifont}
\usepackage[usenames]{color}
\usepackage[rgb]{xcolor}
\usepackage{geometry}
\usepackage{enumitem}

\usepackage{epstopdf}

\usepackage{subfig}

\usepackage{color}

\setlist{itemsep=0pt,parsep=0.5pt,topsep=0.5pt}

\newtheorem{propo}{Proposition}[section]

\newcommand{\daniel}[1]{{#1}}

\newcommand{\eat}[1]{}
\newcommand{\Lump}{\ensuremath{\mbox{\sc LumpState}}}

\newcommand{\lambdacritical}{\lambda_C}
\newcommand{\lambdasaturation}{\lambda_S}

\newcommand{\bpi}{ {\mbox{\boldmath $\pi$}} }

\newcommand{\bsigma}{ {\mbox{\boldmath $\sigma$}} }

\newcommand{\be}{ {\bf e} }

\DeclareGraphicsExtensions{.pdf,.png,.jpg,.eps}

\title{On The Scalability of P2P  Swarming Systems\footnote{The journal version of this paper was published in the Computer Networks journal: de Souza e Silva, E. , Leão, R. M., Menasché, D. S., Towsley, D. (2019). On the scalability of P2P swarming systems. Computer Networks, 151, 93-113.
}}

\author{Edmundo de Souza e Silva, Rosa M.M. Leão, Daniel S. Menasch\'e \\
Federal University of Rio de Janeiro, Brazil \\ \\
Don Towsley \\
University of Massachusetts at Amherst, USA\\}
\date{}

\begin{document}

\renewcommand{\baselinestretch}{1.2} 
\maketitle
\renewcommand{\baselinestretch}{1.75}  

\newcommand{\sadoc}[1]{\textcolor{blue}{[#1 --DAN]}}

\begin{abstract}
One of the  fundamental problems in the realm of peer-to-peer systems is that
of determining their service capacities.   
{In this paper we focus on P2P scalability issues and propose
models to compute the achievable throughput under distinct policies for
selecting both peers and blocks.}
From these models, we obtain novel insights on the
behavior of P2P swarming systems that motivate new mechanisms for publishers
and peers to improve the overall performance. 
In particular,
{we obtain operational regions for swarm system}.
In addition, we show that
system capacity significantly increases if publishers adopt the most deprived
peer selection and peers reduce their service rates when they have all the file
blocks but one. 

\end{abstract}


\section{Introduction}

Peer-to-peer (P2P) swarming systems have been tremendously successful
in disseminating content in the Internet and
major companies have adopted this architecture. 
For instance, \textit{Blizzard entertainment} distributes large files
via the \textit{Blizzard downloader}, which is based on the 
\textit{BitTorrent Opensource}.    
Ubuntu Linux is also available to download via BitTorrent and 
the \textit{Amazon Simple Storage Service} (Amazon S3)  supports the BitTorrent protocol for distributing files.
In addition, recent proposals suggest the use of P2P for distributing
large files in content oriented networks~\cite{ccn}
paving the way towards a Network Swarm Architecture.

P2P architectures have been studied for over a decade and numerous
models have been proposed to address the performance of the approach.
Examples include models
to determine the impact of incentive mechanisms
and  download/upload rates on performance~\cite{fan2006delicate, chow2009bittorrent}, 
fairness~\cite{murai2012heterogeneous} and availability~\cite{e2013interplay}.
In addition, several works have  focused on
block distribution and peer selection strategies~\cite{laurent}, and proposed improvements to 
the standard BitTorrent protocol~\cite{xia_muppala_2010}.

Despite numerous papers in the area, 
scalability issues have only recently been addressed.
A system scales if its throughput, \textit{i.e.},
the  rate at which users complete their downloads,
increases linearly with increasing user population.
P2P systems have been thought to be scalable since each new user joining the system
brings additional resources to it.
Consequently, the total capacity available is expected to increase in proportion to the newly
incorporated resources and, accordingly,
 system throughput should increase linearly and unboundedly with population size. 
However, intrinsic limitations of the P2P architecture to scale in this manner have been observed.
Hajek and Zhou \cite{hajek} showed that, when peers in a swarm and publisher adopt a 
random peer/random useful block policy, the system is unstable, that is,
swarm population grows without bound
if  peer arrival rate $\lambda$ exceeds
the server capacity ($U$) dedicated to that swarm.
Briefly, this instability  occurs because, when two peers meet,
they may not have useful data to share. 
\daniel{Understanding} the stability region of peer-to-peer swarming systems has recently gained attention 
\cite{mathieu, zhu2012stability, hajek, menasche2011implications}, and 
 unstable scenarios have been observed in practice~\cite{murai2012heterogeneous,diego}.

In a peer-to-peer swarming system, each peer makes two decisions before transmitting a block: 
$(a)$  what block to transmit and $(b)$ to whom to transmit it. 
Although the former decision has received some attention in previous works
(for instance, it has been shown that rarest-first block selection and random useful block selection 
yield the same stability region~\cite{hajek}), the implication of the peer selection strategy \daniel{on throughput} 
has not been thoroughly studied (notable exceptions being~\cite{mathieu, menasche2011implications}).  
Previous works have assumed peers choose their neighbors using random peer selection~\cite{nunez2008scaling, zhu2012stability, hajek}.

In this work we  evaluate the impact of different system parameters and
system strategies on attainable throughput.
Our key contributions are:

\textbf{System throughput upper-bound. } First,  we derive an upper bound on the throughput achieved when the  publisher adopts most deprived
peer selection and rarest-first block selection, while peers adopt random peer selection and
random useful block selection. 
\eat{ \textcolor{red}{We MUST be careful here when we talk about bounds, since we still do not prove
anything. May be we should say that we obtain the throughput for large swarm sizes? Or leave the way it is?
see all places when we mention "bounds"} }
The bound is significantly larger than the maximum attainable throughput observed in the scenarios
studied in~\cite{hajek}, where both peers and publishers adopt random peer and random useful
block selection.
This means that system performance \daniel{can substantially} increase if the server adopts a simple policy
that gives priority to those peers that possess the smallest number of blocks in the swarm.

\textbf{Stability region characterization. } 
We identified two regions in which the P2P systems can operate.  
Such regions are characterized by two thresholds, referred to as the critical and saturation thresholds, 
 denoted by 
$\lambdacritical$ and $\lambdasaturation$, respectively, where $\lambdacritical  < \lambdasaturation$. The corresponding
throughputs are given by $\Gamma_C$ and $\Gamma_S$, respectively, where $\Gamma_C = \lambda_C$ and $\Gamma_S  \leq \lambda_S$.  
In the first region,  $\lambda < \lambdacritical$, the system is stable, \emph{i.e.}, download times are finite.
In the second region,  $\lambdacritical < \lambda < \lambdasaturation$, the system is unstable, 
\emph{i.e.}, {delays are infinite but throughput grows as  arrival rate increases.}  
%
In the third region, $\lambda > \lambdasaturation$, the system is unstable and throughput equals $\Gamma_S$.  

\textbf{Benefits of upload throttling, protection of newcomers and admission control. }
Our models suggest  a new very simple and incentive-compatible policy adopted by peers,  
 wherein peers reduce their service capacity when they possess
all blocks but one (see Section \ref{sec:models}).  
By employing this upload throttling  policy, which requires only changes to the protocol implemented by  peers, 
the system can accommodate more users while  remaining stable, specially when near saturation.   
  In addition, we also investigate   policies that involve changes to peers and trackers, 
   wherein  trackers protect newcomers to be preferably served by the publisher (see Section~\ref{sec:results}).  
 Numerical results indicate that such policies, in addition to admission control, can also result 
in significant performance gains.

\paragraph{Paper structure} 
The remainder of this paper is organized as follows.
In Section \ref{sec:discussion} we set the background, discuss
the scalability problem and compare several peer and server policies.
In Section \ref{sec:models} we describe the models used to obtain the results
we present.
Analytical and numerical results obtained from the models are discussed in Sections~\ref{sec:trana} and \ref{sec:results}, respectively.
We show how  system throughput scales with swarm population size and
study the performance gains of the policies we propose.
Related work is presented in Section \ref{sec:related},
Section \ref{sec:limit} summarizes our assumptions and model limitations,
and Section \ref{sec:concl} concludes our work.

\section{Scalability of Peer-to-Peer Systems}
\label{sec:discussion}

\begin{figure}[h!]\center
\includegraphics[width=\columnwidth]{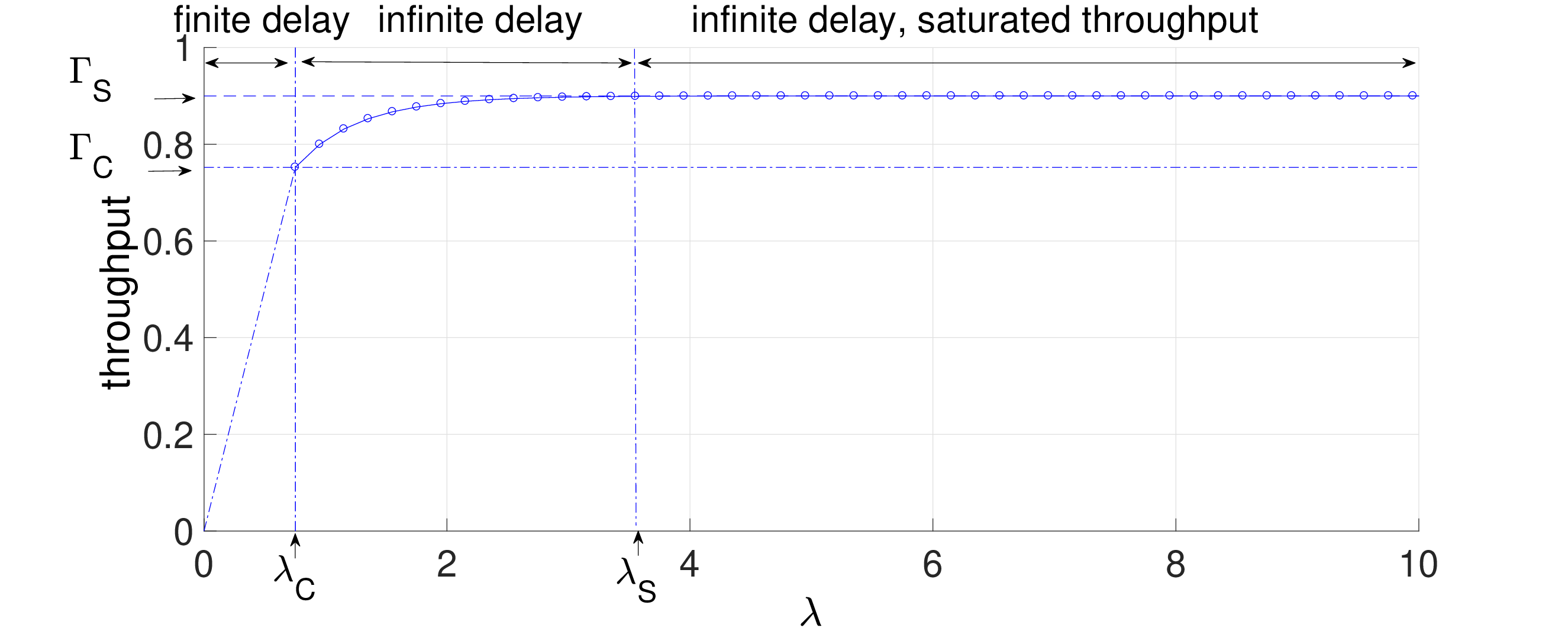}
\caption{System throughput (3 blocks, most deprived peer, rarest first chunk,  $\mu=0.5$, $U=0.3$)} 
\label{fig:limited-thr}
\end{figure}

In a P2P system, every new peer entering a swarm brings additional resources
(transmission capacity and storage) to it since it acts both as a client and as a content 
provider. 
Therefore it is natural to expect  the overall system throughput to increase  
as the number of peers increases.   
However, the throughput growth is not unbounded.
Figure \ref{fig:limited-thr}, obtained from one of the models we developed (Section~\ref{sec:queueing-model}) 
illustrates this well. 
Let $\lambda$ denote the arrival rate of peers to a swarm.  

\paragraph{Illustrative example}  To appreciate the nature of our results through a simple example, Figure~\ref{fig:limited-thr}
shows how the throughput varies as a function of the arrival rate,
for a given peer and server capacities allocated to the swarm.
(The notation is defined in Table {\ref{tab:notation}.)  
Figure~\ref{fig:limited-thr} shows that throughput grows  linearly in  $\lambda$ until $\lambda$ reaches a threshold
  ($\lambdacritical$).  
Download delays are finite in the  region $\lambda < \lambdacritical$ and infinite when $\lambda > \lambdacritical$. 
When  $\lambda=\lambdacritical$, throughput equals arrival rate, $\Gamma_C = \lambda_C$.     
When   $\lambda \geq \lambdacritical$,   throughput  increases until $\lambda$ reaches a threshold $\lambdasaturation$,
 meaning that 
  it is possible to  achieve higher throughput at the cost of infinite delays.
  When  throughput reaches $\Gamma_S$  the 
system  is fully saturated and throughput no longer increases in $\lambda$.  
Briefly, although  available resources increase as the number of peers increases (increasing $\lambda$),
they are not fully utilized.
For instance, this occurs when a peer has no block different from those 
its neighboring peers possess; thus, its service capacity is wasted.

\paragraph{Two fundamental operating regimes} There are two operating regimes observed in the P2P systems of interest to us.  
One is a \textit{stable} operating regime where all peers that arrive acquire
the content in a finite amount of time.  
Associated with this regime is a threshold $\lambdacritical$ such that the system is stable
provided  the peer arrival rate $\lambda$ satisfies $\lambda < \lambdacritical$.
Then,   there is a second regime where 
the maximum achievable throughput,  $\Gamma_S$, referred to as the saturation throughput,  
can be larger than $\Gamma_C$  (see Figure~\ref{fig:limited-thr}). To leverage the possibility of achieving a throughput
larger than $\Gamma_C$, our study suggests that
 it is interesting to implement 
 admission control, so as to achieve a throughput larger than $\Gamma_C$  without incurring infinite download delays.
 

\eat{
In particular, if we let $\Upsilon(\lambda)$ denote the system throughput as a 
function of $\lambda$,
then it appears to be a strictly increasing function of $\lambda$ satisfying 
$\Upsilon(\lambda) = \lambda$ for $\lambda < \lambdacritical$ and
$\lim_{\lambda \rightarrow \infty} \Upsilon(\lambda) = \lambdasaturation > \lambdacritical$.  
Note that $\lambdasaturation$ is a measure associated to the open system.  Our study points out that it is interesting to implement 
admission control, as in a closed system the throughput can be larger than $\lambdacritical$.  
}

\eat{
\textcolor{red}{NOTE: this last limit has to be re-checked.
The limit $KU$ we show for open system depends on the server policy.
The overshoot in the curve occurs only in the closed system and we are able to achieve $\lambdasaturation$.
I believe this is expected, but the sentence we had before is not consistent and has to change.} 
\textcolor{red}{Also recheck the following sentence}.  }

In light of the two operating regimes discussed above, we briefly provide further insight on 
 the throughput curve of Figure \ref{fig:limited-thr}.
Consider a P2P system in which peers leave as soon as they complete their downloads.
The file downloaded by a particular swarm is divided into $K$ equal size blocks.
Recall that $U$ is the server capacity \textit{allocated to the swarm} 
in blocks per time unit. Let  $\mu$ be  the service capacity of each peer, also measured in blocks per time unit.   
\eat{ (theneighbor and block selection policies adopted by the publisher and peers).   }
When peer arrival rates are large relative to $U$ the system will reach,
with high probability, a state in which all peers have all but one of the blocks
(say, block $\mathcal{B}$). 
In this scenario, only the server has block $\mathcal{B}$ and \eat{then} as soon as
a peer obtains the missing block $\mathcal{B}$ from the server it leaves the system without
helping to serve block $\mathcal{B}$ to others.
On the other hand, new peers that arrive are quickly served by their
peers and are able to obtain all blocks but $\mathcal{B}$.
As a consequence, the system behaves approximately like a client server system:
peers depart at  rate $U$ since they do not cooperate to obtain the missing block.
In addition, although  available resources grow with the swarm size, these resources
are not fully~utilized.

We propose a  queueing model  in Section \ref{sec:models} to estimate the 
throughput
taking into account that, when the system is saturated, 
with high probability most peers in the system
eventually obtain all but the rarest block.
We also show that the 
throughput depends on  service policies adopted by the server and peers.

\paragraph{Neighbor and  block selection policies}
In a P2P system, the publisher needs to decide what block to transmit next  and to
which peer.
Likewise, a peer has to decide what peer to contact and which block to request.
In the literature, it has been assumed that peers go through random encounters
and that peers are paired uniformly at
random~\cite{hajek, oguz2015stable, zhou2011stability}.
Nonetheless, if  the publisher can strategically select peers to serve, then  it is possible
to improve the overall system capacity.
Before introducing our models we first describe the policies that we consider.

\paragraph{Random peer selection} 
Throughout this paper, except otherwise noted we assume that peers select a neighbor uniformly at random
at every transmission opportunity to exchange blocks. Such policy is referred to as random peer selection policy.

\paragraph{Random useful peer selection}
We also consider the case where trackers dynamically inform  
each peer $P$ in the set of peers in the swarm of those that are
in need of blocks owned by $P$. 
In this case, peers can select their neighbors uniformly at random among those 
that need the blocks they possess.
We refer to this neighbor selection policy as random \emph{useful} peer selection.

\paragraph{Random useful block selection}
After choosing a neighbor, each peer selects
one of its blocks for transmission to the neighbor.
If the block is selected uniformly at random, the policy is referred to as random useful
block selection.

\paragraph{Rarest first block selection}
If peers have access to a list of the number of replicas of each block, they can build a
rarest-block set containing the indices of the blocks with the least number of copies in
the swarm~\cite{legout:07}.
This set can then be used by peers to select which block to transmit.
This policy is referred to as rarest first block selection.

\paragraph{Most deprived peer selection by publishers} 
The publisher can select its peers and blocks in the same way as the peers.
In addition, the publisher can also select its peers using the most deprived policy.
Under this policy, the publisher prioritizes sending blocks to peers that own the least
amount of blocks among those in the swarm.  
If the arrival rate of peers is large (or the swarm size is large) these peers are likely
to be content-less peers, also referred to as newcomers. 
Table \ref{tab:policies} summarizes these policies.

\begin{table}
\scriptsize
\center
\begin{tabular}{l|l}
\hline
peers & publisher \\
\hline \hline
random peer/random useful block (RP/RUB) & random peer/random useful block  (RP/RUB) \\
random peer/rarest first block (RP/RFB) & random peer/rarest first block (RP/RFB) \\
random useful peer/random useful block (RUP/RUB) & most deprived peer/random useful block (MDP/RUB)        \\
random peer/random useful block (RP/RUB) & most deprived peer/rarest first block (MDP/RFB)        \\
\hline
\end{tabular}
\caption{Neighbor and block selection policies} \label{tab:policies}
\end{table}

\section{Models}
\label{sec:models}

In this section we present models of the P2P systems utilizing  policies presented in the  
previous section.  
  The key goal of the models is to estimate $\lambdacritical$ and $\lambdasaturation$,
for different server and peer policies.

Using our models we show how 
throughput depends on different system parameters. 
Random peer and block selection by peers and publisher have 
been studied in \cite{hajek,  zhu2012stability}.
These works show that the system is stable if and only if $\lambda < U$.
The proposed models allow us to study  simple modifications to the peer and block selection policies. 
For instance, in Section~\ref{sec:trana} we show that  adoption of the most deprived peer selection policy by the publisher can increase throughput.
In addition, in Section~\ref{sec:results} we show that if  peers reduce their service capacity,  maximum achievable throughput
can be further   increased.  
Table \ref{tab:notation} summarizes  notation used in this paper.
\begin{table}
\scriptsize
\center
\begin{tabular}{l|l}
\hline
Variable & Description \\
\hline \hline
\multicolumn{2}{c}{System parameters} \\
\hline
$\lambda$ & peer arrival rate \\
$U$ & the server capacity allocated to a swarm \\ 
$K$ & total number of blocks \\
$\tilde{K}$ &  number of blocks individually accounted by the truncated model\\
$\mu$ & peer service capacity \\
$\mu'$ & reduced peer service capacity \\
$N$ & number of peers in the swarm \\
& (when considering the queuing model, we assume an infinite one-club, so $N=\infty$; \\ 
& in Section~\ref{sec:markovian}, in contrast, we consider a population of fixed finite size $N$) \\
\hline
\multicolumn{2}{c}{Throughput-related variables} \\
\hline
$\Gamma$ & total rate at which peers leave the system (system throughput) \\
$\lambdacritical$ &   critical throughput (when $\lambda < \lambdacritical$, we have finite delays and a finite population of newcomers)  \\
$\lambdasaturation$ & saturation throughput (when  $\lambdacritical < \lambda < \lambdasaturation$ the throughput can still be increased, \\
&  whereas for $\lambda > \lambdasaturation$ we have a saturated system with  an infinite population of newcomers) \\
\hline
\multicolumn{2}{c}{Queuing model variables} \\ 
\hline
$p$ & fraction of blocks served by the publisher that is transmitted to  newcomers \\
& (under MDP/RFB, probability that  publisher serves rarest block to newcomer) \\
$F_i$ & $i$-th queue containing non-gifted peers that have $i$ blocks \\
$G_j$ & $j$-th queue containing gifted peers that have $j$ blocks \\
$n_i$ & random variable denoting the number of non-gifted peers that have $i$ blocks \\
$m_j$ & random variable denoting the number of gifted peers that have $j$ blocks \\
& (those peers have the rarest block and additional $(j-1)$  blocks) \\
$\tilde{m}_{\tilde{K}-1}$ & random variable denoting the number of gifted peers that have $\tilde{K}-1$ or more blocks \\
\hline
\multicolumn{2}{c}{Queuing model rates} \\
\hline
$\gamma_r(i)$ & rate at which non-gifted peers that have $i$ blocks receive the rarest block \\
$\gamma_p(i)$ & rate at which non-gifted peers that have $i$ blocks receive a popular  block \\
$\Psi$ & rate at which the gifted peers serve the rarest block to the one-club members \\
\hline
\end{tabular}
\caption{Table of notation} \label{tab:notation} 
\end{table}



\subsection{Overview of Peers and Publisher Dynamics}
\label{sec:dynamics}

We first consider a publisher policy that gives priority to peers with 
the smallest number of blocks and serves the rarest block to the chosen peers. 
That is, the server adopts the MDP/RFB \daniel{policy} (Table \ref{tab:policies}).
In addition, peers adopt the RP/RUB \daniel{policy, wherein} upon a random peer contact,
a block is  randomly chosen to be downloaded  from among
those blocks that the recipient peer does not have.
We also assume that a peer \textit{reduces} its service capacity  whenever it
 obtains $K-1$ blocks,  changing its service capacity from $\mu$ to $\mu'$
where $\mu' < \mu$. 
The motivation for this rate reduction will be clarified below when we show that
it significantly increases the system throughput.
We refer to Figure \ref{fig:diagrama-1} to describe the model.

%

\begin{figure}[htb]
\center
\includegraphics[width=1\columnwidth]{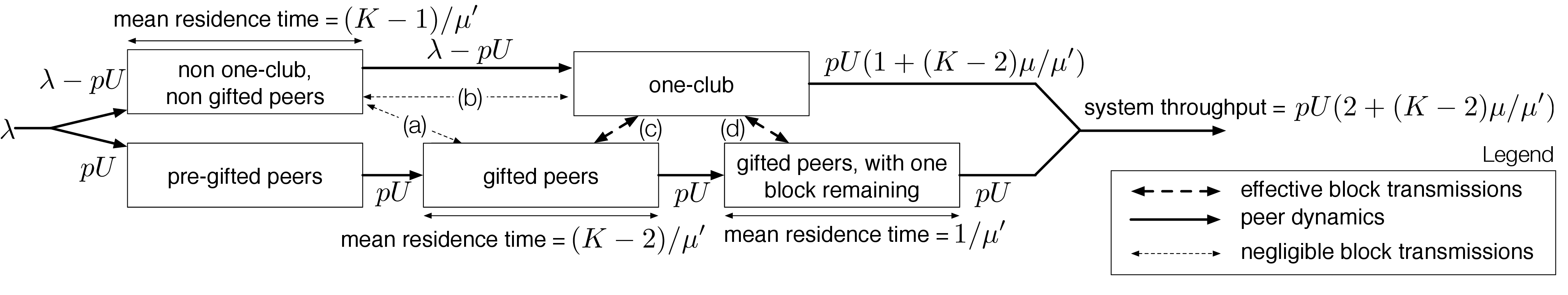}
\caption{Flow dynamics}
\label{fig:diagrama-1}
\end{figure}
In Figure~\ref{fig:diagrama-1} we identify five boxes each representing a set of peers with a given collection of
blocks from the downloaded file.  Solid (resp., dotted) arrows correspond to peer dynamics (resp., block transmissions).  
We assume that the peer arrival rate $\lambda$ is sufficiently large, $\lambda > \lambdasaturation > U$, 
as in Figure \ref{fig:limited-thr}.
In this regime, the system enters a state where an infinite number of peers have all but one of the blocks of the file.  
These peers are said to  belong to the \textit{one-club} set~\cite{hajek} 
\eat{\textcolor{red}{check ref}} 
(see Figure \ref{fig:diagrama-1}). 
\eat{\textcolor{red}{I would prefer to call this set "missing club"...}}

Since the publisher adopts  most deprived peer selection and $U < \lambda$,
a fraction $pU/\lambda$ of peers receive a block from the publisher after arriving to the system.
Note that:  
$(a)$ newcomers are content-less, so they are served with higher priority by the publisher and; 
$(b)$ since $\lambda$ is much larger than $U$,
the publisher finds a newcomer with high probability as soon as it
is ready to serve a new peer.
Parameter $p$ denotes the fraction of blocks served by the publisher to newcomers.  
It captures the fact that  newcomers may not receive all of the publisher capacity available to the
swarm.
Peers that obtain the rarest block are called \textit{gifted peers}. 
The three  boxes in the bottom row of  Figure~\ref{fig:diagrama-1} represent peers in the process of obtaining
the rarest block and those peers that already received~it.  

\textit{Remark}:
$p=1$ corresponds to 
the publisher serving the rarest block to a newly arrived peer, whenever 
 the publisher is available upon the  arrival under consideration.
For this to  occur, it may be necessary   to postpone the  immediate advertisement of newcomers to other peers, i.e., newcomers
may be known 
for some initial time only to the tracker.
Otherwise,  the server and other peers \textit{compete} for service.  A newly arrived peer may obtain
 a popular block from other peers, before obtaining the rarest block from the server, even
when the server is idle upon arrival.  As a result, the competition between server and peers  favors $p < 1$, 
where $p$ depends on $U$ and $\mu$.   We will return to this issue in Section~\ref{sec:results}.

Peers adopt  random peer and random useful block selection policies. 
 We say a block is   \textit{popular}
whenever it is not the rarest block.   
Note that as the \emph{one-club} is assumed to be  large, 
 the download of the vast majority of the  popular blocks involves  peer encounters that occur uniformly  at random between 
 peers not in  the one-club with those in the one-club (arrows $(b), (c)$ and $(d)$ in Figure~\ref{fig:diagrama-1}).
 Since the \textit{gifted} set is 
 relatively small, 
 roughly all popular blocks are served by the one-club population, and a typical peer takes
 on average $1/\mu'$ to download each popular block.    Transmissions of popular blocks
 not involving a member of the one club, such as those represented by arrow $(a)$ in Figure~\ref{fig:diagrama-1}, can be neglected.   
Note also that as peers select their neighbors uniformly at random, a significant number of peer encounters occurs
 between members of the one-club amongst  themselves. 
 Such encounters produce no exchange of blocks.  
 
It follows from Little's result that the expected
 number of peers in each of the boxes shown in Figure~\ref{fig:diagrama-1}  is given by the arrival rate to the box multiplied by
 the corresponding average residence time.  In particular, the expected number of gifted peers (resp., gifted peers with one remaining block to download)
  is $pU(K-2)/\mu'$ (resp., $pU/\mu'$).    Each of these peers serves the one club at rate $\mu$ (resp., $\mu'$). 
  Therefore, the departure rate from the one-club is $pU(K-2)\mu/\mu'+ pU$  (due to arrows $(c)$ and $(d)$ in Figure \ref{fig:diagrama-1}).
  
  Let $\Gamma$ be the system throughput.  Then, the above arguments  imply that 
  \begin{equation}
  \Gamma = \left(\frac{(K-2)\mu}{\mu'}+2\right) pU \label{eq:gammafirst}
  \end{equation} 
 We will revisit the result above in Section~\ref{bounds2}, in light of the queueing network model presented in the next section.  

\subsection{Queueing Network Model}
\label{sec:queueing-model}

Next, we introduce a queueing network model that captures the throughput achieved
by a number of different peer-to-peer swarming policies.  
The model is motivated by the flow dynamics shown in Figure \ref{fig:diagrama-1}.
\eat{ motivate the development
of the queueing network model presented next. }
Figure \ref{fig:mod-filas} illustrates  the  proposed 
 model accounting for a file with $K=3$ blocks (the extension to $K \ge 4$ blocks is straightforward). 
A fundamental  assumption is 
 that there is an infinite number of peers belonging to the \textit{one-club}.   



%
\begin{figure}[htb] \center
\includegraphics[width=0.85\columnwidth]{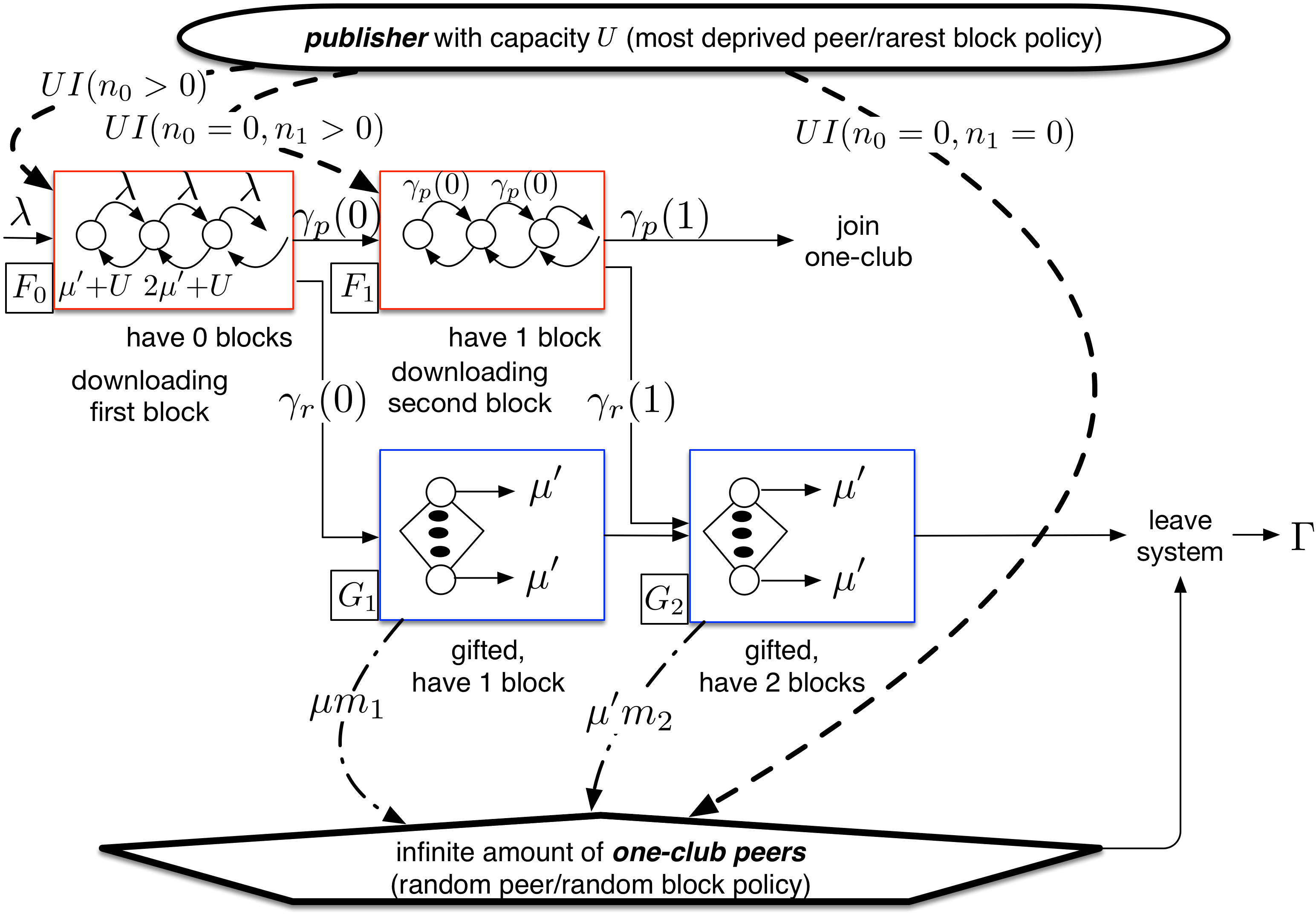}  \\
\caption{Queueing network model ($K=3$).}
\label{fig:mod-filas}
\end{figure}

Referring to Figure \ref{fig:mod-filas},
the  queues in the top characterize peers that do not have the rarest block. These queues, denoted by $F_0, \ldots, F_{K-2}$,  can be 
served both by the publisher and other peers.  
Each of theses queues is modeled as a birth-death process with constant arrival rate
and state dependent service capacity as explained below.  
The first queue (leftmost queue in Figure~\ref{fig:mod-filas}, queue $F_0$) represents newcomers.  Peers in
this queue have no blocks.  
The remaining  queues in the top  row represent peers that have already obtained $1,\ldots, K-2$ popular blocks,
and are waiting to complete the download of an additional block.

The bottom queues, denoted by $G_1, \ldots, G_{K-1}$, represent gifted peers. The  leftmost queue in the bottom row, $G_1$, 
represents all peers that downloaded only 
the rarest block.  The other queues at the bottom represents peers that  possess the rarest block along with $\ell$
popular blocks, $\ell=1,\ldots, K-2$ (queues $G_2, \ldots, G_{K-1}$, respectively).

Newcomers arrive with rate $\lambda$ and join $F_0$.
Since these peers have no blocks they can be served either by the publisher
(that preferentially serves peers with the least number of blocks) or by other peers.
Because the \textit{one-club} is large   by assumption and
encounters occur at random,
the set of all \textit{one-club} members
serves a block to a tagged newcomer at rate $\mu'$.

Let random variables $n_i$ and $m_j$ denote  the number of peers at queues $F_i$ and $G_j$, respectively, for $i=0,\ldots, K-2$ and $j=1, \ldots, K-1$. 
In particular, $n_0$ denotes the number of newcomers. 
 Let $\gamma_r(i)$ and $\gamma_p(i)$ denote  
the rate at which non-gifted peers that have $i$ blocks receive the rarest block and a popular block, 
respectively.  
From the above arguments, 
the rate at which newcomers receive the rarest block is \eat{from the birth-death process:} 
\begin{equation}
\label{eq:rate_gift}
\gamma_r(0) = U (1 - \pi_0(0)) ,
\end{equation}
where $\pi_{0}(0) = \mathbb{P}(n_0=0)$ is the probability that the newcomer queue  is empty.
Note that the rarest block can only be supplied by the publisher,
due to the assumption that the \textit{one-club} is large and the
random peer selection policy adopted by the peers.  Likewise, let $\gamma_p(0)$ denote the rate at
which newcomers obtain a popular block. Then,
\begin{equation}
\label{eq:rate_pop} 
\gamma_p(0) = \mathbb{E}[n_0] \mu' ,
\end{equation}
where $\mathbb{E}[n_0]$ is the expected number of  newcomers.

If a newcomer gets the rarest block (resp., a popular block) from the publisher (resp., from the \textit{one-club}), 
it transitions to queue $G_1$ (resp., $F_1$).
Since peers in $F_1$ have only one popular block, they are eligible
to be served by the publisher, provided that queue $F_0$ is empty (most deprived policy).
Similar to those peers in $F_0$, those in $F_1$ can also be served by the \textit{one-club}.

\textit{Gifted} peers in $G_1$ can only be served by \textit{one-club} members, since the
publisher gives preference to those peers that do not have the rarest block.  
Recall that $m_j$ is the number of gifted peers that have the rarest block in 
addition to $j-1$ blocks.   
Queue $G_j$ is then modeled as an M/M/$\infty$ queue with aggregate departure rate $\mu' m_j$, $j=1, \ldots, K-1$.
Once peers in $G_j$ obtain a popular block, they move to queue $G_{j+1}$ to get 
an additional (popular) block, for $j=1, \ldots, K-2$. 
Finally, after downloading the rarest block and $K-2$ popular blocks peers move to
queue $G_{K-1}$, which models the peers with $K-1$ blocks, one of them being the rarest.
After obtaining an additional block, they leave the system.

The average number of customers at queue $G_j$, $j=1, \ldots, K-1$, is given by
\begin{align}
\mathbb{E}[m_j]&= \sum_{l=0}^{j-1} \gamma_r(l) / \mu', \quad j=1,2, \ldots, K-1  \label{eq:meanm1} 
\end{align}
as queues $G_j$, $j=1, \ldots, K-1$, are M/M/$\infty$ queues with arrival rate $\sum_{l=0}^{j-1} \gamma_r(l)$ and 
mean residence time $1/\mu'$ (see Figure~\ref{fig:mod-filas}). 

If a peer in $F_{j-1}$ gets a block from the publisher, it moves to queue $G_j$.
Otherwise, it moves to $F_j$, for $j=1,\ldots, K-2$.

\paragraph{Publisher service rate approximation} We should note that the service capacity of queue $F_i$ ($i=1,2, \ldots, K-2$) depends on the states of queues
$F_0, \ldots, F_{i-1}$.
When queue $F_i$ is in state $s$, i.e., when there are $s$ customers at that queue, its service capacity  is given by $s \mu' + U I[n_0=0, \ldots, n_{i-1}=0]$,
where $I[\mathcal{E}]=1$ if predicate $\mathcal{E}$ is true and 0 otherwise. 
As an approximation,
we break this dependency by using the independent stationary value $\mathbb{P}(n_i=0)$ of each queue $F_l$, $l < i$.
Therefore, the aggregate  rate at which peers in $F_i$ are served is approximated as $\mathbb{E}[n_i] \mu'+ U \prod_{l=0}^{i-1} \pi_0(l)$.
From this, we obtain 
the rate at which peers in queue $F_i$ move to $G_{i+1}$:
\begin{equation}
\label{eq:rate_gift_j}
\gamma_r(i) = U \left( { \prod_{l=0}^{i-1} \pi_0(l) } \right) (1 - \pi_0(i)) ,
\end{equation}
where the product in parentheses is defined as $1$ when $i=0$.
The rate at which peers in $F_i$ move to $F_{i+1}$ is:
\begin{equation}
\label{eq:rate_pop_j}
\gamma_p(i) = \mathbb{E}[n_i] \mu'.
\end{equation}

\paragraph{Service rate at which one-club is served by gifted} Once the arrival rate at each queue is obtained we compute the rate at which 
the \textit{one-club} is served.
We observe that peers in queues $G_1, G_2, \ldots, G_{K-1}$ possess the
rarest block and, therefore, they can serve peers in the \textit{one-club}.  
In particular,
each tagged peer in $G_1, G_2, \ldots, G_{K-2}$  serves the one-club  at rate $\mu$. Peers in $G_{K-1}$ serve the one club at rate $\mu'$.
(Refer to the dashed lines in Figure \ref{fig:mod-filas}.)
Let $\Psi$ be the rate at which the \textit{gifted} peers serve the rarest block
to the \textit{one-club} members. Then, 
\begin{eqnarray}
\Psi &=&  \sum_{j=1}^{K-2} \mu \mathbb{E}[m_{j}] + \mu' \mathbb{E}[{m_{K-1}}],  \label{eq:psi11} \\
     &=& \sum_{j=1}^{K-2} \frac{\mu}{\mu'} \sum_{l=0}^{j-1} \gamma_r(l) + \sum_{j=0}^{K-2} \gamma_r(j), \label{eq:psi21} \\
     &=& \frac{\mu}{\mu'} \sum_{j=0}^{K-3} (K-2-j) \gamma_r(j) + \sum_{j=0}^{K-2} \gamma_r(j), \\
     &=& \frac{\mu}{\mu'} \sum_{j=0}^{K-2} \left( { (K-2-j) + \frac{\mu'}{\mu} } \right) \gamma_r(j) 
\label{eq:serv_one-club}
\end{eqnarray}
%
%
where $\gamma_r(j)$ is given by \eqref{eq:rate_gift_j}, and~\eqref{eq:psi21} follows by replacing~\eqref{eq:meanm1}  into~\eqref{eq:psi11}.

\paragraph{System throughput}
Let $\Gamma$ be the total rate at which peers leave the system (i.e., the system throughput).
Then,
\begin{align}
\label{eq:Gamma1}
\Gamma &= \sum_{i=0}^{K-2} \gamma_r(i) + \Psi + U \prod_{i=0}^{K-2} \pi_0(i).
\end{align}
The first term in the summation in~\eqref{eq:Gamma1} corresponds to the departure rate of the \textit{gifted} peers, 
  whereas the second and third terms correspond to the 
rates at which the \textit{one-club} is served by  \textit{gifted} peers and by the
publisher, 
respectively.

\subsection{Model Truncation} \label{sec:truncation}


To find the critical throughput $\Gamma_C$ using the model proposed in the
previous section requires the solution of a fixed-point problem encompassing
$2(K-1)$ equations (see Section~\ref{sec:meanfield}).    
%
%
To reduce the computational complexity when $K$ grows large,  in what follows
we consider an approach to truncate the model.
The proposed truncated model captures the evolution of peers while they download their  first few blocks 
before either entering the
one-club or getting the rarest block and then  downloading the remaining blocks.  
Model accuracy depends on the number of first few blocks considered.  
The accuracy increases with the number of blocks considered,
at the cost of increased computational complexity to solve the model.

Let $\tilde{K}$ denote the number of stages  taken into account when tracking the download of gifted peers.   The first stage (stage zero) corresponds to newcomers that have zero blocks, the $i$-th stage corresponds to peers that downloaded $i$ blocks and are downloading their $(i+1)$-th block, $i=1, \ldots,\tilde{K}-2$, and the last stage  corresponds to the download of the last $K-\tilde{K}+1$ blocks.   Non-gifted peers start from the same first stage, and evolve through additional $\tilde{K}-2$ stages before entering the one-club.  
In general, we use symbols with tilda to denote quantities related to the truncated model.

If $K=\tilde{K}$ the truncated model described below is equivalent to the
non-truncated model.
The truncated model comprises $2 (\tilde{K}-1)$ queues.
In the remainder of this paper, when $K \ge 4$ 
we consider an instance of the model which 
consists of six queues, i.e., $\tilde{K}=4$, and is illustrated in Figure~\ref{fig:mod-filastruncated}.    In our experiments this was sufficient to achieve accurate results.   

%

\begin{figure}[htb] \center
\includegraphics[width=0.85\columnwidth]{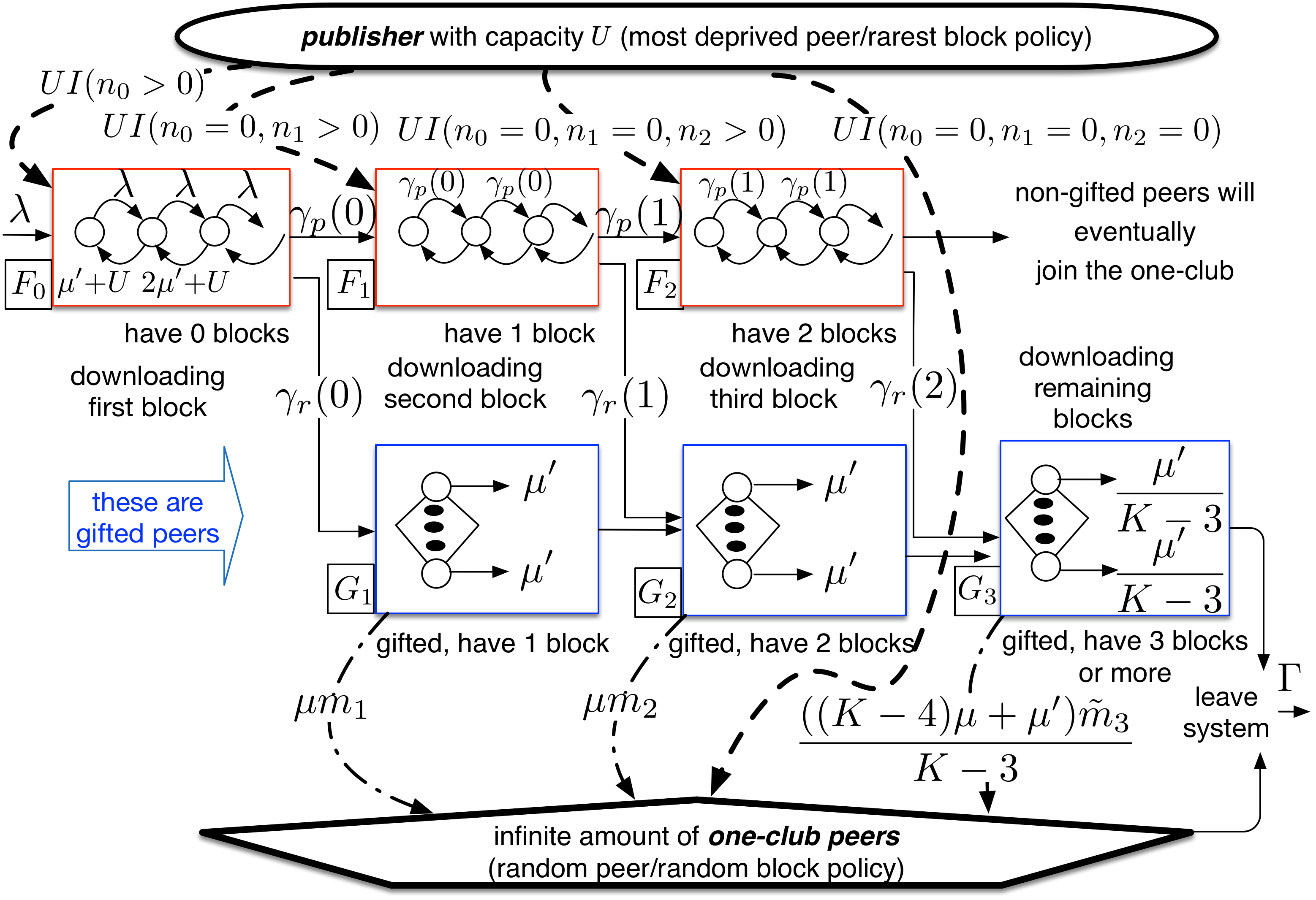}  \\
\caption{Truncated queueing network model ($\tilde{K} =4$ and $K \ge \tilde{K}$).}
\label{fig:mod-filastruncated}
\end{figure}

Let $F_i$ be the $i$-th queue containing  non-gifted peers, $i=0,1,2, \ldots, \tilde{K}-2$ and let $G_j$ be the $j$-th queue containing
gifted peers, $j=1,2,3, \ldots, \tilde{K}-1$.	Recall that $m_j$ denotes the number of gifted peers that have exactly $k$ blocks.  In the truncated model, we denote by $\tilde{m}_j$ the number of gifted peers in queue $G_j$, $\tilde{m}_j = m_j$, $j=1,\ldots,\tilde{K}-2$, and further   	let $\tilde{m}_{\tilde{K}-1}$ denote the number of gifted peers that have $\tilde{K}-1$ \emph{or more} blocks.  
The truncated model differs from the model presented in the previous section in two ways described below.


\textbf{$(a)$ Non-gifted peers with more than $\tilde{K}-2$ blocks are not
explicitly taken into account.} 
We assume that the probability that non-gifted peers with more than $\tilde{K}-2$ popular blocks obtain the
rarest block from the publisher is negligible.  
This is true since the publisher gives priority to peers with the least number of blocks. 
   Let $K \ge 4$. If a peer reaches $F_{\tilde{K}-2}$ and then
obtains one popular block,    it further downloads  additional $K-\tilde{K}$ popular blocks and 
 eventually joins the \textit{one-club}. 
Non-gifted peers with more than $\tilde{K}-2$ popular blocks are not explicitly represented in the truncated model. 
These peers have no effect on the model, since with high probability
they eventually join the \textit{one-club}, which is infinite in size.  

\textbf{$(b)$ Gifted peers with more than $\tilde{K}-2$  blocks are combined into a single class.} 
The rightmost  queue at the bottom of Figure~\ref{fig:mod-filastruncated}  
represents all peers that have the rarest block along with $\tilde{K}-2$ or more popular blocks.  
Since $G_{\tilde{K}-1}$ models peers that have $\tilde{K}-1$ or more blocks and because this is an M/M/$\infty$ queue,
the expected number of peers that have the \textit{rarest} block plus
$\tilde{K}-2, \ldots, K-2$ popular blocks are all identical. Recall that    $\tilde{m}_{\tilde{K}-1}$  is the number of peers at queue $G_{\tilde{K}-1}$ and 
\begin{equation}
\label{eq:meanmtilde}
\mathbb{E}[\tilde{m}_{\tilde{K}-1}] = \sum_{l=0}^{\tilde{K}-2} \gamma_r(l) \frac{K-\tilde{K}+1}{\mu'}.
\end{equation}
A fraction $1/(K-(\tilde{K}-1))$ of $\tilde{m}_{\tilde{K}-1}$ have $(K-1)$ blocks and serve the \textit{one-club} 
at rate $\mu'$ and
a fraction $(K-\tilde{K})/(K-(\tilde{K}-1))$ have $\tilde{K}-1, \ldots, K-2$ blocks and serve the \textit{one-club} 
at rate $\mu$. 


Recall that $\Psi$ is the rate at which the \textit{gifted} peers serve the rarest block
to the \textit{one-club} members. Then, following the same rationale as in~\eqref{eq:psi11}-\eqref{eq:serv_one-club},
\begin{eqnarray}
\Psi &=&  \sum_{j=1}^{\tilde{K}-1} \mu \mathbb{E}[m_{j}] + \frac{(K-(\tilde{K}+1))\mu + \mu'} {K-\tilde{K}+1} \mathbb{E}[\tilde{m}_{{\tilde{K}}-1}]  \label{eq:psi1a} \\
     &=& \sum_{j=1}^{\tilde{K}-1} \frac{\mu}{\mu'} \sum_{l=0}^{j-1} \gamma_r(l) + \frac{(K-(\tilde{K}+1))\mu + \mu'} {\mu'} \sum_{j=0}^{\tilde{K}-1} \gamma_r(j) \label{eq:psi2} \\
     &=& \frac{\mu}{\mu'} \sum_{j=0}^{\tilde{K}-2} (\tilde{K}-1-j) \gamma_r(j) + \frac{(K-(\tilde{K}+1))\mu + \mu'} {\mu'} \sum_{j=0}^{\tilde{K}-1} \gamma_r(j) \\
     &=& \frac{\mu}{\mu'} \sum_{j=0}^{\tilde{K}-1} \left( { (K-2-j) + \frac{\mu'}{\mu} } \right) \gamma_r(j) 
\label{eq:serv_one-cluba}
\end{eqnarray}
In what follows, we will use the expression of $\Psi$ derived above to assess system throughput.


\paragraph{System throughput}
The system throughput under the truncated model follows the same rationale as in~\eqref{eq:Gamma1}, and is given by  
\begin{equation}
\label{eq:Gammatruncated}
{\Gamma} = \sum_{i=0}^{\tilde{K}-1} \gamma_r(i) + \Psi + U \prod_{i=0}^{\tilde{K}-1} \pi_0(i).
\end{equation}
The rationale for the  terms in the equation above is presented below~\eqref{eq:Gamma1}.  

In what follows, we will use~\eqref{eq:Gammatruncated} to obtain bounds on the system throughput (Section~\ref{sec:trana}), 
as well as to numerically 
assess how the throughput varies as a function of different system parameters (Section~\ref{sec:results}).

\subsection{Model Summary} 

Table~\ref{tab:modelsum} summarizes the rates associated with each of the six queues that comprise the proposed model.
As shown in the third column of Table~\ref{tab:modelsum}, 
all queues have a component of their service capacity that scales with respect to the number of customers
 in that queue.  Such a component captures the self-scaling property of peer-to-peer swarming systems. 
   In addition, queues $F_j$, $j=0,1,2, \ldots, \tilde{K}-2$, also have a constant 
 service capacity component,  due to the publisher.
 
 
 As all queues  explicitly captured by the model have a self-scaling component,  they are by construction
 stable.  Nonetheless, the one-club population is  assumed to be 
  infinite in size and remains so 
 if the arrival rate is greater than the system departure rate.  
 For this reason, if $\lambda > \Gamma$ the system is unstable and peers experience infinite delays.  Otherwise, the one-club
 will eventually vanish and peers  experience finite delays.   These two regimes are further discussed in Section~\ref{overview}
 (see also  Appendix~\ref{app:opensys}).
 
 The third column in Table~\ref{tab:modelsum}  shows the departure rate at each of the possible
 states at each queue.  It accounts for the stochastic nature of the system. The fourth column,
 in contrast, presents only averages and  accounts for the simplifying assumption of independence 
between queues.  In the fourth column,   we replace the event  denoting that  queues  $F_0, \ldots, F_{j-1}$ are empty, 
$\cap_{l=0}^{j-1} \{n_l = 0\}$, by the corresponding product of probabilities $\prod_{l=0}^{j-1}\pi_0(l)$.
 
 Finally, note that equating the second and fourth columns of Table~\ref{tab:modelsum} the values
 of  $\mathbb{E}[n_j]$, $\mathbb{E}[m_j]$ and $\mathbb{E}[\tilde{m}_{\tilde{K}-1}]$ immediately follow.  In particular,
 from the last two lines we obtain \eqref{eq:meanm1} and \eqref{eq:meanmtilde}.

\begin{table}\begin{footnotesize}
\begin{tabular}{l|l|l|l}
\hline
queue & arrival rate & departure rate from state $s$ & average departure rate  \\
\hline
\hline
\multicolumn{4}{c}{Newcomers queue} \\
\hline
$F_0$ & $\lambda$ & $I(s>0)U + s \mu'$ & $(1-\pi_0(0))U + \mathbb{E}[n_0] \mu'$  \\
\hline
\multicolumn{4}{c}{Intermediary non-gifted peers queues ($j=1,2, \ldots, \tilde{K}-2$)} \\
\hline
$F_j$ & $\gamma_p(j)+(1-\pi_0(j))U\prod_{l=0}^{j-1} \pi_0(l)$ & $I(s > 0  \textrm{ and } \cap_{l=0}^{j-1} \{n_l = 0\}) U + s \mu'$ &   $(1-\pi_0(j))U\prod_{l=0}^{j-1} \pi_0(l) + \mathbb{E}[n_j] \mu'$  \\
\hline
\multicolumn{4}{c}{Gifted peers queues ($j=1,2, \ldots, \tilde{K}-2$)} \\
\hline
$G_j$ &  $\sum_{l=0}^{j-1} \gamma_r(l)$ & $s \mu'$ & $\mathbb{E}[m_j] \mu'$ \\ 
$G_{\tilde{K}-1}$  &   $\sum_{l=0}^{\tilde{K}-2} \gamma_r(l)$  & $s \mu'/(K-(\tilde{K}-1))$  & $\mathbb{E}[\tilde{m}_{\tilde{K}-1}] \mu'/(K-(\tilde{K}-1))$   \\
\hline 
\end{tabular}\end{footnotesize}
\caption{Queueing model summary} \label{tab:modelsum}
\end{table}

\section{System Throughput Analysis} \label{sec:trana}

The goals of this section are  to  a) obtain bounds on the system throughput, 
b) validate the  model introduced in the 
previous section,  and c)  indicate the potential
advantages of admission control.   
To these aims, we first analyze the model proposed in the previous section  subject 
to increasing arrival rates.
While the arrival rate is within the system capacity region ($\lambda \leq \lambda_C$),
peers experience finite delays (Section~\ref{bounds}).  
If the arrival rate is further increased, peers  experience infinite delays  and 
  $\Gamma \leq \lambda$ 
   (Section~\ref{bounds2}).


Then, to validate the proposed model, in Section~\ref{sec:markovian} we  introduce a finite state fixed 
population Markov model, corresponding to a detailed representation of the system. The fixed population model is amenable to exact numerical solution for small  populations (e.g.,  swarms with up to  35 peers are considered in Section~\ref{sec:illustrfix}).  
In Section~\ref{sec:results} we will use this Markov model  to numerically validate the throughput
predicted by our proposed model, comparing  predicted results  against those obtained through  
exact numerical solution of the fixed 
population Markov model.  In addition, in Section~\ref{sec:results} using this model we  will observe that throughput initially increases and then decreases, as the population size grows, motivating the benefits of admission control.   

\subsection{Overview}  \label{overview}

Table~\ref{sec:regimes} summarizes  different system regimes  captured by the open queueing model.
When $\lambda=0$,  the six queues in the open queueing model are empty. The 
assumption that the system is initialized with a very large one-club implies that the throughput equals  the publisher 
service capacity, $\Gamma=U$, and
the one-club will eventually vanish.   

As $\lambda$ increases,  throughput increases until 
$\Gamma=\lambda=\lambdacritical$.  Although we are not able to prove the existence and uniqueness of the critical throughput value,
 we observed a unique value  of $\lambda$, $\lambda=\lambdacritical$,
 in   all the numerical experiments, for which the queueing model yields
$\Gamma=\lambda$. 
 The throughput $\Gamma_C = \lambda_C$  approximates  the behavior of the system with a fixed population size, wherein a new arrival
 occurs immediately after every departure. 
 
 Referring back to Figure~\ref{fig:limited-thr}, when $\lambda \leq \lambdacritical$ the curve shows that  system throughput
 equals its arrival rate. This is because Figure~\ref{fig:limited-thr} shows  throughput due to arrivals of newcomers.  
 Nonetheless, as the proposed model  accounts for a large one-club, and the dynamics of the one-club is out of the scope of the model, at steady state we may have $\Gamma > \lambda$.  
 If we also account for the contribution of the  initially large one-club towards the throughput, the throughput 
 curve  
remains above the line $x=y$  up to reaching the point $\lambda=\lambdacritical$ (see Appendix~\ref{app:opensys}).

It follows from the discussion above that when  $\lambda \leq \lambdacritical$, we have $\Gamma \ge \lambda$, which implies that 
all arrivals are served and additional 
  departures occur from the one-club.
  As our model predicts that the one-club eventually vanishes in this regime, we refer to it as \emph{finite delay regime}.
  In contrast, when $\lambda  > \lambdacritical$,  we have $\Gamma \le \lambda$,  implying that 
  some  arrivals  join the one-club,  which in this case grows unboundedly. For this reason, we refer to this 
  regime as \emph{infinite delay regime}.  If the arrival rate is further increased, the system fully saturates and the  throughput never surpasses  $\Gamma_S$.      
   Additional  numerical results illustrating the different system regimes are presented in Appendix~\ref{app:opensys}.

\begin{table}
\begin{tabular}{lll}
\hline
regime & throughput & description \\
\hline
$\lambda=0$ & $U$ &  all departures occur from the one-club (which eventually vanishes) \\
$\lambda < \lambdacritical$  & $\ge \lambda$ & all arrivals are served and additional \\
& &  departures occur from the one-club (which eventually vanishes) \\
$\lambda = \lambdacritical$ & $\Gamma_C=\lambdacritical$ & critical regime, which approximates \\
& & the behavior of the system with a fixed population size \\
$ \lambdacritical < \lambda < \lambdasaturation$ & $\leq \lambda$ & saturated system, wherein new arrivals  experience infinite delays \\
& & (and one-club grows unboundedly). Note: $\lambdasaturation \ge \Gamma_S$ \\
$ \lambda \ge \lambdasaturation$ & $\Gamma_S$ & $\Gamma_S$ is the maximum throughput achievable by the open system \\ 
\hline
\end{tabular}
\caption{Overview of system regimes as captured by the open queueing model.} \label{sec:regimes}
\end{table}

Figure~\ref{fig:diagram} complements Table~\ref{sec:regimes} and summarizes the 
two operating regimes observed in  P2P systems
of interest to us.  

\begin{figure}[h!]\center
\includegraphics[width=\columnwidth]{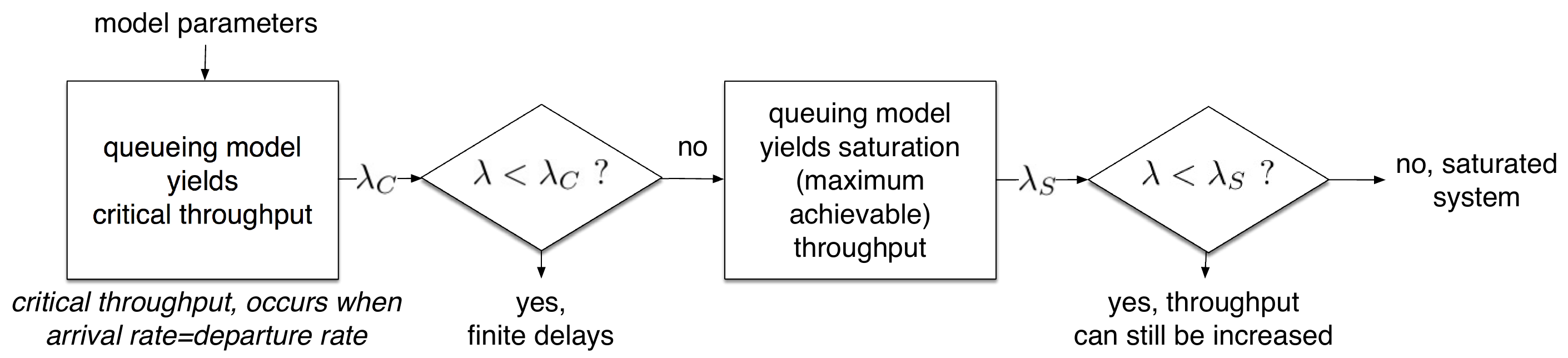} 
\caption{System throughput regions: (a) when $\lambda < \lambdacritical$, finite delays (model assumes $\infty$ one-club); (b) $\lambdacritical < \lambda < \lambdasaturation$, infinite delays (model assumes $\infty$ one-club, and a server always busy serving newcomers).}  \label{fig:diagram} 
\end{figure}

\subsection{Critical  Throughput $\Gamma_C$}  \label{bounds}

 Recall that the critical throughput    $\Gamma_C = \lambdacritical$ is defined such that for
  $\lambda  \leq \lambdacritical$
   system departure rate equals its  arrival rate, i.e., $\Gamma=\lambda$, accounting  for throughput due to arrivals of newcomers.  
We formulate and solve a fixed point problem to obtain $\Gamma_C$.  The iterative algorithm is described in Appendix~\ref{sec:appiterative}.


To validate the critical throughput obtained using the proposed model,  
we introduce 
a detailed fixed size population model in Section~\ref{sec:markovian}.  
Under such a closed model,   every departure is immediately  followed by an arrival, which corresponds to letting 
the departure rate equal the arrival rate in the open population model.   We numerically compare the results 
obtained using the two models in Section~\ref{sec:results}.


\subsection{Saturation  Throughput $\Gamma_S$} 

\label{bounds2}

Next, we use the proposed  model to estimate the 
throughput when the publisher is able to use all
its bandwidth on newcomers and is always busy. 
We derive the saturation throughput using the queueing model,  and present
 two  alternative derivations to obtain the same result in Appendix~\ref{sec:additionalderivs}.

\subsubsection{Queueing Network Model} \label{sec:qnetmode}

\begin{propo}
\label{prop:propo1}
When the server adopts the MDP/RFB policy, peers adopt the RP/RUB policy and
all the publisher   service capacity is devoted to newcomers, 
the system throughput is limited by
\begin{equation}
\label{eq:open-lim} \Gamma_S =\left(\frac{(K-2) \mu}{\mu'}  + 2\right)U
\end{equation}
\end{propo}

\emph{Proof: } We use the queueing network model to obtain \eqref{eq:open-lim}.  We consider a saturated system,
wherein  the service capacity of the publisher is devoted to newcomers.  
Therefore, $\gamma_r(0) = U$, and $\gamma_r(i) = 0$ for $i=1,2, \ldots, K-2$.
This means that peers in $F_1, F_2, \ldots, F_{K-2}$  will join the \textit{one-club}.
In addition, the arrival rate to queues $G_1, G_2, G_3, \ldots, G_{K-1}$  is $U$, since the
server only serves  newcomers.  Then,  \eqref{eq:serv_one-club} reduces to
\begin{equation}
\Psi = \frac{\mu}{\mu'} (K-2) U + U  \label{eqgammnew}
\end{equation}
Substituting~\eqref{eqgammnew} into \eqref{eq:Gamma1} yields
\begin{equation}
\Gamma_S = \frac{\mu}{\mu'} (K-2) U + 2U = \left(\frac{\mu}{\mu'} (K-2) + 2\right) U 
\end{equation}
Note that $\pi_0(0)$ can be made arbitrarily  close to zero by increasing the arrival rate $\lambda$.  Therefore,
 for a large enough $\lambda$ all assumptions of the proposition hold, and  
  the result follows. $\Box$

%

\subsubsection{Discussion}


\begin{itemize}

\item Zhou and Hajek \cite{hajek} showed that the limiting throughput is $U$ for
the random peer/random useful block selection policy.
Proposition \ref{prop:propo1} indicates that, if the publisher adopts the most deprived peer and 
rarest first block policy,  the largest peer arrival rate that the system can support is $(((K-2)\mu/\mu') + 2)U$. 
That is, with a minor change in the publisher policy, the achievable throughput can be 
significantly increased.

\item The larger the number of blocks the larger is the maximum achievable  throughput, which 
 quantifies the advantage of dividing a file into small blocks. 
However, one cannot reduce the block size towards zero, due to  inherent overhead introduced
by headers associated with each block.  Future works consists of using Proposition~\ref{prop:propo1} to cope with
the trade-off between decreasing the block size and
increasing the overhead of headers and other control data that is transmitted per block.
\eat{\textcolor{red}{I added this sentence. It is clear what to do, since this is similar
to the optimum packet size problem of the 80's, but I have not done it. Please see
what you think.}} 
\item The proposition states a counter-intuitive result.
If peers \textit{reduce} their upload rate once they obtain all but one of the blocks,
the system throughput increases inversely to $\mu'$.  
Reducing $\mu'$ saves peer bandwidth, increases throughput and is incentive-compatible. 
Note that equation \eqref{eq:open-lim} does  not impose any limit to $\mu'$.
If $\mu'$ decreases towards zero, at some point the throughput may decrease. 
\eat{In particular, if $K=2$ and $\mu'=0$ the system degenerates to a client server model
since only the publisher will be able to serve.} 
In Section~\ref{sec:results}, we indicate through a few examples 
that $\mu'$ can be made more than one order of magnitude smaller than $\mu$, which is sufficient for
large performance gains.
\eat{\textcolor{red}{PLease check and modify. You may cut this last comment if it is not clear.}}
\end{itemize}

\subsection{Evaluating the Throughput Attained by  Fixed Size Populations}
\label{sec:markovian}

Next, we develop a detailed Markov model that allows us to obtain throughput as a function
of the number of peers in the swarm for different server and peer policies (Table \ref{tab:policies}).
%
%
The model implements  fundamental characteristics of the policies studied
here and is used for validation purposes to support the findings of this work.
Due to the size of its state space, solving it numerically
is limited to a relatively small number of blocks and moderate population sizes.
Nevertheless, the results clearly support our main conclusions. \eat{ and provide the basis
for an approximation that we propose in Section \ref{sec:queueing}.   }
Next, we present an overview of the Markov model.  For a detailed description  
see Appendix~\ref{app:moddetails}.

We model a fixed size population system (closed system).
As such, 
whenever a peer leaves the swarm, a newcomer immediately arrives.
Therefore, it approximates the behavior of swarms whose populations are approximately
constant and allows us to study  system bottlenecks as population size grows.
Although closed and open systems exhibit different behavior (e.g., ~\cite{adam}), 
the closed system \eat{is helpful to give}  provides insights into the stability region of the corresponding
open system.  In addition, the closed system can be used to assess the asymptotic throughput of real
swarms, through controlled experiments~\cite{diego}.  


Recall that $N$ is the number of peers in the swarm.
%
A na\"ive choice of state description consists of  tracking the identities of the blocks that each peer possesses.
Clearly, this leads to a state space explosion and is intractable even for
small peer populations and numbers of blocks. 
However, the model has symmetries that can be leveraged  to  
\textit{lump} the state space (see Appendix~\ref{app:lumped}).
Let $\Omega$ denote the model state space.
One such symmetry allows us to characterize each state  
$\bsigma \in \Omega$ as a vector $\bsigma = (\sigma_1, \ldots \sigma_M)$
where $\sigma_i$ is the number of peers with signature $i$ and $M=2^{K}-1$ 
is the total number of signatures.
Other symmetries allow for additional reductions of the state space cardinality of about one order
of magnitude.
Due to space limitations we briefly discuss such less evident symmetries, using a simple example.
Assume the file contains $3$ blocks, namely $\mathcal{B}_1$, $\mathcal{B}_2$ and $\mathcal{B}_3$.
Each of these blocks may become the rarest. 
Suppose that $R$ peers have all blocks except block $\mathcal{B}_1$ and $N-R$ peers have only the rarest block.
It is not difficult to verify that this state can be lumped with similar states in which the
rarest block is $\mathcal{B}_2$ or $\mathcal{B}_3$.
Our Markov model takes into account  states that can be lumped before their generation  (see Appendix~\ref{app:lumped}).

\subsubsection{Illustrative Example} \label{sec:illustrfix}

Figure \ref{fig:comp_polices} plots the throughput obtained from the detailed Markov model for different publisher
and peer policies in Table \ref{tab:policies}.\footnote{The source code  to reproduce the plots presented in this paper,  implemented using the Tangram II tool~\cite{tangram09}, is available at \url{https://tinyurl.com/p2pscale}.
}
Although not all possible combinations of policies  in Table~\ref{tab:policies} are shown in  Fig.~\ref{fig:comp_polices},  
\eat{and only a small set of parameter values are presented in Fig.~\ref{fig:comp_polices} to keep it readable,}
 the conclusions we obtain from the figure are valid in a broader set of settings.
In Figure~\ref{fig:comp_polices} we let $K=3$, $U=1$ and $\mu=\mu'=0.5$.  

In what follows, we assume that peers leave the system
immediately after completing their downloads.  Appendices~\ref{app:lingering} and~\ref{app:additionalpolic} consider the case where peers remain in the system
as seeds after download completion.

%
\begin{figure}[htb] \center
 \includegraphics[scale=0.6]{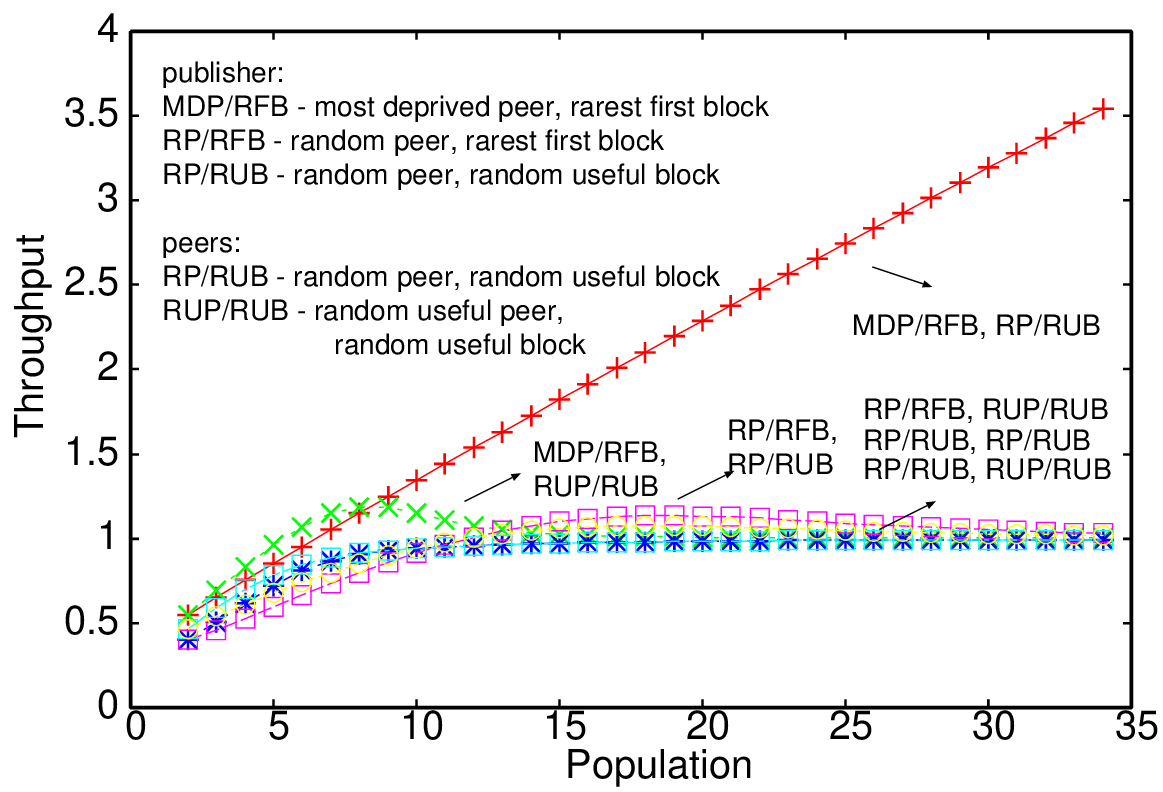}  \\
\caption{Comparing different policies under the fixed-size population model.}
\label{fig:comp_polices}
\end{figure}
From Figure~\ref{fig:comp_polices} it is evident that the largest throughput values are obtained when
the publisher adopts the MDP/RFB and peers use the RP/RUB policies, for moderate to high swarm population sizes.
This is true for a wide range of parameter values.  
Therefore, we choose to focus our results on this policy in Section~\ref{sec:results}.
The advantages of the MDP/RFB policy adopted by the publisher will be quantified
in what follows.

Figure \ref{fig:comp_polices} shows, in particular, that the adoption of RP/RUB by peers leads to larger throughputs 
than RUP/RUB.  If peers adopt the random useful peer policy (RUP),
 one-club members  prioritize the service of 
 gifted peers.   This, in turn,  leads to  a reduction of the mean residence time of the latter, and a corresponding
  system throughput reduction.

Another important feature shown in Figure~\ref{fig:comp_polices}  concerns the maximum throughput  achievable in  
the  closed system.  Note that the throughput of MDP/RFB, RP/RUB surpasses the threshold $\Gamma_S= KU$.  
This is because the boundary conditions imposed by the closed-system naturally prevent the one-club to grow, which in some settings
translates into increased block diversity and a consequent increase in throughput.   Such observation 
is well aligned with  
cautionary tales discussed in~\cite{adam}, which should be kept under consideration when contrasting open and closed systems.  
In summary, the closed system is instrumental
to validate the critical throughput $\lambdacritical$, as we will show in the upcoming sections, as well as to motivate the
study of admission control, as considered in Section~\ref{sec:results}. In what follows, we also indicate its applicability
to  the transient analysis  of the system.

\subsubsection{Transient Analysis}  \label{sec:transientanaly}

Next, we provide insight into transient system behavior.
To this aim, we consider the  fixed population model under the same setup as described in the previous section.
We consider a file with size $K=3$ blocks and a population of 15 peers. 
Assume that initially all peers have no blocks.   Peers adopt the RP/RUB policy, whereas the publisher policy is varied
according to the experimental goals.


Let the random variable $T$ denote  the time it takes until $90$\% of the peer
population has all  blocks except the rarest.
Figure~\ref{fig:transient}(a) shows the cumulative distribution function (CDF) of $T$.
We see that \eat{, at $t=50$ } $\mathbb{P}(T \leq 50) > 0.9$ for all the policies considered.  
For comparison purposes, we note that by $t=50$ we observed in our numerical experiments that 
an average of approximately $25$ (resp., $50$) peers left the system, when $U=0.5$ (resp., $U=1.0$).

\begin{figure}[htb]
   \centering
\includegraphics[scale=0.45]{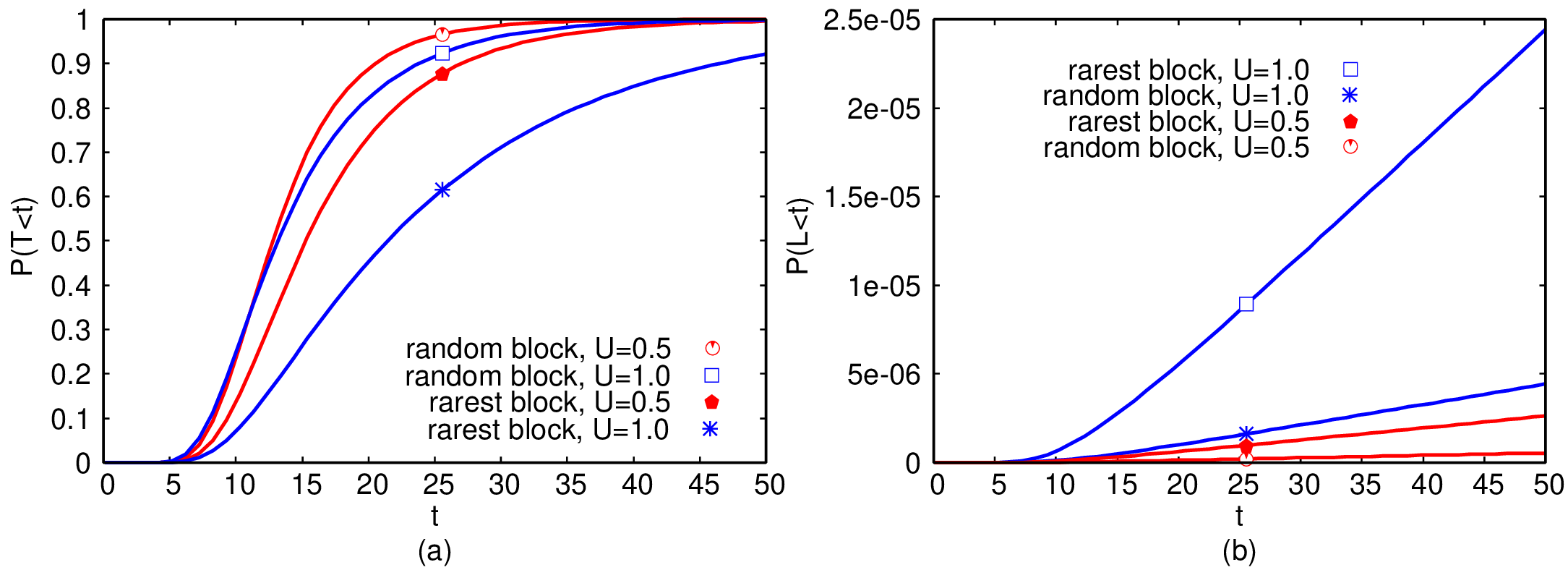}
   \eat{ \subfloat[\ ]{
   \includegraphics[scale=0.35]{figs/transient-1}
   \label{fig:transient-1}
   }
   ~
   \subfloat[\ ]{
   \includegraphics[scale=0.35]{figs/transient-2}
   \label{fig:transient-2}
   } } 
\vspace{-0.25in}
\caption{Peers evolving with time in a swarm: distribution of time until $(a)$ all peers enter the one-club and $(b)$ half population leaves the one-club.  The publisher selects peers uniformly at random and its block selection policy varies. Peers adopt the random peer/random block policy ($\mu=1.0$).}
\label{fig:transient}
\end{figure}

Once the system reaches the 
 state where most peers have all but the rarest block,
it takes  considerable \eat{amount of } time to leave that state.
This is shown in Figure~\ref{fig:transient}(b).
We assume that at $t=0$ all peers have all blocks but the rarest (except a single
peer that has no blocks).
Let  random variable $L$ denote  the time it takes until $50$\% of the peers leave
the one-club, $i.e.$, until less than 8 peers have all blocks except the rarest.
Figure~\ref{fig:transient}(b) 
shows the CDF of $L$.
At $t=50$, the probability that less than 8 peers have all blocks except the rarest 
is less than $10^{-4}$ for all policies considered. 

\begin{figure}[h!]
\hspace{-0.23in} 
\includegraphics[scale=0.28, angle=-90,origin=c]{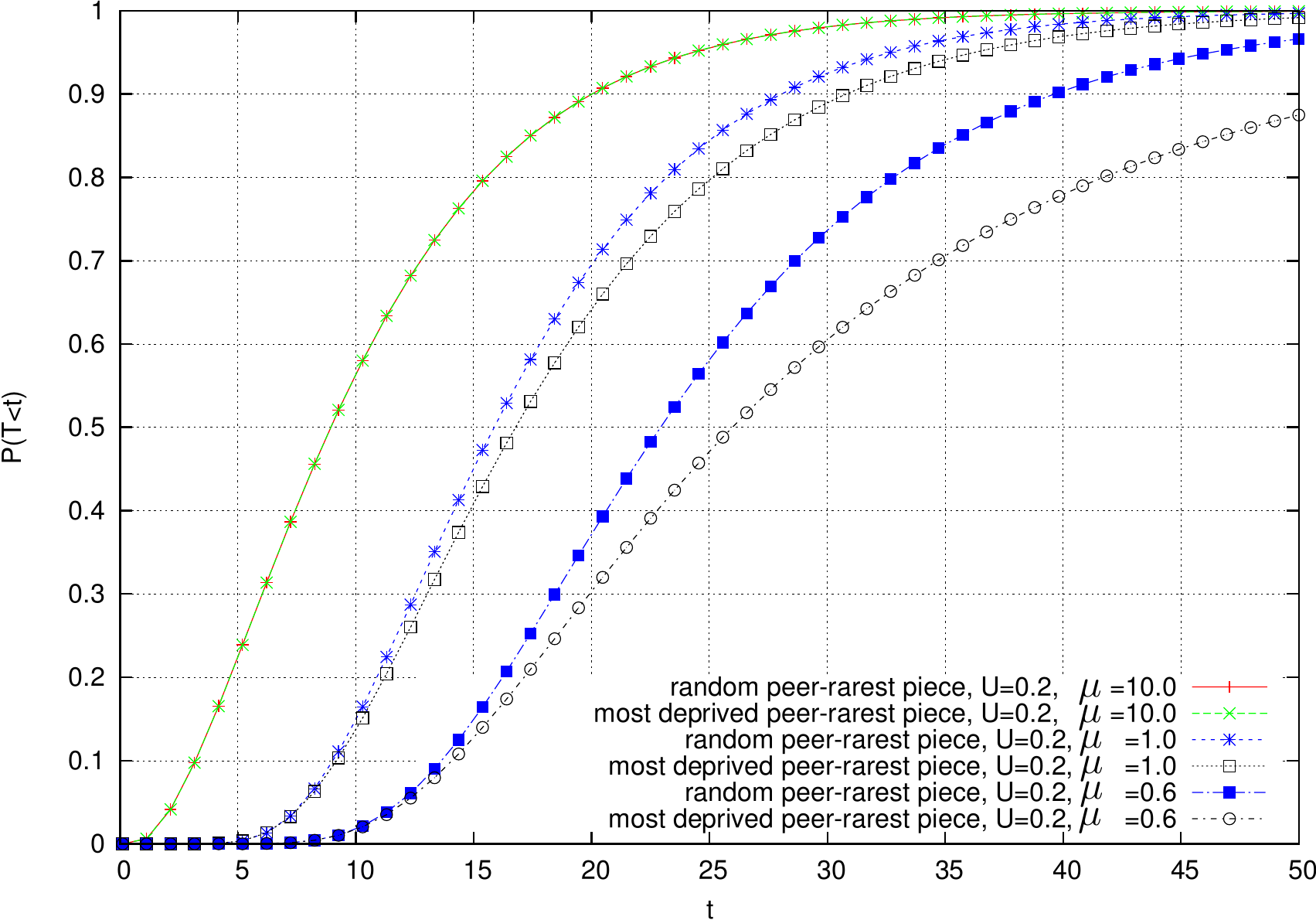}  
\vspace{-0.45in}
\includegraphics[scale=0.28, angle=-90,origin=c]{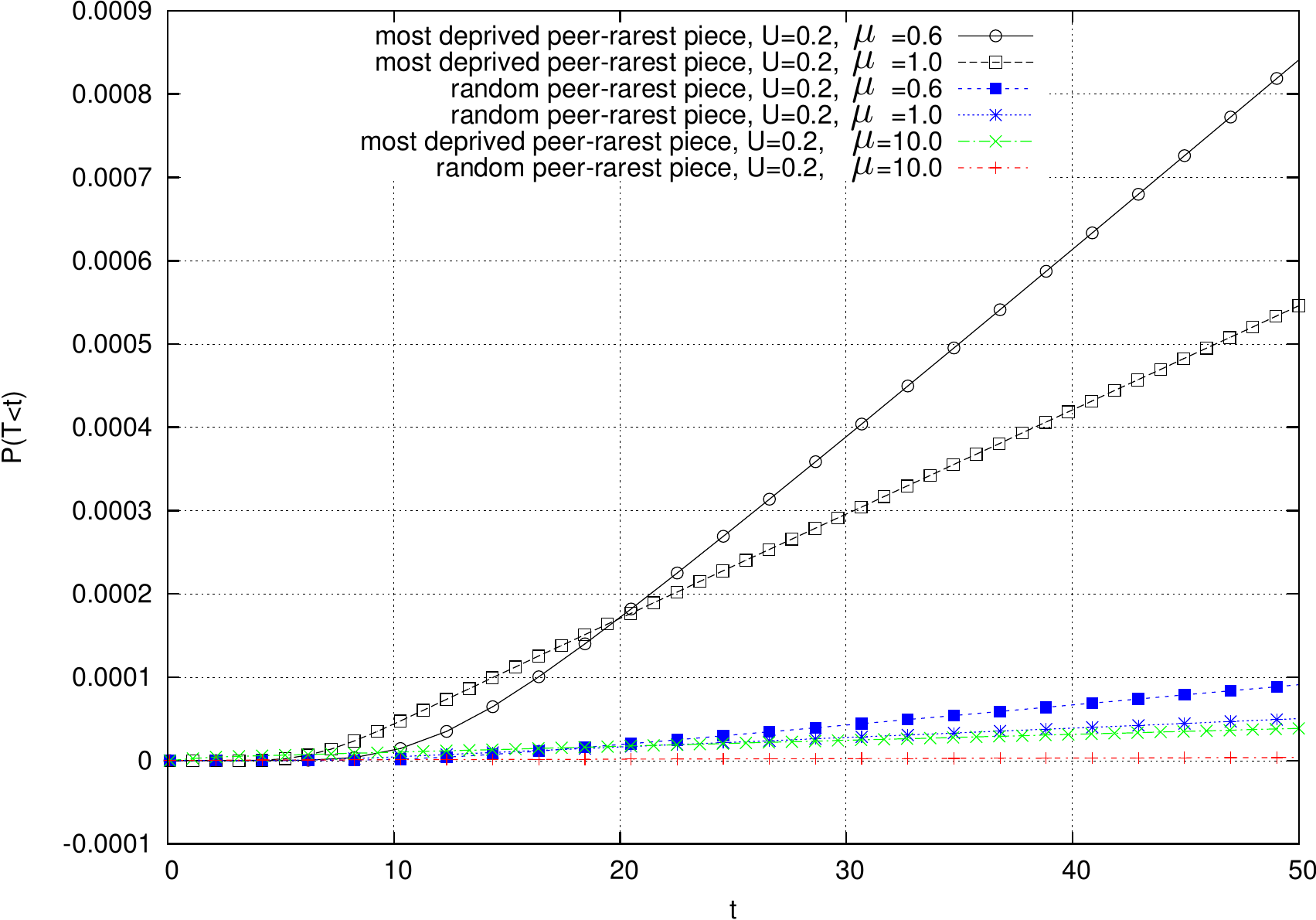}  
\begin{tabular*}{1.0\textwidth}{@{\extracolsep{\fill} }cccc}
& (a) time to enter one-club \quad &\quad  (b) time to leave one-club  & \\
\end{tabular*}
\caption{Transient analysis with most-deprived peer selection.}
\label{figmostdepr}
\end{figure}

In Figure~\ref{figmostdepr} we illustrate  gains obtained when the publisher adopts the 
most-deprived peer policy.
The larger the difference between $U$ and $\mu$, the more significant are 
the gains of using the most-deprived peer policy as opposed
to the random peer selection. 
This is expressed both in terms of  time to enter  a state where 90\% of the population 
is in the one-club  starting from
an empty system (Figure~\ref{figmostdepr}(a)), as well as the time to leave the state where all peers are in the one-club, 
and reach a state where less than half 
of the population has all blocks except one (Figure~\ref{figmostdepr}(b)).

Using the fixed size population model, we are able to compute transient metrics which would otherwise  be very costly to
 obtain using  an open model.  The larger state space cardinality of the open model, together with the fact that the system may
  be unstable under the settings of interest,  are two of the challenges that  in practice allow the open
   model to only be solved 
  through simulations.  With the fixed size population model, we were able to numerically solve the Markov model and
    obtain the transient results 
presented in this section.


\section{Numerical Results}
\label{sec:results}

In this section we present numerical results obtained with the  queueing network model and 
the detailed Markov 
model.   The  results presented in this section complement emulations with real BitTorrent clients present in~\cite{diego},
which showed
that the missing piece syndrome can occur in practice, and that the assessment of  throughput must account for
it. 
Our goals are to:
$(a)$ use the detailed Markov model to validate the queueing network model;
 $(b)$ show how the throughput varies with different model parameters and 
$(c)$ motivate the proposed server and peer policies.\footnote{Appendix~C contains additional results about peers that reside in the system after 
completing their downloads.}


\subsection{Validating Critical Throughput $\lambdacritical$} \label{sec:validation}

We first  use the detailed Markov model to validate the critical throughput $\lambdacritical$ estimates obtained
from the  queueing network model. 
We consider two scenarios, $U \le \mu$ and $U > \mu$.
In what follows, we let $K=3$ and $\mu=\mu'$. 


In Figures~\ref{validating}(a) and~\ref{validating}(b) we consider  scenarios where $U \le \mu$
and $U > \mu$, respectively.   
In these figures we present throughput results obtained using three approaches: 
$(a)$   numerical solution of the detailed  Markov model  for  population sizes varying from 2
to 30 users, using the GTH solution method~\cite{grassmann1985regenerative}, $(b)$ simulations of the detailed  Markov  
model for  population sizes ranging from 2 to 250 users, and $(c)$  asymptotic throughputs computed using the 
queueing network model.
As a sanity check, we verify that  the solutions obtained  using the GTH method  
lies within the confidence interval of the simulations.
For all values of $U$ and $\mu$, as population increases,
 throughput increases until it reaches a maximum  
and then it decreases and tends approximately to  $\lambdacritical$.
From the figures we observe that the queueing model presented in  
Section~\ref{sec:models} provides a good estimate of $\lambdacritical$,  
the maximum relative error 
 being equal to 10\%. 




\begin{figure}[htb]
\hspace{0.1in}
\includegraphics[width=0.47\textwidth]{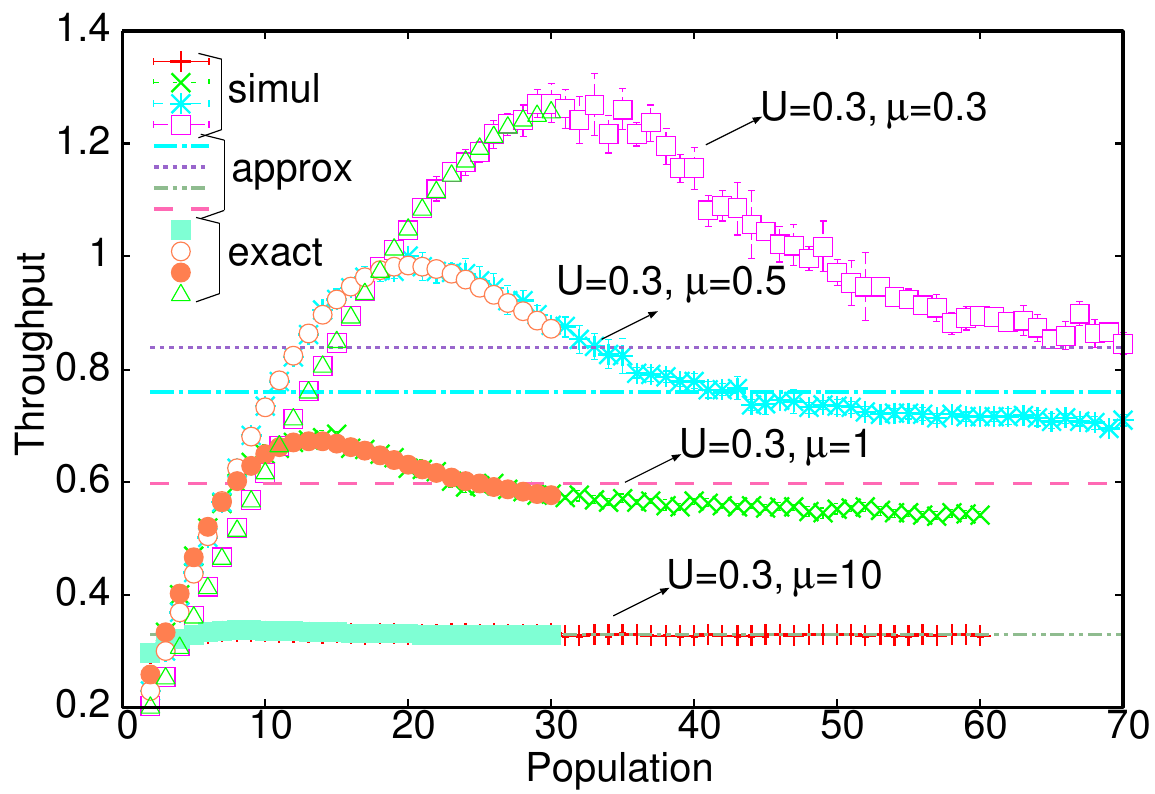} 
\hspace{0.1in}
 \includegraphics[width=0.47\textwidth]{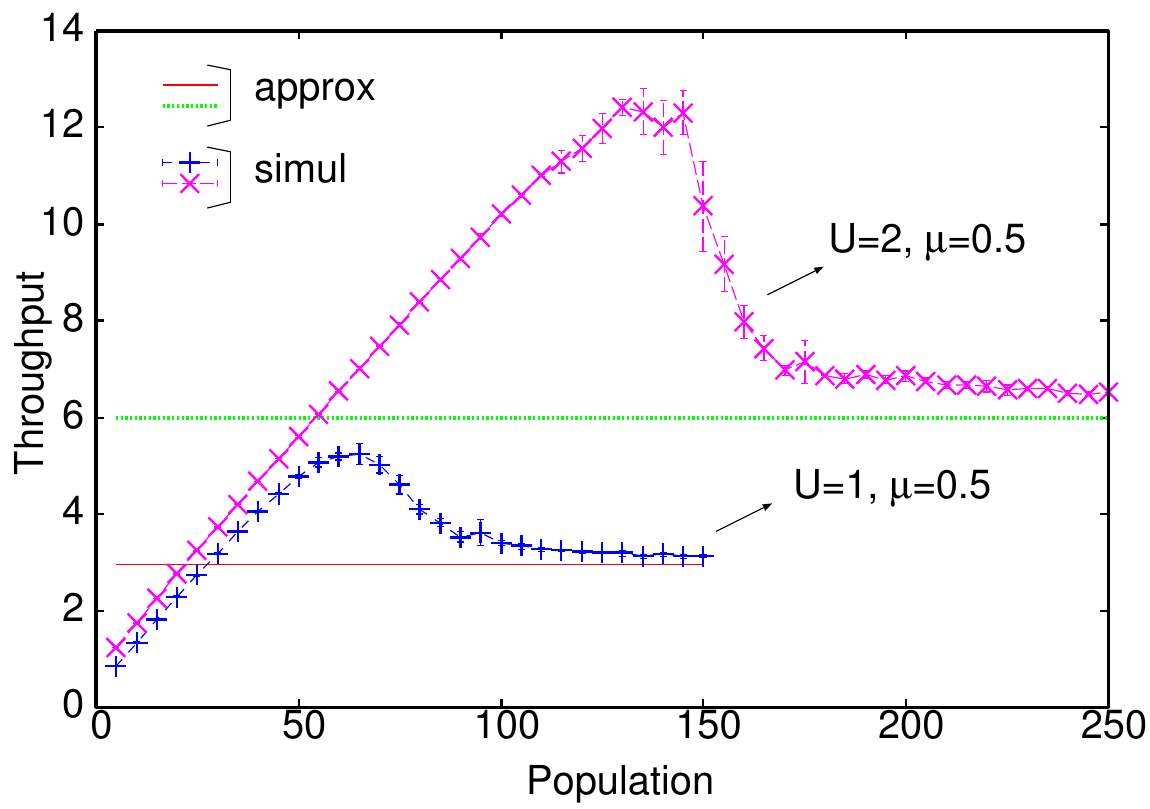} 
 \hspace{0.1in}
\center
\begin{tabular*}{0.8\textwidth}{@{\extracolsep{\fill} }ll}
$\quad \quad $ (a)  $U \le \mu$ ($U=0.3$) & (b)  $U > \mu$ ($\mu=0.5$)  
\end{tabular*}
\caption{Validating the queuing network model, $K=3, \mu=\mu'$. } \label{validating}
\end{figure}

\subsection{Impact of Varying Service Capacity of the Server}

Figure~\ref{throughputKU} illustrates how throughput varies with 
$K$ and $U$.  
These results were obtained from the queueing network model of 
Section~\ref{sec:models}, letting $\mu=\mu'=1$.  
The figure shows that the model captures the fact that the throughput 
is approximately linear in $K$ and $U$, for constant $\mu$.
This occurs because the majority of the server transmissions are for 
newcomers.  
Each newcomer that is turned into a gifted peer serves, on average, $K-1$ one-club peers before
leaving the system. 
Thus, for each block served by the publisher, roughly $K$ peers leave the system 
($K-1$ from one-club plus one gifted).     
As the server rate is $U$ blocks per time unit, the system
throughput grows linearly with $K$ and $U$.  

\begin{figure}[htb]
\hspace{-0.3in}
\centering
\includegraphics[scale=0.4]{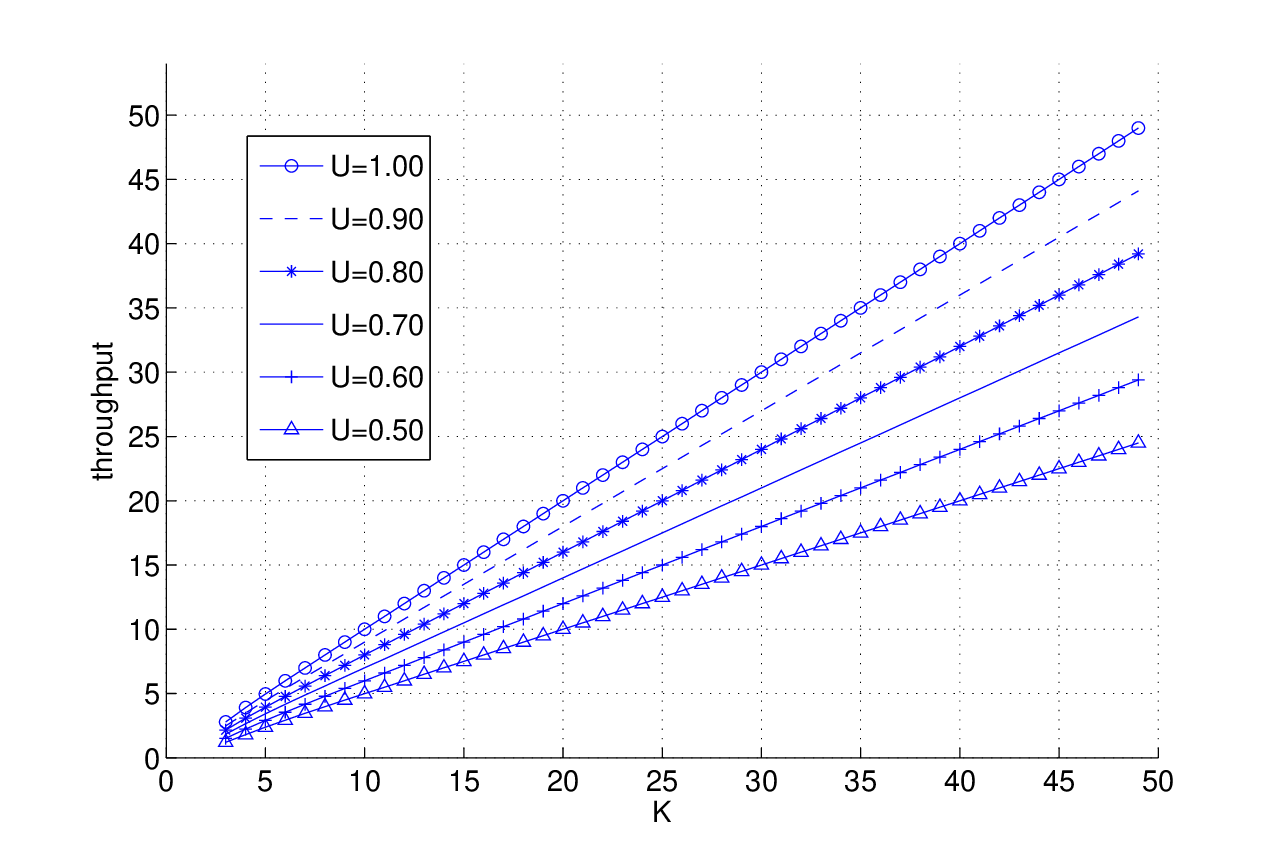}
\hspace{-0.3in}
\includegraphics[scale=0.4]{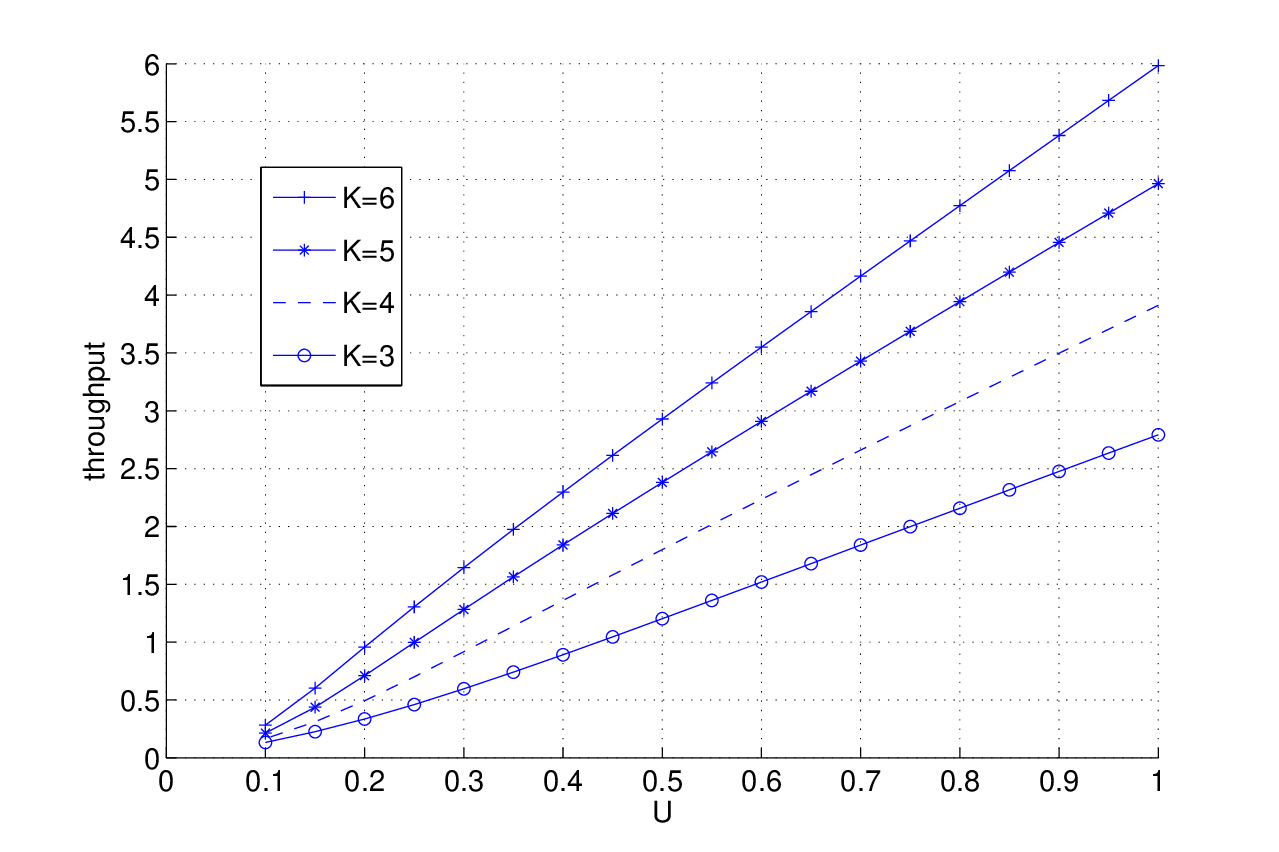}  
\center
\begin{tabular*}{1.0\textwidth}{@{\extracolsep{\fill} }llll}
& (a) Throughput as a function of $K$ & (b) Throughput as a function of $U$  &
\end{tabular*}
\caption{The throughput varies linearly with $U$ and $K$ ($\mu=\mu'=1$). } 
\label{throughputKU}
\end{figure}

\subsection{Uplink Throttling Alleviates the Missing Piece Syndrome}
\label{sec:reduced-rate}

In this section we study system throughput when peers that have collected 
all blocks but one  
serve with rate  $\mu'$, where $\mu' < \mu$. 
\eat{The one-club peers only serve the gifted peers.} 
The motivation to decrease the upload rate of the one-club peers is 
to keep the gifted peers in the system longer serving the rarest block. 
This policy is motivated by the results presented in 
Section~\ref{sec:dynamics}, which   indicate that the throughput can increase if
the upload rate of the one-club peers decreases. 

Figures~\ref{fig:modif1}  and~\ref{fig:modif2}  show the increase in throughput 
when the one-club peers reduce their upload rate from $\mu$ to $\mu'$.
The figures show that throughput increases as  $\mu'$ decreases.
The proposed policy is very simple and  incentive-compatible,  allowing \eat{is strongly in accordance with 
the incentive of} peers to increase system performance and save bandwidth.

Figure~\ref{fig:modif2}  indicates that  the queueing network model  predicts an unbounded   throughput increase  as the upload
rate of  peers that have all blocks except one decreases, i.e., as 
$\mu' \rightarrow 0$.   
This is because the queueing network model 
assumes an infinite one-club.  
As $\mu'$ decreases, gifted peers  remain longer in the system,  contributing to an increase of the departure rate from the infinite one-club.   
Nonetheless, if we consider finite populations  
 the throughput decreases as  $\mu'$ approaches zero. 
In particular, if  $K=2$ and $\mu'=0$, the throughput
 equals  $U/2$, since all blocks will be served only 
by the server (the system degenerates to a client server system).

The results presented so far were obtained using the  queueing network model.   
Figure~\ref{fig:modif2}(b), obtained using the detailed Markov model of the closed-system, 
 shows how   throughput varies as a function of  the population size. 
We note that the throughput is significantly larger when $\mu'<\mu$ 
for a population greater than 10 users,
and as the population grows the throughput gains   increase. \eat{
As a general rule, it is quite advantageous to reduce the upload rate of the one
club peers because it increases significantly the stability region of the system.} 


\begin{figure}[htb]
\centering
\subfloat[\ $\mu=1, U=1$]{
\includegraphics[scale=0.4]{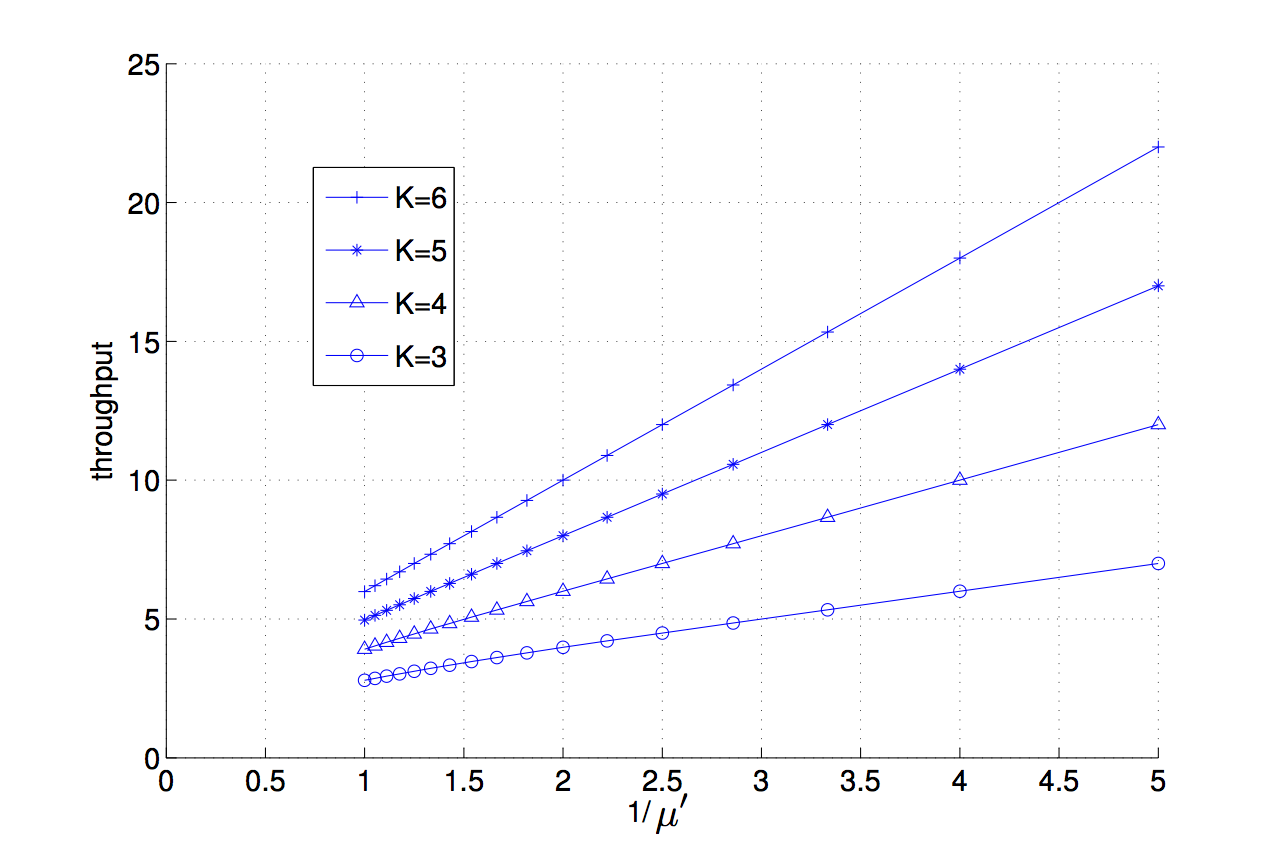} 
}
\subfloat[\ $\mu=1$ ]{
\includegraphics[scale=0.4]{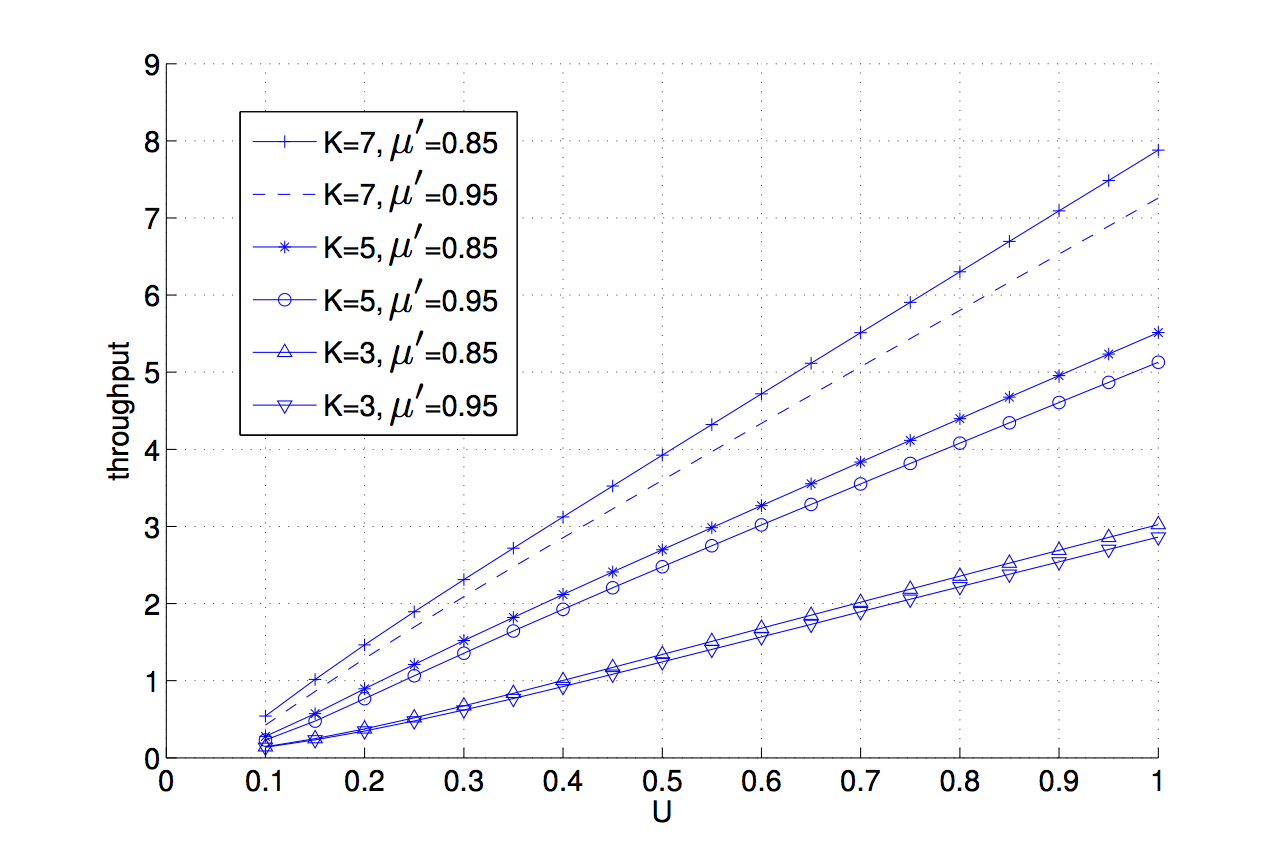}  
}
\caption{Throughput when  policy $\mu'<\mu$ is adopted by the one-club peers  ($K$ varying between 3 and 7): (a) as $1/\mu'$ and (b) $U$ increase, the throughput increases roughly linearly.} 
\label{fig:modif1}
\end{figure}

\begin{figure}[htb]
\centering
\subfloat[\  $K=3,\mu=1,U=0.5$]{
\includegraphics[scale=0.4]{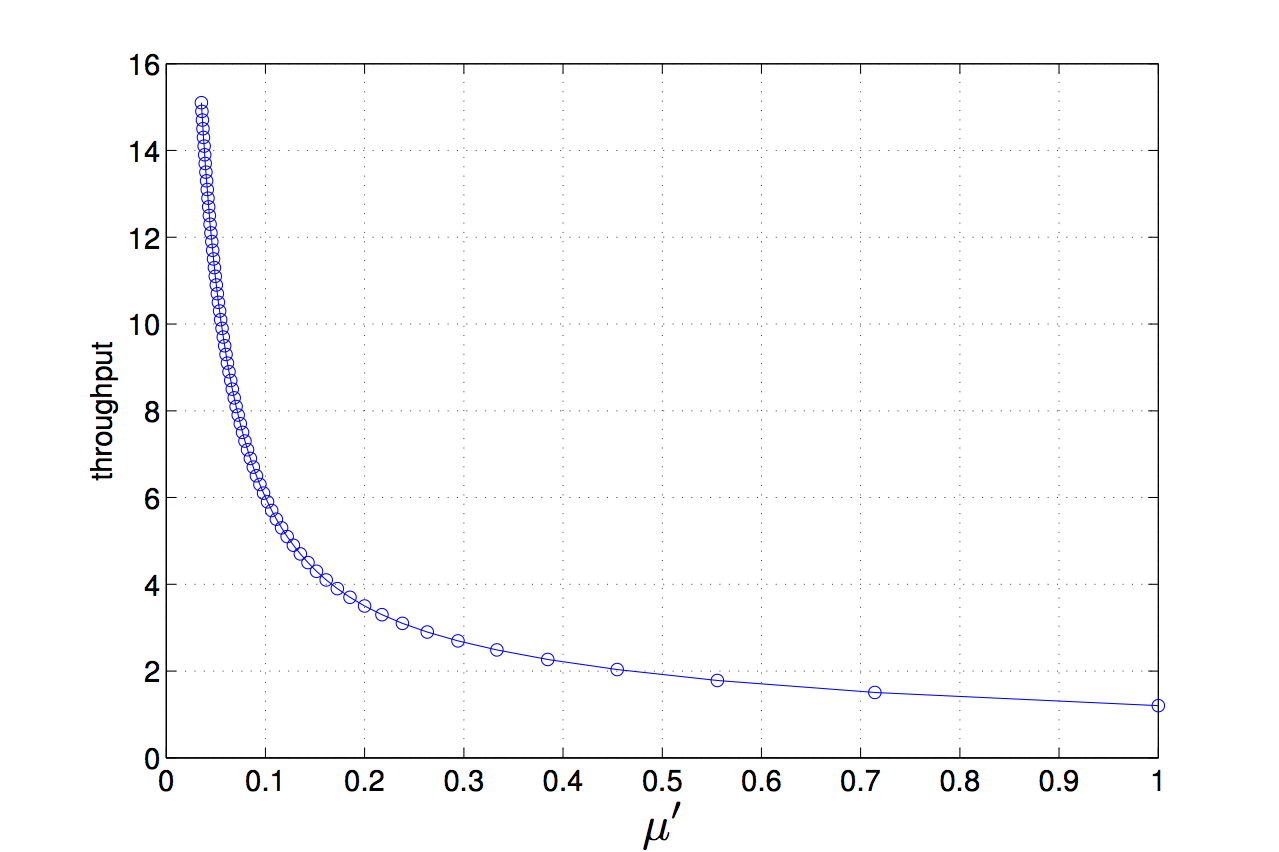} 
}
\subfloat[\ $K=3,\mu=1,U=0.2$]{
\includegraphics[scale=0.4]{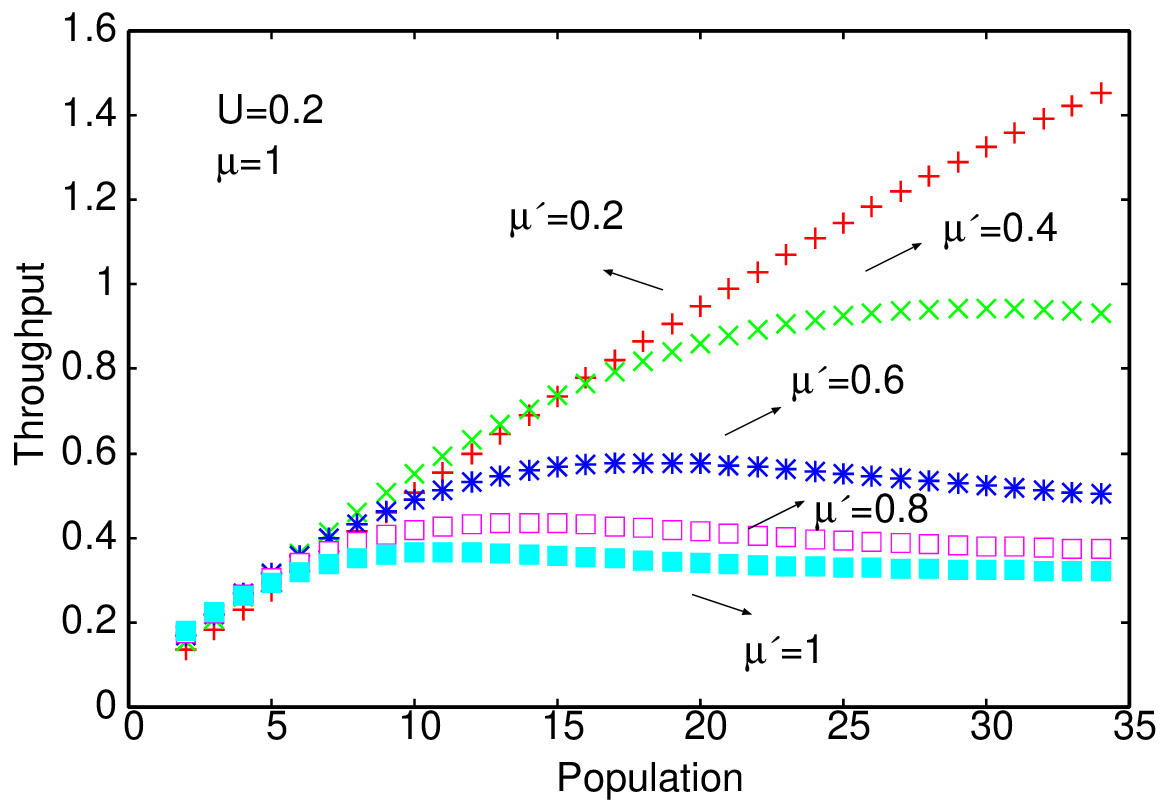}
}
\caption{Throughput when  policy $\mu'<\mu$ is adopted by the one-club peers ($K=3$).}
\label{fig:modif2}
\end{figure}
 
\subsection{Shielding Newcomers Can Further Increase System Throughput} \label{sec:shield}

In this section we show scenarios wherein  {\em hiding} newcomers  
from other peers can significantly improve system throughput. 
Under the most-deprived peer policy, 
the publisher \emph{competes} against the one-club members to serve  newcomers. When $\mu=\mu'$, given a large one-club, 
the publisher
 serves a newcomer with 
probability  $U/(U+\mu)$.  
Suppose the server rate $U$ is much smaller  than the peer upload rate $\mu$. 
In this case, newcomers receive, with high probability, 
the most popular blocks from the one-club members before  obtaining the  rarest
block from the publisher.  
The idea of \emph{shielding a newcomer} is to prevent the newcomer from receiving the most popular blocks before it
receives  the rarest one.
This policy is easily implemented by the tracker: it should  not announce  
newcomers to other peers. 

\begin{figure}[h!tb]
\hspace{0.1in}
\includegraphics[scale=0.4]{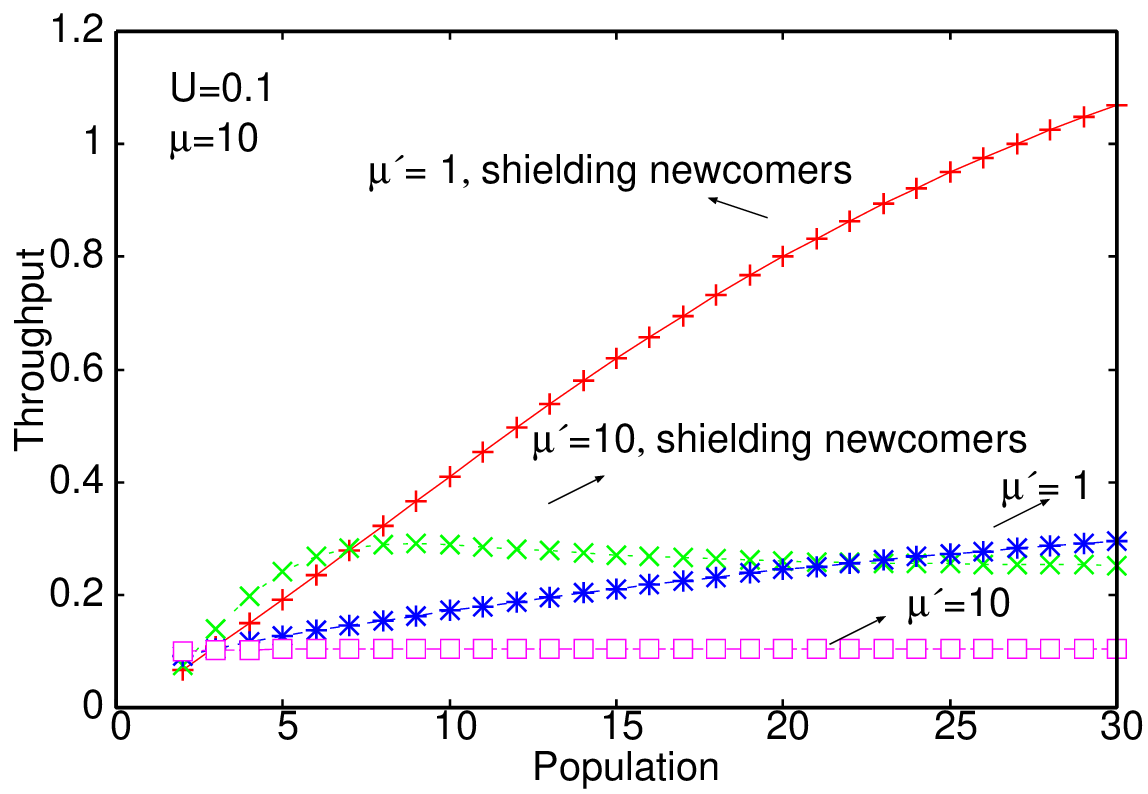}	
\hspace{0.1in}
\includegraphics[scale=0.4]{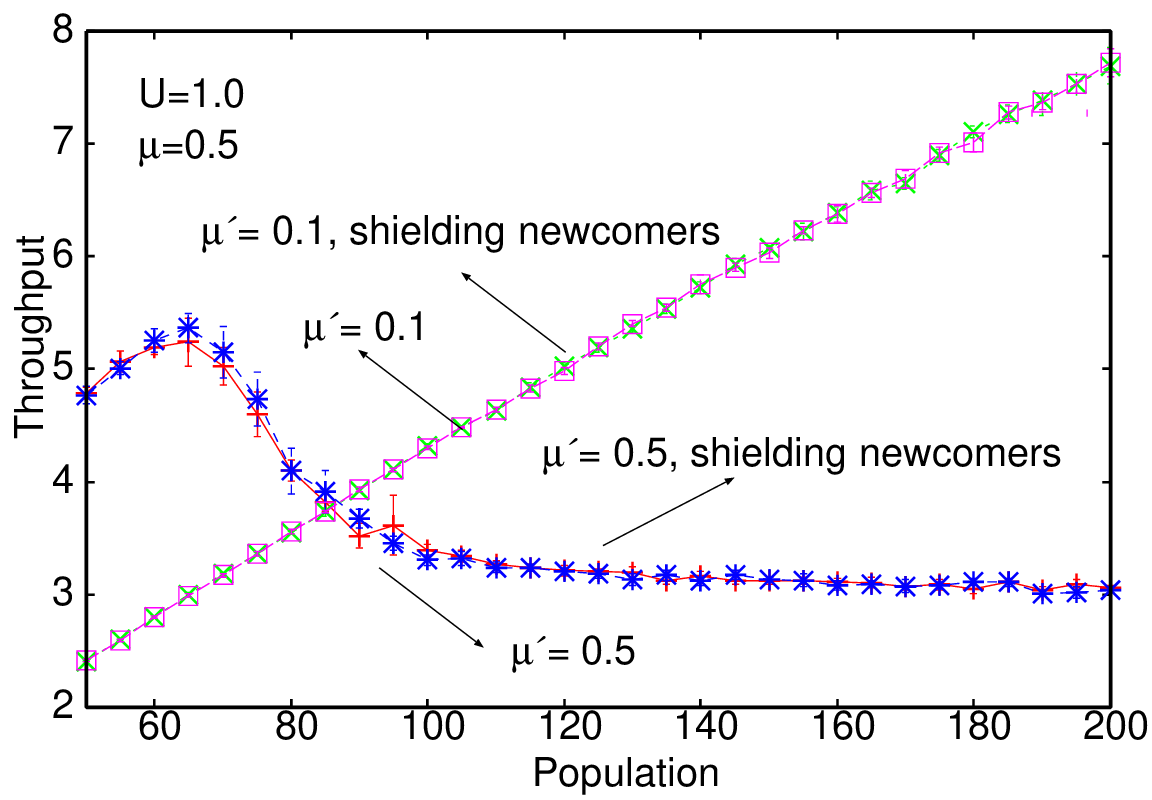}
\center
\begin{tabular*}{0.6\textwidth}{@{\extracolsep{\fill} }ll}
(a) $U < \mu$ & (b) $U > \mu$  
\end{tabular*}
\caption{Throughput when combining the policies of shielding newcomers and setting $\mu'<\mu$.} 
\hspace{-0.3in}
\label{shielding}
\end{figure}

Figure~\ref{shielding}, obtained using the detailed Markov model of the fixed-size population system, 
shows system throughput as a function of population size, 
when $U<\mu$ and $U>\mu$, for $K=3$. 
In Figure~\ref{shielding}(a),  
we consider four scenarios: $(i)$ $\mu=\mu'=10$, $(ii)$ $\mu=10$ and $\mu'=1$,
$(iii)$ $\mu=\mu'=10$, shielding newcomers and $(iv)$ $\mu=10$ and $\mu'=1$, 
shielding newcomers. 
In scenario $(i)$, the one-club peers serve newcomers before they receive
the rarest block from the server.
In this case, system throughput  $\Gamma$ converges to the server service capacity, $U$.
In the second scenario, one-club peers reduce their upload rate and 
throughput increases when compared to scenario $(i)$, as expected 
(see Section \ref{sec:reduced-rate}). 
The third scenario implements the policy of shielding newcomers.
The throughput $\lambdacritical$ is approximately equal to $0.25$.  
For a small population size, the throughput reaches the  value of $\Gamma_S$ established in 
Proposition~\ref{prop:propo1}.
Scenario $(iv)$ combines both policies: the upload rate of one-club peers is reduced and newcomers are shielded. 
Throughput  significantly increases when these policies are used in combination.  
When the policies are used separately, the throughput is approximately $0.3$ 
for a population
of 30 users.  When both policies are used, the throughput 
exceeds $1.0$. 

Figure~\ref{shielding}(b) presents the throughput when $U>\mu$. 
The figure shows that, in this case, the policy of shielding newcomers
does not help to improve throughput.
When  server capacity $U$ is greater than the upload
rate of the one-club peers $\mu$, the server  transmits the rarest block to
newcomers at a higher rate than the rate at which they receive the most popular
blocks.
It is unnecessary to {\em hide} the newcomers from the one-club peers because
the former will spontaneously receive the rarest block from the latter before
receiving other blocks from the remaining peers. 
However, throttling the upload capacity for one-club peers continues to
be very effective.

\subsection{Assessing the Benefits of Admission Control}

Next, we present the benefits of admission control.  
As shown in Figure~\ref{validating} 
(fixed-size population model), as  population size grows, the
throughput initially increases and then decreases.  
This behavior suggests that it may be beneficial to exert 
control over the size of the population that participate in the swarm.

Figure~\ref{fig:admissioncontrolb} confirms the potential benefits of admission
control.  The curves in Figure~\ref{fig:admissioncontrolb} were obtained using
the
model described in Appendix~A.  We considered an infinite
population of peers, joining a system under admission control.  In contrast to
Section~\ref{sec:markovian}, we allow the population size to vary over time.
We let  $\lambda=6$ peers/s, $K=3$, $U \in \{2, 0.3\}$ blocks/s and
$\mu=\mu'=0.5$ blocks/s. Publisher and peers adopt MDP/RFB and RUP/RUB
policies, respectively.  For each scenario considered, we simulated 10 runs,
where each run consisted of 100,000 events, and computed    95\% confidence
intervals.   
  
Figure~\ref{fig:admissioncontrolb} shows that  as the population cap
increases, throughput increases and then decreases.   When the population cap
equals 1 (extreme left of Figure~\ref{fig:admissioncontrolb}), the throughput
roughly equals  $U/K$.  Indeed, the throughput is slightly  less than $U/K$ as
when a peer leaves the system, it takes on average $1/\lambda$ for the next
peer to arrive and start receiving service.  As the population cap increases,
the throughput increases, and reaches its peak at $N=15$ (resp., $N=5$) for
$U=2$ (resp., $U=0.3$).  If the  population cap is further increased, the
overhead due to encounters which do not translate into useful transmissions
plays a more significant role, and the throughput decreases reaching  the
asymptotic value of $U$.   Such asymptotic value is a consequence of two
facts.  First, note that peers adopt  the random useful peer selection (RUP),
which implies that gifted peers are rapidly served by the one club and then
leave the system (see Section~\ref{sec:illustrfix}).  Second,  results shown in
Figure~\ref{fig:admissioncontrolb} do not consider  the shielding of newcomers
(see Section~\ref{sec:shield}).   
    Therefore, the publisher competes with the one-club to serve newcomers, who end up being served 
      most of the time by latter.
    The one-club quickly builds up, and is served at rate $U$ for a large enough  population size.


\begin{figure}[h!]
\hspace{-0.15in}
\includegraphics[width=0.57\columnwidth]{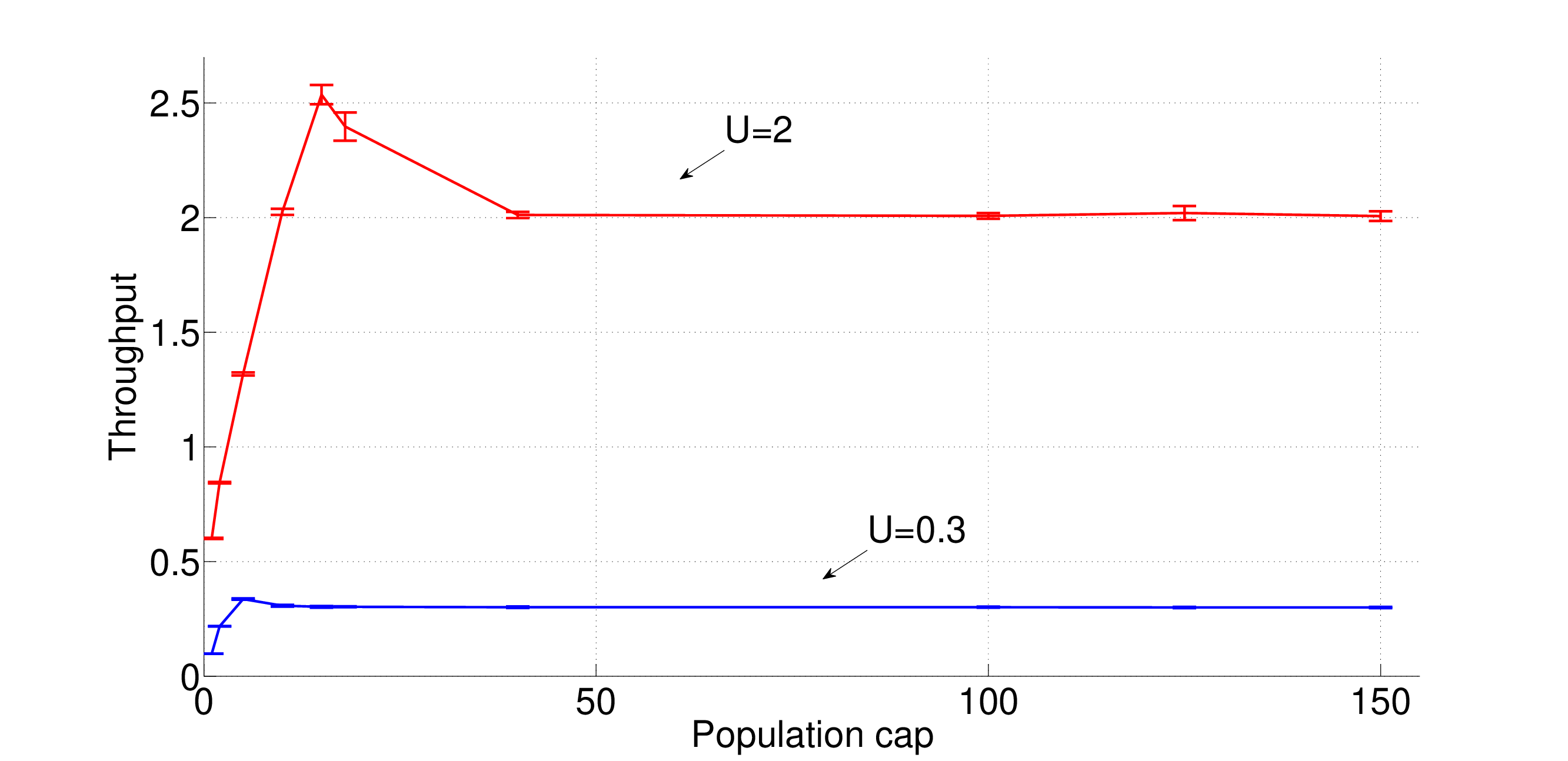}  
\hspace{-0.65in}
\includegraphics[width=0.57\columnwidth]{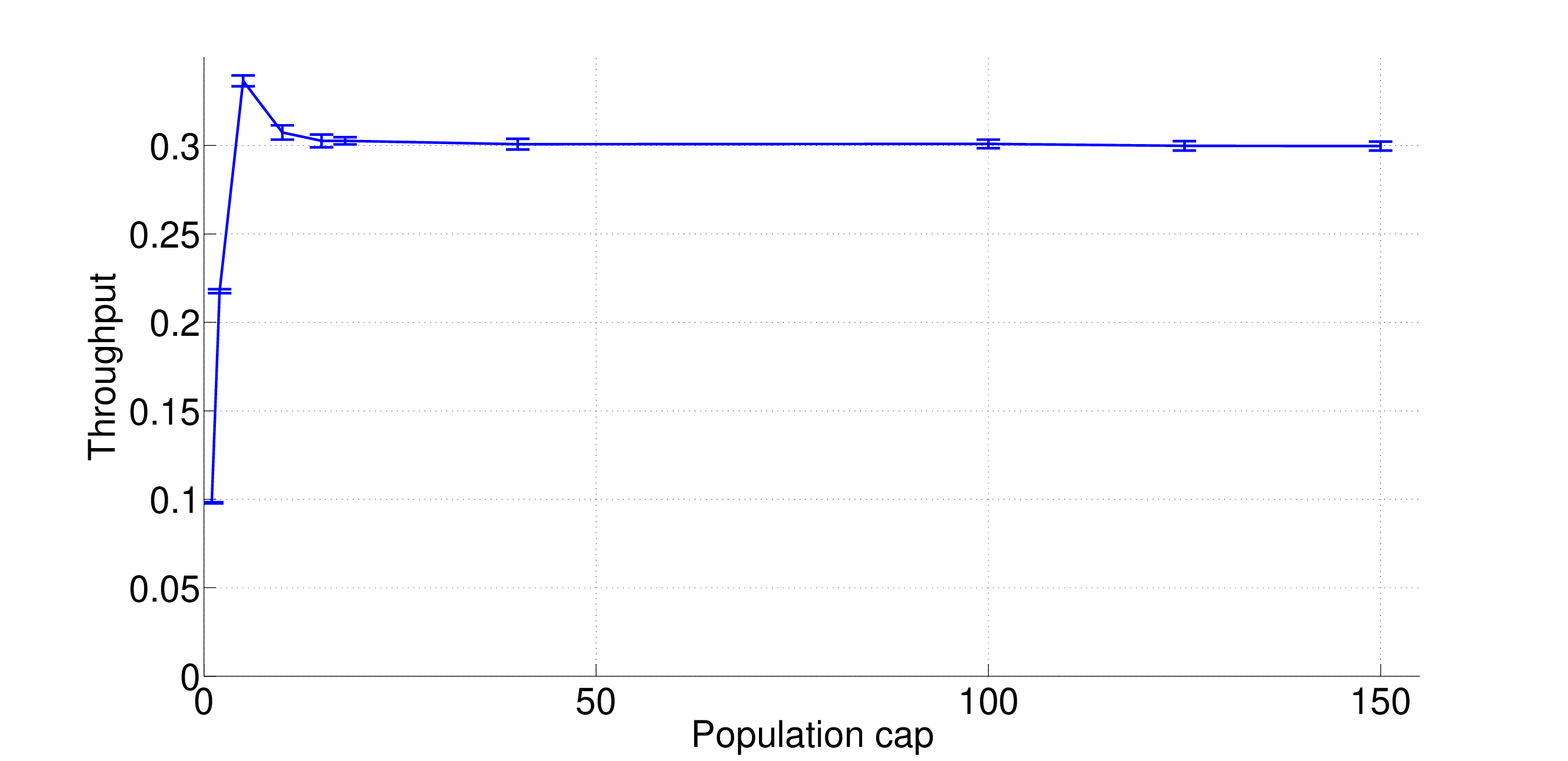}  
\begin{tabular*}{1.0\textwidth}{@{\extracolsep{\fill} }cccc}
& (a) $\mu=\mu'=0.5$, and $U \in \{ 2,0.3\}$  \quad &\quad  (b) $\mu=\mu'=0.5$, and $U =0.3$  & \\
\end{tabular*}
\caption{Illustrating the benefits of admission control under the MDP/RFB and RUP/RUB policies.}
\label{fig:admissioncontrolb}
\end{figure}

Figure~\ref{fig:admissioncontrolbrr} shows the benefits of admission control in the  same setting as described above, except that  
 publisher and peers adopt MDP/RFB and RP/RUB policies, respectively.  Note that the gains due to admission control
are now more pronounced.  
Still, the asymptotic throughput does not increase substantially when contrasted against Figure~\ref{fig:admissioncontrolb}. 
When $U=0.3$, the competition between the publisher and the one-club  to serve newcomers is responsible 
for the asymptotic throughput remaining close to $U$.  When $U = 2$, the peer arrival rate, set to $\lambda=6$,  is not large
enough to make the server use all its  capacity to serve newcomers.  In this setting, we observe that roughly 30\% of the server bandwidth
is devoted to newcomers, whereas the remainder is used to serve the one-club.  
If we increase the arrival rate to $\lambda=60$, in contrast,
we obtain a throughput of 6.67 $\pm 0.0908$, for a population cap of 200 peers.   In the latter case,
virtually all the service capacity of the server is used to serve newcomers and, as discussed in Section~\ref{sec:illustrfix}, 
it should come with no surprise
that the   obtained throughput is larger than $\lambdasaturation=K U$ due to the effects of admission control.   

The results in this section indicate the importance of the core strategies introduced  in this paper, 
namely the adoption of MDP/RUB (resp., RP/RUB) by the server (resp., the peers), the 
reduction of 
the service capacity of the one-club, the shielding newcomers and admission control. 
  Under different settings presented above, 
disabling any of these mechanisms may have
detrimental effects on performance.

 
\begin{figure}[h!]
\hspace{-0.35in}
\includegraphics[width=0.57\columnwidth]{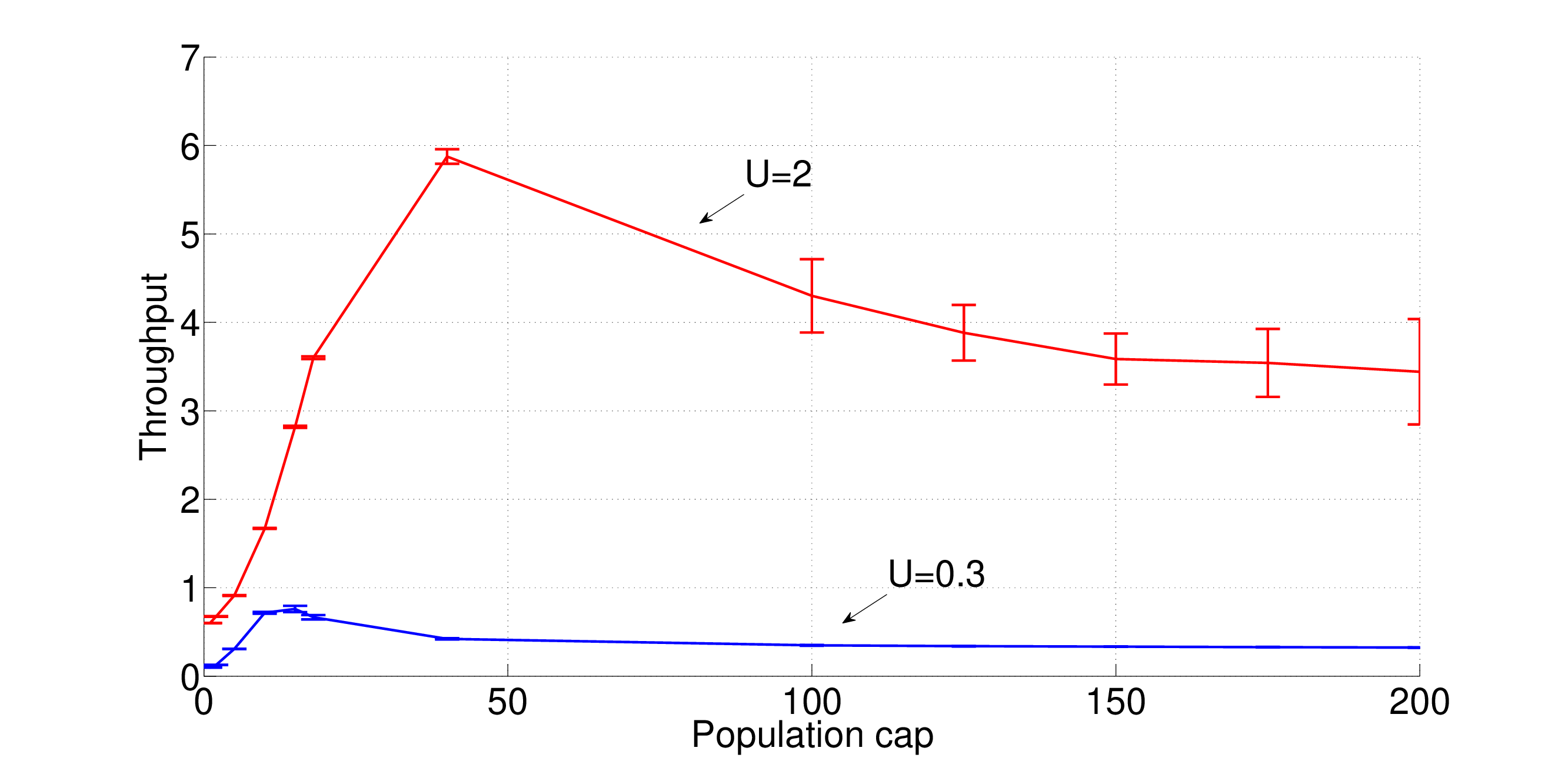}  
\hspace{-0.35in}
\includegraphics[width=0.57\columnwidth]{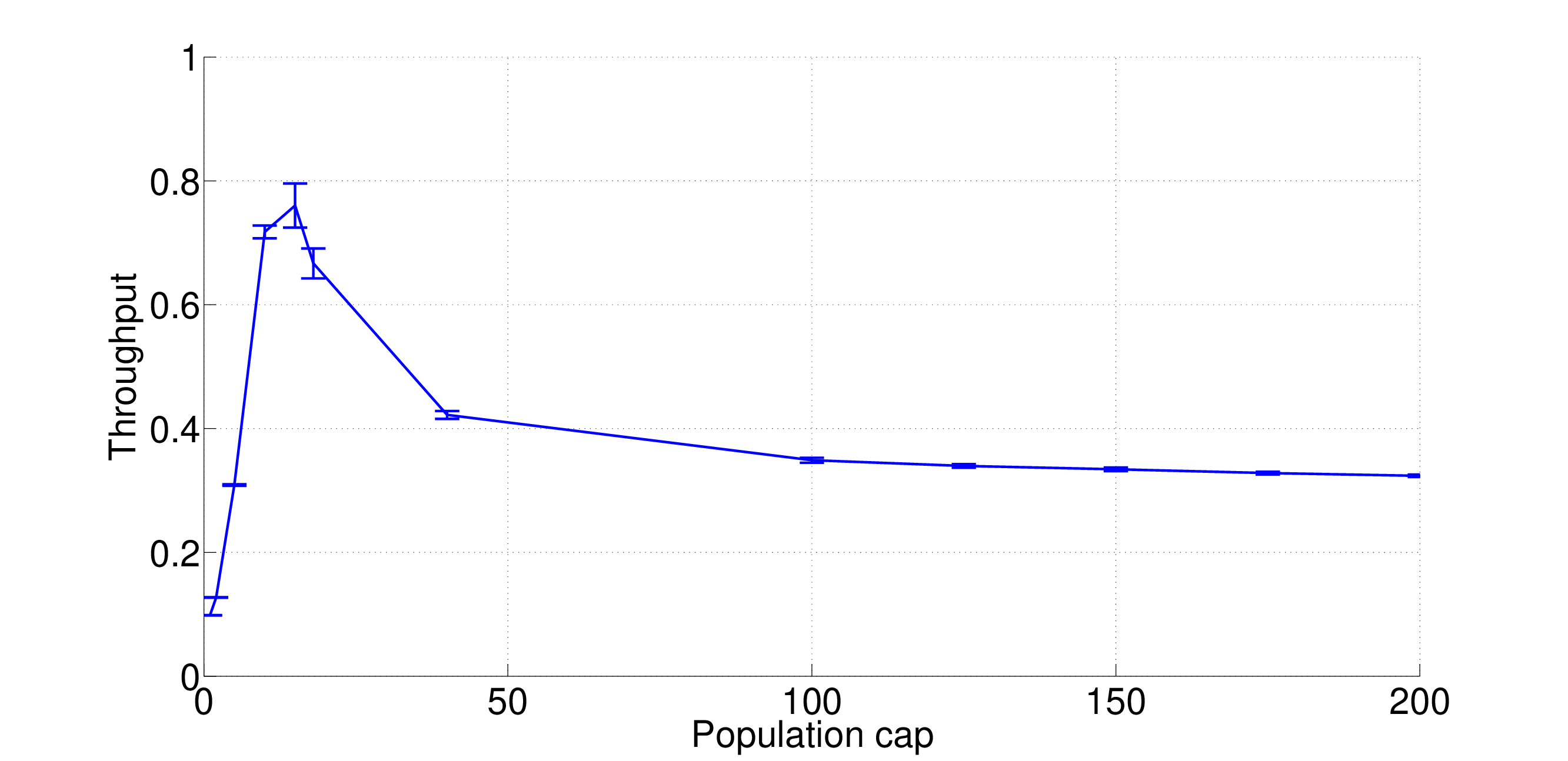}  
\begin{tabular*}{1.0\textwidth}{@{\extracolsep{\fill} }cccc}
& (a) $\mu=\mu'=0.5$, and $U \in \{ 2,0.3\}$  \quad &\quad  (b) $\mu=\mu'=0.5$, and $U =0.3$  & \\
\end{tabular*}
\caption{Illustrating the benefits of admission control under the MDP/RFB and RP/RUB policies.}
\label{fig:admissioncontrolbrr}
\end{figure}

\section{Related Work}
\label{sec:related}

There  is  a  vast  literature  on the  stability  and  throughput  of
peer-to-peer swarming systems focusing  on its relations with multiple
swarms and bundling~\cite{zhu2014stability, e2013interplay, zhu2013stable},
self-sustainability~\cite{ciullo2012stochastic},
real-time  content  dissemination~\cite{zhu2013tree, baccelli2013can},
coding~\cite{westphal2014stable} and system 
design~\cite{altman2013predicting,hwangjoint,mathieu,oguz2015stable,zhu2014stability,bilgen2017stable}.   
Nonetheless, we were  unable to find  any previous  work that  accounted for  the  
\emph{missing piece syndrome}  when  computing  the  \emph{throughput}  of  peer-to-peer
swarming systems.

In this paper, a number of illustrative results considered small swarms with less than
one hundred peers.  In some cases, we indicated that such small number of peers
suffices to reach the swarm asymptotic throughput.  In addition, it has been reported
in related work that swarms with small 
populations are common in the wild. According to~\cite{e2013interplay},
 most of the swarms are very small: approximately 73\% of the swarms measured by~\cite{e2013interplay}
   are formed by less 
 than 10 peers and 58\% have less than 5 peers.

In previous  works, authors assumed  that either peers  and publishers
adopted  random-peer selection~\cite{nunez2008scaling},  files  had at
most two blocks~\cite{norros2011stability} or swarming systems behaved
differently   as   compared   to   the   system   analyzed   in   this
paper~\cite{leskela2010interacting}.  The most-deprived peer selection
policy     was      first     proposed     by      Bonald     
\emph{et al.}~\cite{bonald2008epidemic}.  As indicated  in this paper, if the
publisher  adopts  the  most-deprived  peer  selection  strategy,  the
throughput of  the swarm can increase  even if the  remaining peers do
not change their strategies.

Yang and  de Veciana were  the first to  consider a closed  system to
analyze  the   transient  increase   in  throughput  after   a  flash
crowd~\cite{veciana-PEV2006}.   They  also  considered  an  idealized
fluid model to study the  steady state.  In their seminal paper, Yang
and de Veciana did not account for the fact that a block might become
rare and its retrieval turn into a system bottleneck.

In the  peer-to-peer literature, fluid models  have been traditionally
used   to    study   system   
performance~\cite{massoulie2008coupon,qiu2004modeling,simatos,paganini-ferragut-2016,ferragut2015queueing, gaidamaka2016application,anjum2017survey} 
and scheduling strategies~\cite{zhang2010optimal}.   
The  importance  of taking  into 
account the  fact that the file  is divided into  finite blocks rather
than  considering  the  fluid  limit was  indicated  
in~\cite{mathieu,zhu2014stability}, who discovered that under the Markovian framework introduced by Massoulié and Vojnovic~\cite{massoulie2008coupon} the free random distribution of file pieces suffers 
 from the so called missing piece syndrome. 
 Note that, for tractability purposes, Massoulié and Vojnovic~\cite{massoulie2008coupon} considered a model with built in symmetries across peers.   Then, Hajek and Zhu~\cite{zhu2014stability} showed that  when seeding rate is scarce, symmetry breaking  occurs.   Such symmetry breaking plays a key role in the system,  leading to one piece becoming very rare.   In this paper, we also account for this symmetry breaking and its consequences.

    The missing piece phenomenon can be avoided by the presence of a sufficient number of seeds, but this solution requires some altruism of peers in the form of staying in system after completing  their own downloads (see Section~\ref{sec:limit}).
    It is an important challenge to find fully distributed protocols that would guarantee stability even 
     with non-altruistic peers. Reittu~\cite{reittu2009stable}
      invented the first one, which was later proven to work~\cite{oguz2015stable}.
      None of those works focused on assessing system throughput.  
\emph{In this  paper,  we compute  the throughput  of
swarming systems} accounting for the fact that the file is divided into
finite  blocks,  and  use  our  model  to  motivate  novel  scheduling
strategies.

Paganini and Ferragut~\cite{paganini-ferragut-2016} combined M/M/$\infty$
and M/M/1 queues to model the download progress in P2P networks.  The former 
captures the self-scaling characteristics of the system, while the  latter is used to 
model the over-provisioned regime wherein peers quickly download the file and leave the system 
before having a chance to cooperate with other peers.  In this paper, in contrast,
we also combine M/M/$\infty$ and M/M/1 queues, 
 but differently from~\cite{paganini-ferragut-2016} we use the latter
to characterize the under-provisioned regime wherein the publisher becomes the system bottleneck.

Scheduling  strategies to  improve system  throughput usually  rely on
some sort of altruism. To improve system throughput, in previous works
it has been proposed that  peers reside in the system after completing
their downloads~\cite{zhu2014stability}, barter  for content that they
were not interested  in~\cite{zhu2014stability} or refrain from taking
advantage of all contact opportunities~\cite{oguz2015stable}.  
Similar in spirit, in~\cite{bilgen2017stable}  the authors 
 propose a provably-stable  incentive-compatible strategy, group suppression, which prevents the increase of the {\em one-club}.    
 The idea consists of temporarily refraining certain peers from uploading blocks.
 In~\cite{modesup}, the authors propose a variation of group suppression, referred to as mode suppression,
 wherein  peers may  suppress the transmission of the most popular piece. 
 Whereas group suppression is proven to be stable for  files with up to three blocks~\cite{bilgen2017stable}, 
 in~\cite{modesup} the authors
 prove that mode suppression is stable  for any number of blocks.  
 In this paper, we show that  a simple  and incentive-compatible
 strategy, which consists of reducing the service capacity of the peers
 that  have  all  but  one  block,  can   significantly  improve  system
 throughput.    An experimental  reality check supporting the improvement in system throughput  is reported in~\cite{diego}.

It is well known 
 that in swarming systems non work-conserving strategies  may perform better than work-conserving ones~\cite{zhang2010optimal}.
 In essence, group suppression, mode suppression and capacity throttling are different incarnations of non work-conserving
 strategies.  
  The authors of~\cite{modesup} identify an interesting open problem, which consists of determining, among non work-conserving 
  strategies, those that  minimize sojourn time for stable protocols.  We envision that the throughput results obtained with the 
 models
  presented in this paper constitute  a first step  towards a better understanding of the impact of piece and peer selection 
  strategies
  on  sojourn times.



\section{Assumptions and Limitations}
\label{sec:limit}

Next, we discuss the key simplifying assumptions adopted in this paper, and some of the related limitations.

\textbf{Constant peer arrival rate: } in the proposed open system model, we assume that peers arrive according to
a Poisson process with constant arrival rate.  For contents whose arrival rates vary over time,
the constant arrival rate approximation must be applied at short windows of time.  In a number of scenarios of interest,
 the steady state
throughput
is   reached after a few peers leave the system (see Section~\ref{sec:transientanaly}), allowing us to rely on  asymptotic values
 when analyzing the throughput over finite time intervals.  


 
 \textbf{Constant peer mean upload capacity: } 
in most of our analysis and simulations we assumed a constant 
mean upload capacity.
 It is straightforward to adjust our simulations to account for a distribution of upload capacities across peers. 
In~\cite{chow2009bittorrent}, the authors considered an analytical model embracing heterogeneous upload capacities. Nonetheless, they did not account for the missing piece syndrome.  Our aim here, in contrast, is to  provide a simple model that accounts
for the missing piece syndrome.   Accounting together for the missing piece syndrome and heterogeneous   upload capacities would make the model more complex, and as in any modeling framework we need to trade between simplicity and generality.  

\textbf{Peer pairing occurs uniformly at random: } 
 peer pairing in real BitTorrent involves a number of mechanisms, including tit-for-tat (TFT).   In this paper, to simplify the analysis we ignore those mechanisms and consider the simplest possible peer pairing, which is uniform at random.  Still, experimental results indicate that the insights obtained through the simplified model are reflected in emulated experiments using the real Bittorrent client~\cite{diego}.

\textbf{Push or pull strategies: }
we assume that the mean time between contacts of peers 
  for opportunities  to  
 upload a packet (i.e. push) are  characterized by the mean time to upload a block.  Alternatively, the model wherein peers contact others for opportunities to    
  download  a block  (i.e.  pull) is mathematically equivalent to the one considered in this paper, as further discussed in~\cite{hajek}.  In the two cases, we assume that the contact itself is instantaneous, as the upload or download rates are captured through the time between contacts.  By assuming that pairing among peers occurs uniformly at random,  most control information  flows between peers and the publisher together with the tracker.

  \textbf{Scalability of peer selection policies: }  next, we indicate simple strategies that may increase the scalability of publisher decisions.  Consider a publisher that adopts  random useful peer selection, most deprived peer selection or random useful block selection.  Newcomers are, by definition, most deprived. If newcomers favor contacts with the publisher,  they will naturally simplify the coordination required by the publisher to prioritize transfers of useful blocks to useful peers or to most deprived peers.       In particular, we envision that in scenarios of practical interest most gains obtained by the use of   most deprived peer selection are due to  new peers being prioritized, and actually book-keep how many blocks each peer owns is not necessary.     A detailed analysis of that matter is out of the scope of this paper.

  \textbf{Scalability of block selection policies: } 
strict rarest first block selection  is typically not scalable and BitTorrent users sample their neighbors to identify the rarest block across a neighborhood. In  a swarm wherein the missing piece syndrome is present, it should be straightforward to determine the missing piece (rarest block).  In any case, selecting the rarest block out of 
the neighboring peers should suffice in a number of scenarios of practical interest~\cite{legoutrf}, in particular if peers can count on network coding~\cite{niu2010topological}.  

\textbf{Peers leave immediately after completing downloads: } in BitTorrent, there are no incentives for peers to remain in the system after completing their downloads.  In essence,  making swarms independent from each other builds scalability and robustness but
precludes incentives for cooperation across swarms after peers complete their downloads.  Cooperation across swarms requires some sort of mechanism to translate contributions in a swarm into rewards in another, which would cause interdependencies among swarms and breaking their self-sustaining nature.  We briefly analyze the system accounting for peers that altruistically linger as seeds in Appendix~\ref{app:lingering}.

\textbf{Peer churn: }  we assume that peers remain in the system before completing their downloads.  If peers have a deadline and leave the system in case their downloads do not complete by the deadline, the one-club does not grow unboundedly and the system is always stable.  
We briefly analyze the system accounting for peers that may abandon the system before completing their downloads  in Appendix~\ref{app:churn}.

\textbf{Usefulness of shielding newcomers: } 
our numerical results indicate that the shielding of newcomers  is unnecessary when the effective service capacity of the server  is larger than peer capacity.  The effective service  capacity is the capacity dedicated to a swarm.
The effective capacity of the server may be small, for instance, when the server is
serving a multitude of swarms, as opposed to serving only a few swarms.  
%
%
%
%
%
 In addition, focusing exclusively on the  case wherein publishers have large capacity goes counter the idea that anyone can publish content using Bittorrent.  Home users, for instance, may have small servers and we envision that the shielding of newcomers may be particularly helpful in those scenarios.
According to \cite{bittyrant}, around 70\% of peers have uplink capacity smaller
than 100 KB/s. 

\section{Conclusions}
\label{sec:concl}


Due to their ability to scale, robustness and efficiency, P2P systems are
responsible for a significant portion of today's Internet traffic and 
constitute the basis for new architectures such as content centric networking~\cite{ccn}.
Although P2P systems are very popular, their fundamental limitations are yet to be
fully understood.   In particular, the throughput of such systems can be impacted by the missing piece syndrome, but that effect has not been considered in previous works.

In this paper, we present new results to quantify the throughput of P2P systems when
the effective service capacity of the publisher is small compared to the arrival rate of peers.   
We evaluate the impact of different system parameters and system strategies
on attainable throughput
through the use of
models. 
Using those models, we  derive a new upper bound on the throughput achieved when the publisher adopts most deprived peer selection and rarest-first block selection.
 Our models also suggest a new very simple and incentive-compatible policy,   wherein peers reduce their service capacity when they possess all blocks but one.   By employing this upload throttling policy, the system can accommodate more users while remaining stable, specially when near saturation.

One of the ultimate goals of P2P swarming systems is to support very high loads (e.g., flash crowds) counting with scarce service capacity from publishers (e.g., home users).
This, in turn, is  the setup wherein the missing piece syndrome is most likely to occur.  
Experimental results recently indicated that the missing piece syndrome may indeed occur in real BitTorrent swarms~\cite{diego}.  In this paper, we complement the experimental evidence presented in~\cite{diego} with a  foundational theory to assess the factors that impact swarm throughput under the missing piece syndrome.  
Taken together, we believe that these works advance the state of the art towards 
understanding the fundamental limitations of P2P swarming systems and achieving feasible throughput goals.




\section{Acknowledgments}

E. de Souza e Silva, Rosa M. M. Leão
and Daniel S. Menasché are partially supported by grants from
CNPq, FAPERJ and FAPESP. Don Towsley is partially supported by grants
from NSF.

\bibliographystyle{plain}
\bibliography{scalability}

\renewcommand{\baselinestretch}{1.5}

\appendix
\section*{Appendices}

\section{Open System Throughput}

\label{app:opensys}


Our goal in this appendix is to show how system throughput varies as a function of 
 arrival rate, as predicted by the open queueing network
model.  Figures \ref{fig:mu03} and \ref{fig:mu1} 
show   system throughput as a function of the peer
arrival rate, for $K=3, U=0.3, \mu=\mu', \mu \in \{ 0.3, 0.5, 1.0, 10.0 \}$. 
We used the queueing network model of Section~\ref{sec:queueing-model}, with 4 queues, to obtain the results
plotted in the figures.   As discussed in Section~\ref{sec:models}, 
when considering 4 queues we get rid of $F_2$ and $G_2$ in Figure~\ref{fig:mod-filas}, and queue
 $G_3$ corresponds to gifted peers that have 2 blocks.

With 4 queues, the queueing network model comprises the following set of equations
\begin{align}
\lambda_0 &= \lambda \label{eq:lambda0lambda} \\
\lambda_1 &= \gamma_p(0) = \mathbb{E}[n_{0}] \mu \\
\pi_{0}(i)^{-1} &= \sum_{s=0}^\infty \lambda_i^s \left(\prod_{l=1}^s (U + \mu l)\right)^{-1}, \quad i \in \{0,1\}   \label{eq:pi0i}\\
\pi_{s}(i) &=   \pi_{0}(i) \lambda_i^s \left(\prod_{l=1}^s (U + \mu l)\right)^{-1}, \quad i \in \{0,1\}   \label{eq:pisi} \\
\gamma_r(i) &= U(1-\pi_{0}(i)) \left(\prod_{l=0}^{i-1} \pi_0(l) \right), \quad i \in \{0,1\} \label{eq:gammari} \\
\mathbb{E}[n_{i}] &= \sum_{l=0}^{\infty} \pi_{l}(i) l, \quad i \in \{0,1\}   \label{eq:eni} \\
\Gamma &= U+ 2\gamma_r(0) + \gamma_r(1) \label{eq:gammau2}
\end{align}
~\eqref{eq:pi0i}-\eqref{eq:pisi} correspond to the steady-state solution of the birth-death processes associated 
with queues $F_0$ and $F_1$, with corresponding arrival rates $\lambda_0$ and $\lambda_1$,  while
the additional equations are taken directly from Section~\ref{sec:models}.  
The set of equations above is solved given $\lambda$ and $\mu$. First,
   \eqref{eq:pi0i}-\eqref{eq:eni} are evaluated  for $i=0$, then for $i=1$, and finally 
 \eqref{eq:gammau2} is evaluated.

\begin{figure}[h!]\center
\includegraphics[width=0.6\columnwidth]{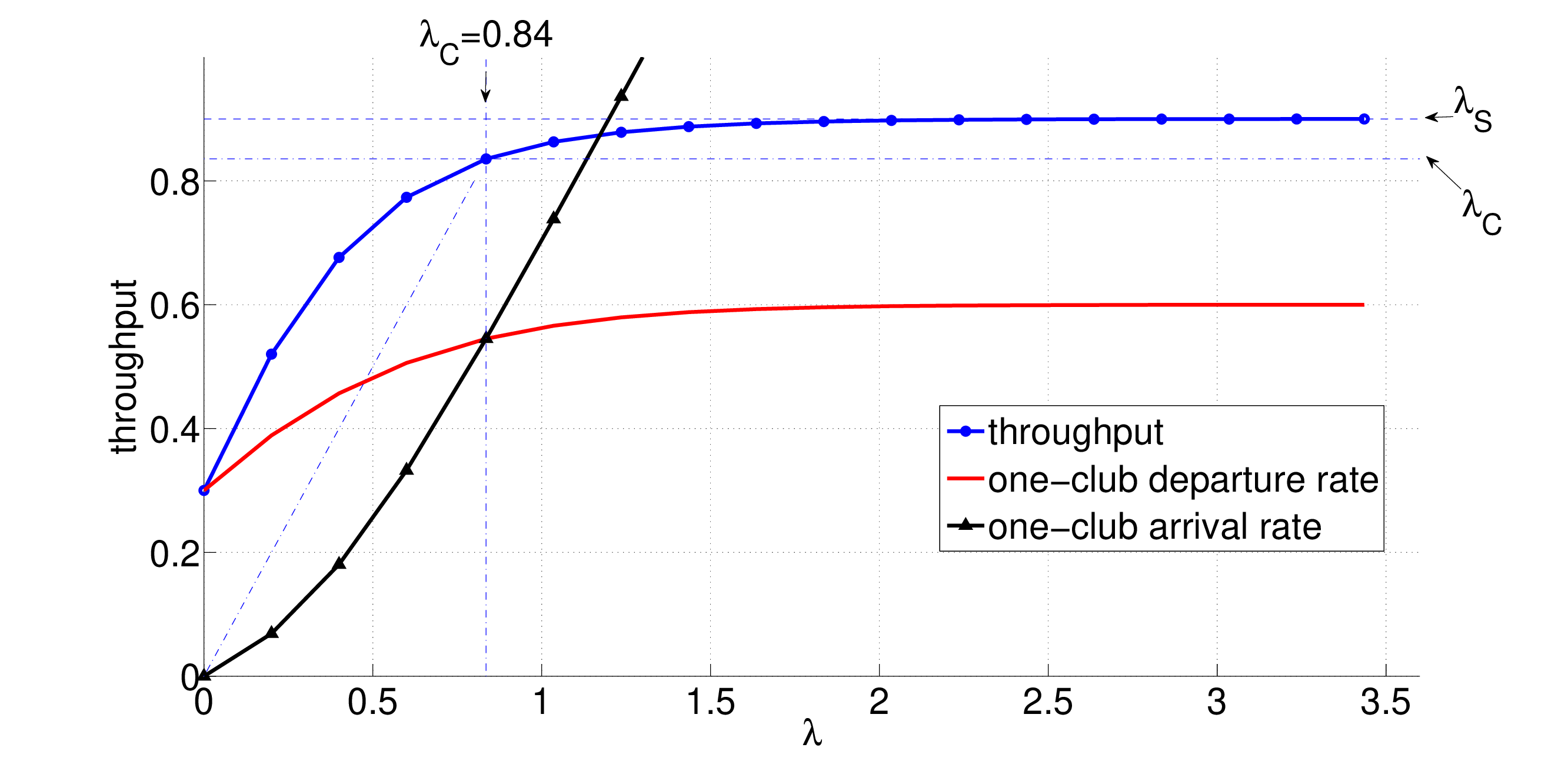}
\\
(a)
\\
\includegraphics[width=0.6\columnwidth]{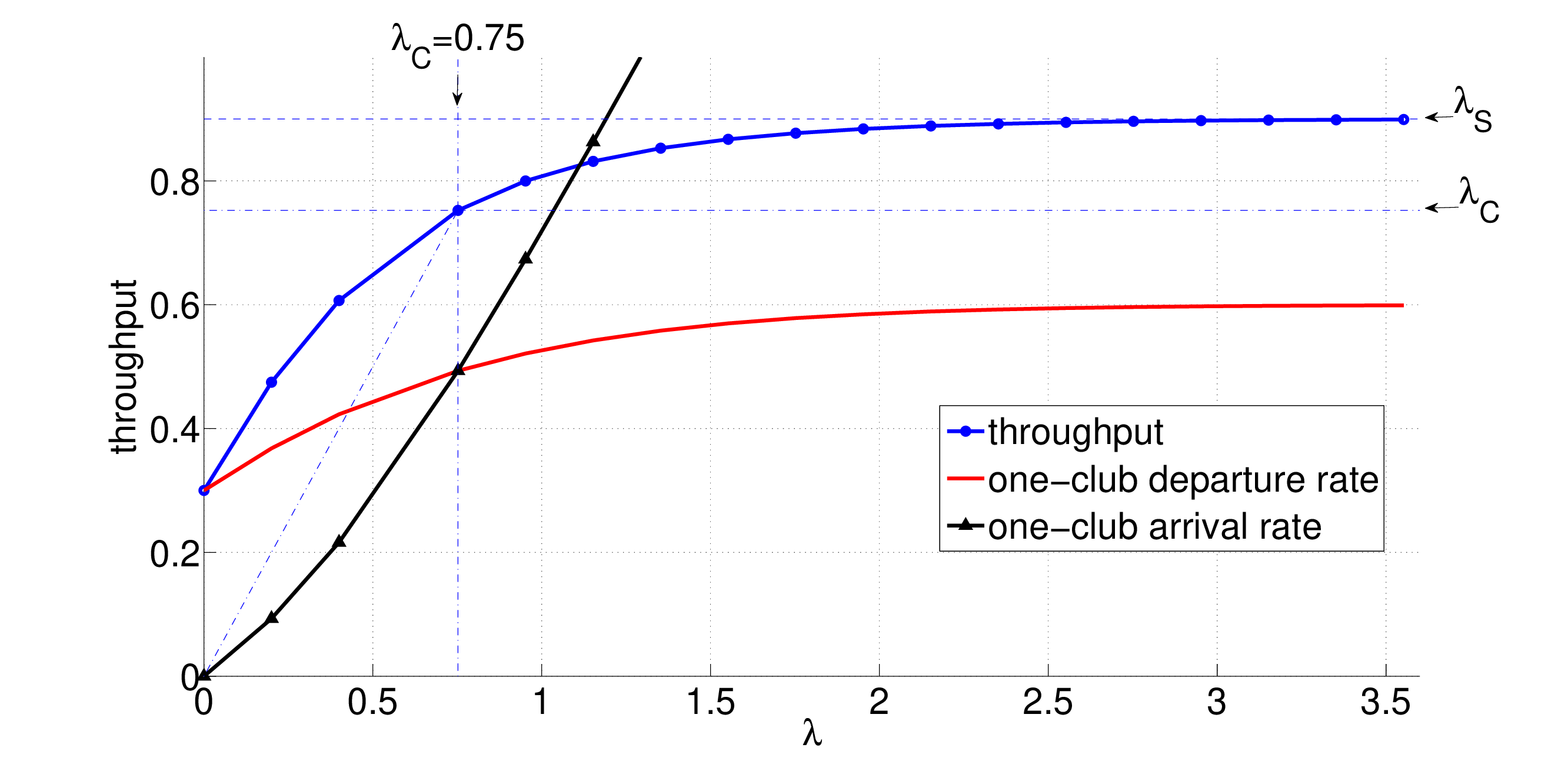}
\\
(b)
\caption{System throughput: $K=3$, $U=0.3$, (a) $\mu=0.3$ and (b) $\mu=0.5$}
\label{fig:mu03}
\end{figure}


\begin{figure}[h!]\center
\includegraphics[width=0.6\columnwidth]{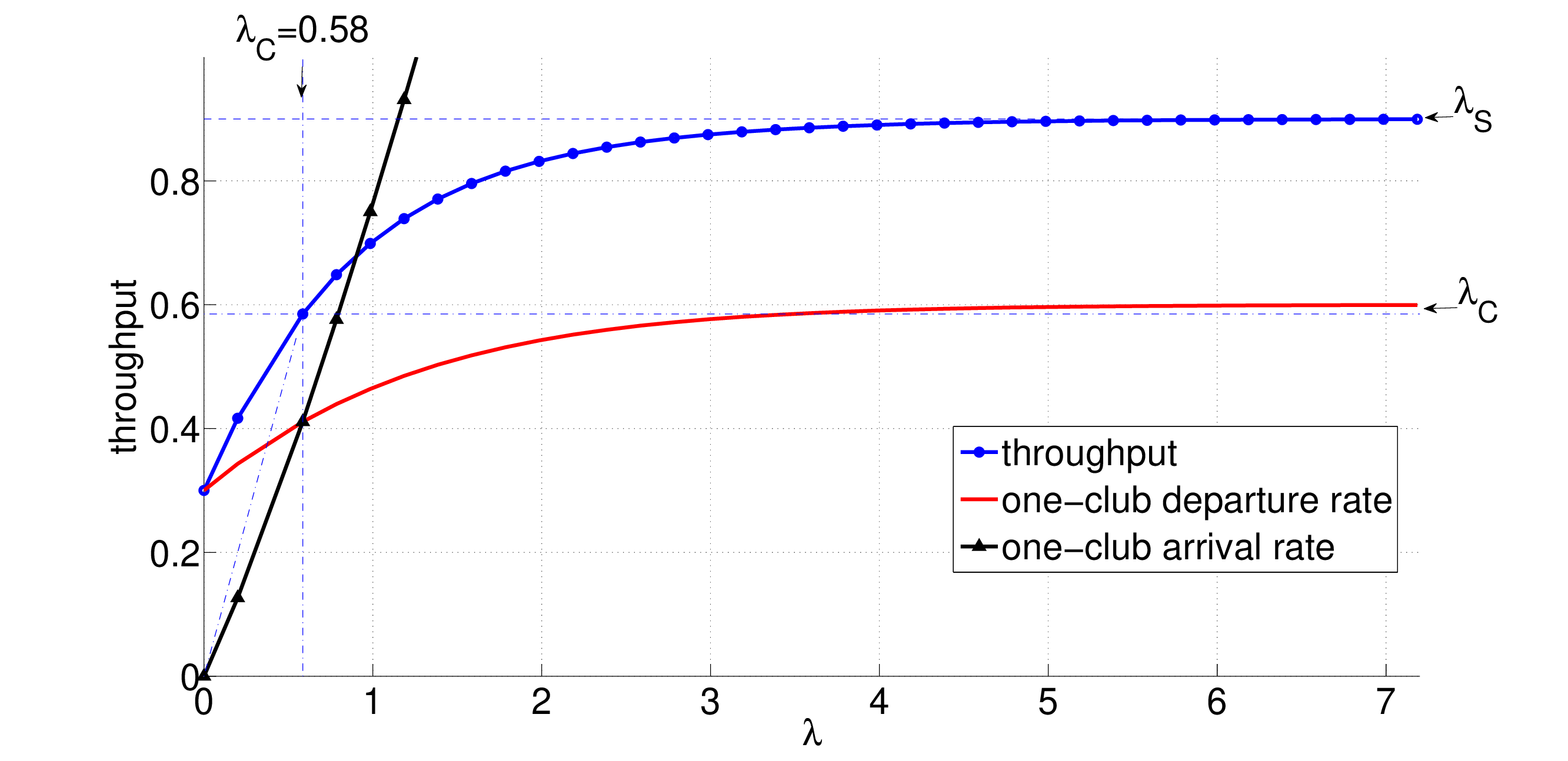}
\\
(a)
\\
\includegraphics[width=0.6\columnwidth]{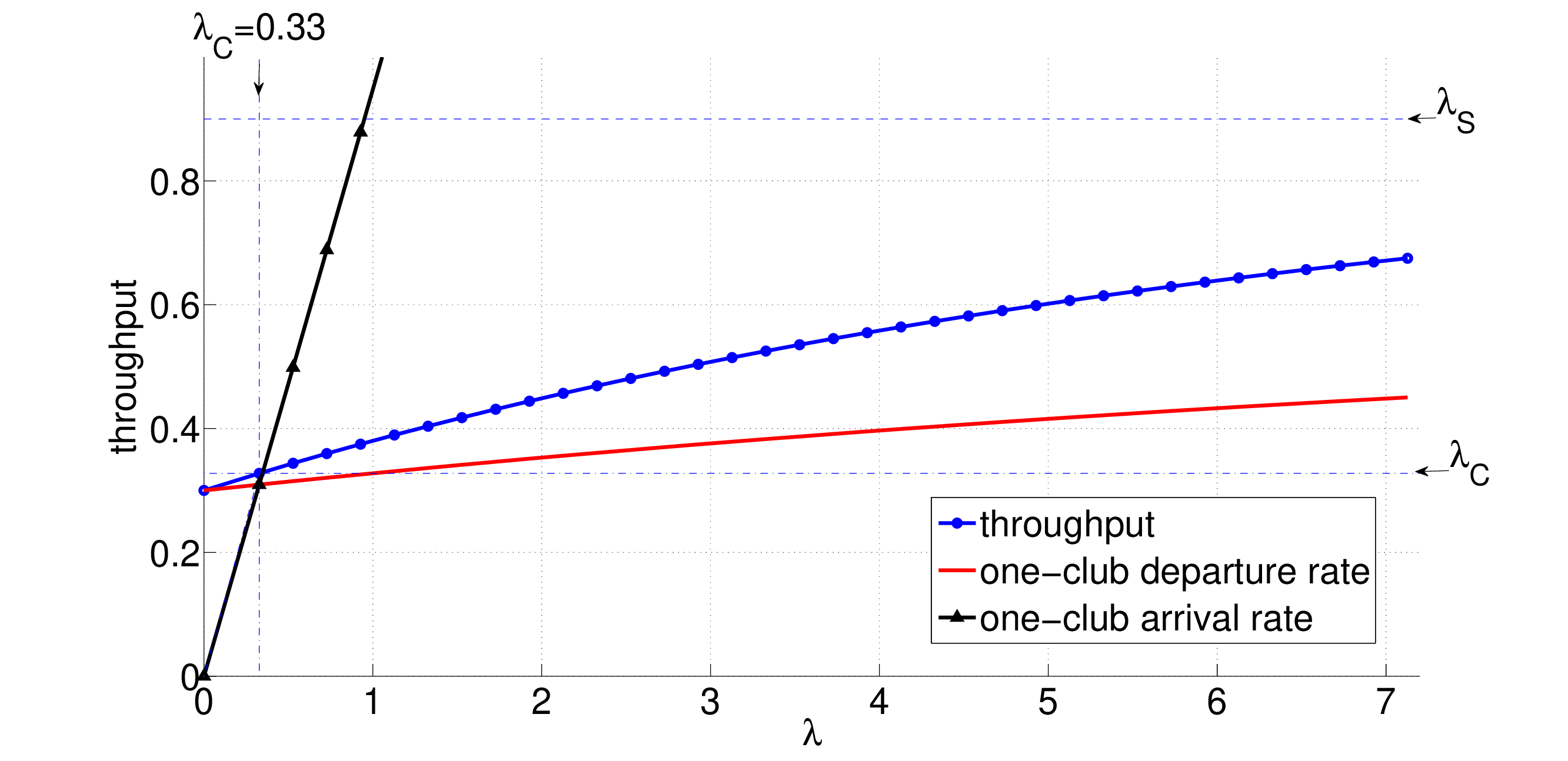}
\\
(b)
\caption{System throughput: $K=3$, $U=0.3$, (a) $\mu=1.0$ and (b) $\mu=10.0$}
\label{fig:mu1}
\end{figure}

When searching for $\lambdacritical$,  we solve~\eqref{eq:lambda0lambda}-\eqref{eq:gammau2} with the additional 
constraint $\lambda=\Gamma$.   To this aim,  we search for the solution using
 the nonlinear  interior-reflective Newton method described in \cite{coleman1996interior}, and implemented by the
  \texttt{lsqnonlin} function in MATLAB. Alternatively, the value of $\lambdacritical$ is inferred by increasing $\lambda$, 
  computing the corresponding throughput $\Gamma$,
   and 
  searching for the critical value for which $\lambda=\Gamma$.

Note that the departure rate from the one-club is given by $\mu \mathbb{E}[m_1+m_3]+ U \pi_{0}(0) \pi_{0}(1)=2\gamma_r(0) + \gamma_r(1)+ U \pi_{0}(0) \pi_{0}(1)$,
while the arrival rate to the one club is given by $(\lambda-(1-\pi_0(0))U) - U \pi_0(0)(1-\pi_0(1)) = \lambda - U(1-\pi_0(0)\pi_0(1))$.
In the figures, the blue line (marked with circles) corresponds to  the system throughput (eq.~\eqref{eq:gammau2}), the red line (plain line) 
corresponds to  the arrival rate of peers to the one-club,
and the black line (marked with triangles) corresponds to  the departure rate of peers from one-club.

In Figures~\ref{fig:mu03} and~\ref{fig:mu1}
   the system throughput converges to $KU$.   When $\lambda=0$, throughput equals $U$, due to departures from the one-club.
   As $\lambda$ increases, throughput increases due to the collaboration between peers.  
   In this region ($\lambda < \lambdacritical$), the departure rate  from the one-club is larger than the arrival rate to the one-club, 
   which favors a reduction in the 
   one-club size.     
When $\lambda=\lambdacritical$, we observe that the one-club departure and arrival rates equal each other, i.e.,   
the black  curve (arrival rate of peers to the one-club) crosses the red curve 
(departure rate of peers from the one-club) when $\lambda=\lambdacritical$. 
For $\lambda >  \lambdacritical$, the throughput still increases until  saturation.  When $\lambda =\infty$,
 the throughput equals $\lambdasaturation=KU$. 

Comparing Figures~\ref{fig:mu03} and~\ref{fig:mu1}, we observe that as $\mu$ increases $\lambdacritical$ decreases.  
This is because
we are assuming $\mu=\mu'$. An increase in the service capacity of peers causes a decrease in the residence
time of gifted peers, which in turn contributes to a reduction in the departure rate from the one-club.


\section{Iterative Fixed Point Algorithm To Compute $\Gamma_C$}

\label{sec:appiterative}

Next, we present an iterative process to   approximate the critical throughput attained by  the  population of peers. 
The process consists of increasing the arrival rate up to reaching its critical value.  
Let $\Gamma^{(i)}$ and $\lambda^{(i)}$ be the value of $\Gamma$ and $\lambda$ 
computed at the $i$-th iteration.     

Assume that the  arrival rate at the first iteration $\lambda^{(0)}$ is given.
The steady state solution for queue $F_0$ is obtained by constructing a simple birth-death process 
with parameter values dependent on $U$ and $\mu'$.
After computing $\mathbb{E}[n_0]$ and $\mathbb{P}(n_0=0)=\pi_0(0)$,  
equations~\eqref{eq:rate_gift} and~\eqref{eq:rate_pop} yield the arrival rates to queues $G_1$ and $F_1$, respectively.
Similarly, arrival rates to the remaining queues are computed using~\eqref{eq:rate_gift_j} and~\eqref{eq:rate_pop_j}, 
and  
system departure rate $\Gamma^{(0)}$  is determined using  \eqref{eq:Gamma1}.
    Then, we let $\lambda^{(1)} = \Gamma^{(0)}$ and repeat
the process.  
The final solution is obtained by iterating
\begin{equation}
\lambda^{(n)} = \Gamma^{(n-1)} \label{eq:iterative}
\end{equation} 
until convergence is achieved. Let $\tilde{n}$ be the minimum value for which
$\lambda^{(\tilde{n})} = \Gamma^{(\tilde{n}-1)}$. Then,
\begin{equation}
\lambdacritical = \lambda^{(\tilde{n})}  \label{eq:criticalalgo}
\end{equation}

\section{Alternative Derivations of $\Gamma_S$} \label{sec:additionalderivs}

\subsection{Flow Dynamics}

The simple result presented in Proposition~\ref{prop:propo1}  
can also be directly obtained by observing the flow dynamics shown
in Figure~\ref{fig:diagrama-1}.
Recall that based on Figure~\ref{fig:diagrama-1} we derived an upper bound on the system throughput, 
given by~\eqref{eq:gammafirst}.  Setting $p=1$ in ~\eqref{eq:gammafirst} we obtain 
Proposition \ref{prop:propo1}.

It is worth contrasting the simple flow dynamics diagram in Figure~\ref{fig:diagrama-1} with the queueing network model
 in Figure~\ref{fig:mod-filas}.  The two leftmost boxes in Figure~\ref{fig:diagrama-1} are captured by queue $F_0$. 
 The remaining two boxes in the bottom layer in Figure~\ref{fig:diagrama-1}, in turn, are captured by queues $G_1$, $G_2$ and $G_3$ in 
 Figure~\ref{fig:mod-filas}.  As indicated above, the expressions for $\Gamma_S$ obtained using 
 Figures~\ref{fig:diagrama-1} and~\ref{fig:mod-filas} are in agreement with each other.

\subsection{Mean Field Approximation} \label{sec:meanfield}

Next, we consider a deterministic mean field fluid approximation to the queueing network model.
 The approach is similar, for instance, to the one considered by Qiu and Srikant~\cite{qiu2004modeling}, which was also
 influenced by a Markov model~\cite{veciana2}.  

Let $x_i(t)$ (resp., $y_i(t)$)  be the number of peers at queue $F_i$ (resp., $G_i$).  
Let $\eta_i$ be the fraction of peers from queue $F_i$ that are served by the one-club, i.e.,
 $1-\eta_i$ is the fraction of peers from queue $F_{i}$ that move to queue $G_{i+1}$, $i=0,1,2, \ldots, \tilde{K}-2$. To simplify presentation, we consider the truncated model with $\tilde{K}=4$ (extension for larger values of $\tilde{K}$ is straightforward).  For $\tilde{K}=4$,  the fluid model is given by   
\begin{eqnarray}
\frac{dx_0}{dt}&=&\lambda - x_0 \mu'  -U I(x_0>0) \\
\frac{dx_1}{dt}&=& (x_0 \mu' + U I(x_0 > 0)) \eta_0 - (x_1 \mu' + U I(x_0=0, x_1>0))\\
\frac{dx_2}{dt}&=& (x_1 \mu' + U I(x_0 = 0, x_1>0)) \eta_1 - (x_2 \mu' + U I(x_0=0, x_1=0, x_2 > 0))\\
\frac{dy_1}{dt}&=& (x_0 \mu' + U I(x_0 > 0)) (1- \eta_0) - y_1 \mu'\\
\frac{dy_2}{dt}&=& (x_1 \mu' + U I(x_0 = 0, x_1 > 0)) (1-\eta_1) + y_1 \mu' - y_2 \mu'  \\
\frac{dy_3}{dt}&=& (x_2 \mu' + U I(x_0 = 0, x_1 =0, x_2>0)) (1-\eta_2) + y_2 \mu' - y_3 \frac{\mu'}{K-3}   
\end{eqnarray}
The mean field approximation is obtained from the fluid model replacing the 
joint event  denoting that  queues  $F_0, \ldots, F_{j-1}$ are empty, and queue $F_j$ is non-empty, 
$I(\cap_{l=0}^{j-1} \{x_l = 0\}, \{x_j > 0\})$, by the corresponding product of 
probabilities $\prod_{l=0}^{j-1} \pi_0(l) (1-\pi_0(j))$ (see also Section~\ref{tab:modelsum}).
In the saturated regime, we have $\pi_0(0)=0$.   In steady state, and under saturation, 
\begin{equation}
\eta_0= \frac{\lambda-U}{\lambda}, \eta_1=\eta_2=1, \quad x_0 = x_1= x_2 = \frac{\lambda-U}{\mu'}, \quad y_1 = y_2= \frac{U}{\mu'}, y_3=(K-3)\frac{U}{\mu'} 
\end{equation}
System throughput is given by
\begin{align}
\Gamma_S &= \mu \left(y_1 + y_2 \right) + \left( \frac{K-4 }{K-3}\ \mu + \frac{1}{K-3} \mu'\right) y_3 + U  \label{eq:thrmeanf} \\
&= (2+(K-4))\mu U/\mu' + 2 U \\
&= \left(\frac{(K-2)\mu}{\mu'}+2\right) U  \label{eq:eqopenlimvalue}
\end{align}
The first  term in~\eqref{eq:thrmeanf} corresponds to  departures due to service from peers in queues $G_1$ and $G_2$
 to peers in the one-club.  Peers in $G_1$ and $G_2$ serve at rate $\mu$.  
The second term in~\eqref{eq:thrmeanf} corresponds to the departures due to services from peers in queue $G_3$.
  A fraction $({K-4})/({K-3})$ of those peers serve at rate $\mu$, whereas the remaining fraction serves at rate $\mu'$.  
   The third term corresponds to the departures of gifted peers.  After algebraic manipulation, the resulting
 expression~\eqref{eq:eqopenlimvalue} equals~\eqref{eq:open-lim}.

\section{Markov Model Details} \label{app:moddetails}

We model a swarm as a continuous-time Markov chain with state space $\Omega$ and infinitesimal generator $Q$.  Let $\mathcal{F}=\{1,\ldots,K\}$ and $\mathcal{C}$ be the set of subsets of $\mathcal{F}$.  

Each user has a signature, defined as a set containing element  $i$ if the user has block $i$ and 0 otherwise, $i=1, 2, \ldots, K$.  As users leave the system as soon as they obtain their last block, each user has one of $2^{n}-1$ signatures.

Let $\sigma_C$ be the number of peers with signature $C$, where $C \in \mathcal{C} \setminus \mathcal{F}$. 
State $\bsigma \in \Omega$ is characterized by the number of users with each signature, $\bsigma=(\sigma_\emptyset, \sigma_{\{1\}}, \ldots, \sigma_{\mathcal{F} \setminus \{K\}})$.  Note that it is possible to lump the state space, but to simplify presentation in this appendix we consider the unlumped state space.  
Let $\be_C$ denote the vector with the same dimension
as $\bsigma$, with a one in position $C$ and other coordinates
equal to zero.

Let $\Gamma_s(C,C')$ and $\Gamma_p(C,C')$ be the aggregate transition rate of peers of type $C$ to type $C'$ due to service from the server   and from other peers, respectively.  We assume peers adopt the random peer, random useful block  selection, whereas the publisher strategy is varied. In some cases, it will be convenient to make explicit the block which is received by a peer with signature $C$,   replacing $C'$ by $C'_j$ when block $j$ is received.  After a peer with signature $C$ receives block $j$,  $j \notin C$, the number of peers with signature  $C'_j$ increases by one.  
Recall that as soon as peer completes its download a new one arrives (closed system).
Then, 
\begin{equation}
C'_j = \left\{
\begin{array}{ll}
C \cup \{j\}, & \textrm{if }  |C| < K-1  \\
\emptyset, & \textrm{otherwise}
\end{array}
\right.
\end{equation}

Next, we characterize  the positive elements of $Q$.
 When a peer that has all blocks except $j$ gets block $j$, its signature transitions from 
$\mathcal{F} \setminus \{j\}$ to $\mathcal{F}$.  Then, it immediately leaves the system and another peer, with signature $\emptyset$, arrives.  Then, the rate at which the system transitions from state 
$\bsigma$ to $\bsigma - \be_{\mathcal{F} \setminus \{j\}} + \be_\emptyset$ is,
\begin{equation}
q_{\bsigma,\bsigma - \be_{\mathcal{F} \setminus \{j\}} + \be_\emptyset} =  \Gamma_s(\mathcal{F} \setminus  \{j\}, \emptyset) +  \Gamma_p(\mathcal{F} \setminus  \{j\},\emptyset), \textrm{ for } j = 1, 2, \ldots, K
\end{equation}
When a peer that needs more than one blocks gets block $j$, its signature transitions from $C$ to $C \cup \{j\}$.  
The corresponding rate at which the system transitions from state 
$\bsigma$ to state $\bsigma - \be_{C} + \be_{C \cup \{j\} } $ is 
\begin{equation}
q_{\bsigma,\bsigma - \be_{C} + \be_{C \cup \{j\} } } = \Gamma_s(C,C \cup \{j\}) + \Gamma_p(C, C \cup \{j\}),  \textrm{  for all  } C \in \mathcal{C},  |C| < K-1
\end{equation}

Let $\pi_\sigma$ be the steady state probability of state $\sigma$. The vector of  steady state probabilities  is denoted by $\bpi$.  \eat{, where  $\pi Q = 0$.}

Let $\lambda$ be the throughput of the Markov model. 
The  throughput is given as a function of 
 $q_{\bsigma,\bsigma - \be_{\mathcal{F} \setminus \{j\}} + \be_\emptyset}$
 as follows, 
\begin{equation}
\lambda = \sum_{\sigma \in \Omega} \sum_{j=1}^{K} \pi_\sigma q_{\bsigma,\bsigma - \be_{\mathcal{F} \setminus \{j\}} + \be_\emptyset} 
\end{equation}

Let $\mathcal{N}'_S$ be the set of neighbors of each of the peers with signature $S$, $i.e.,$
 the candidate peers to which a peer with signature $S$ can potentially transfer content. Except otherwise noted, we assume 
$|\mathcal{N}'_S|=N-1$.  

Let $\mu_S$ be the service capacity of peers with signature $S$.  When considering homogeneous peers, 
we let $\mu_S=\mu$ for all $S \in \mathcal{C}$.  When studying the special policy wherein peers with all blocks 
except one reduce their service capacity to $\mu'$, we let $\mu_S = \mu$ if $|S| < K-1$ and $\mu_S=\mu'$ if $|S| = K-1$.  

For $C \in \mathcal{C}$ and $j=1, 2, \ldots, K$, 
we have
\begin{equation}
\label{eqrr}
\Gamma_p(C,C'_j) =
\left\{ 
{
\begin{array}{ll}
{\sigma_C} \left( { \sum_{S: j \in S \setminus C} \frac{{\mu}_S \sigma_S}{|S-C|  |\mathcal{N}'_S| }  } \right),  
                           &  j \notin C \\
0,                         &  \textrm{otherwise}
\end{array}
}
\right.
\end{equation}
If the tracker does not announce all newcomers to other peers, equation~\eqref{eqrr} still holds, except for $C=\emptyset$.  In this case, $|\mathcal{N}'_\emptyset| = N-1$ but $|\mathcal{N}'_C| \leq N-1$ for $C \neq \emptyset$ (see Section~\ref{sec:models}).  

In what follows, we characterize $\Gamma_s(C,C')$ for the different strategies considered in this paper,
\begin{itemize}
\item \emph{random peer, random block}:  the publisher allocates capacity $U \sigma_C/N$ to serve peers with signature $C$, and each of the useful blocks is transferred with same probability. Then,
\begin{equation}
\Gamma_s(C,C'_j) = \left\{
\begin{array}{ll}
 \frac{U \sigma_C}{N (K-|C|)},& j \notin C \\
0, & \textrm{otherwise}
\end{array}
\right.
\label{eqrandomrandom}
\end{equation}
\item \emph{random peer, rarest block}: let $\mathcal{R}_C$ be the set of less replicated  blocks among those which are useful for a peer with signature $C$. Then, replacing   $K-|C|$ by  $|\mathcal{R}_C|$ in \eqref{eqrandomrandom} we obtain
\begin{equation}
\Gamma_s(C,C'_j) = \left\{
\begin{array}{ll}
 \frac{U \sigma_C}{N |\mathcal{R}_C|}, & j \notin C, j \in \mathcal{R}_C \\
 0, & \textrm{oterwise}
\end{array}
\right.
\label{eqmostdeprmodel}
\end{equation}
\item \emph{most deprived peer, rarest block}:  let $\mathcal{M}$ be the set of signatures of most deprived peers.  Then, replacing   $N$ by  $\sum_{C:C \in \mathcal{M}}\sigma_C$ in \eqref{eqmostdeprmodel} we obtain
\begin{equation}
\Gamma_s(C,C'_j) = \left\{
\begin{array}{ll}
 \frac{U \sigma_C}{\left(\sum_{C: C \in \mathcal{M}} \sigma_C\right) |\mathcal{R}_C|},&  C \in \mathcal{M} , j \notin C, j \in \mathcal{R}_C  \\
 0, &  \textrm{otherwise}
\end{array}
\right.
\end{equation}
\end{itemize}

\section{Lumped Model Details}

\label{app:lumped}

We present the algorithm used to generate the  lumped state space. Algorithm~\ref{alg:lump} takes as input a state of the unlumped
state space, and generates as output the corresponding state in the lumped version of the model.  To this aim, it sorts the block 
according to their number of replicas (line 1),  and reorders the block identifiers based on this ordering (line 4).  The new signatures are generated (line 12) and the
new state is computed  (line 15).

\eat{ Table~\ref{lumpedandunlumped} illustrates the gains obtained through lumping.    }

Note that 
\eat{not all possible combinations of signatures are reachable from the initial state, 
as to generate this table we assume that the publisher adopts 
the random peer, rarest first block policy.  Hence, } the number of states in the unlumped version of the model
is ${2^K-2 +N\choose N}$,  $i.e.$, the number of ways of dividing $N$ indistinguishable 
users into $2^K-1$ groups, where each group corresponds to a signature.  The number of states in the lumped model is 
up to an order of magnitude
 smaller than the number of states in the unlumped model. 
For instance, for $K=3$ and a population of $N=20$ users, lumping decreases the number of states from 230,230 to 46,163 (see Table~\ref{lumpedandunlumped}).

\begin{algorithm}[h!]
\caption{$\Lump$} \label{alg:lump}
\begin{algorithmic}[1]
{\small{
\INPUT state $\bsigma$ in unlumped state space $\Omega$
\OUTPUT state $\bsigma'$ in lumped state space $\Omega'$
\STATE $s \leftarrow $ list of block identifiers, sorted by number of replicas
\STATE $i \leftarrow 1$
\WHILE {$i \leq K$} 
\STATE $n(s(i)) \leftarrow i$, $i \leftarrow i + 1$
\ENDWHILE
\STATE $\mathcal{C} \leftarrow$ set of subsets of $\{ 1, \ldots, K\}$
\STATE $\mathcal{C} \leftarrow \mathcal{C} \setminus \{ \mathcal{F} \}$ 
\WHILE {$\mathcal{C} \neq \emptyset$} 
\STATE remove signature $C$ from $\mathcal{C}$, $C' \leftarrow \emptyset$
\STATE $j \leftarrow 1$
\WHILE {$j \leq K$} 
\STATE if ($j \in C$) $C' \leftarrow C' \cup \{ n(j) \}$
\STATE $j \leftarrow j + 1$
\ENDWHILE
\STATE $\sigma'_{C'} \leftarrow \sigma_C$
\ENDWHILE
\STATE  output $\bsigma'$ is the lumped state
}}
\end{algorithmic} \label{alg:main}
\end{algorithm}

\begin{table}[h!]
\[
\begin{array}{l|l|l}
\hline
\textrm{number of peers ($N$)} & \textrm{number of states } & \textrm{number of states }  \\
& \textrm{(lumped model)} &  \textrm{(unlumped model)} \\
\hline
 5 &  127    &    462        \\
 6 &  243   &      924    \\
 7 &  429   &      1716     \\
 8 &  728   &       3003   \\
 9 & 1174    &       5005  \\
10& 1836   &       8008    \\
11&   2772   &     12376      \\
12&  4086   &        18564   \\
13 &  5868   &       27132    \\
14 &  8268   &        38760  \\
  15 &11418  &        54264  \\
  16 & 15525  &     74613   \\
  17 &20775    &   100947     \\
  18 &27445    &   134596    \\
  19& 35787    &    177100   \\
  20 & 46163 &      230230    \\
\hline
 \end{array} \]
\caption{File with $K=3$ blocks. Illustrating the gains due to lumping.} \label{lumpedandunlumped}
\end{table}

\section{Delayed and Premature Departures}

Next, we consider delayed departures (peers lingering as seeds) and premature departures (peer churn).

\subsection{Delayed Departures: Peers Lingering as Seeds}

\label{app:lingering}

We consider the case where peers remain in the system as seeds after completing their downloads.   Let $K=2$, $U=1$ and $\mu=1$.
Figure~\ref{fig:scale}(a) shows that when $\gamma \le 1.2$ the throughput increases linearly as the population increases.  In this 
case, the system is stable, in accordance to ~\cite{hajek}.  The situation is not as simple when $\gamma \ge 1.5$.  When $\gamma=\infty$, Figure~\ref{fig:scale}(b) shows that the throughput reaches an asymptote when the population grows,  for the reasons explained in this paper.  It also indicates that if the publisher replaces the random piece policy by the rarest piece policy, the gains are negligible, which again is in accordance to ~\cite{hajek}.  For more details about this scenario, refer to~\cite{menasche2012stability}.

\begin{figure}[H]
\center
\includegraphics[angle=-90,scale=0.42]{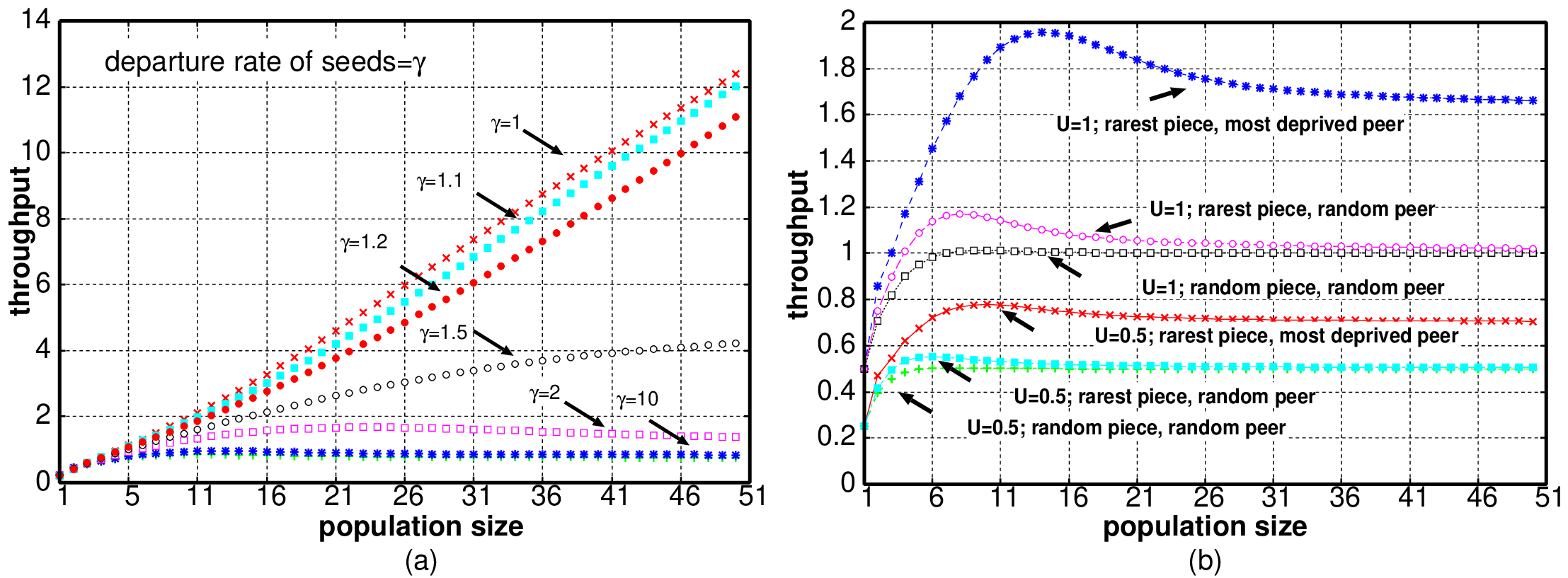}
\caption{Scalability when peers linger in the system after completing download: $(a)$ \emph{peers} remain in the system as seeds for an average of $1/\gamma$ after completing \emph{download}; $(b)$ \emph{peers} depart immediately.}
\label{fig:scale}
\end{figure}

 \subsection{Premature Departures: Peer Churn}
 \label{app:churn}

 Next, we consider the scenario  wherein peers can prematurely abandon the system before concluding their downloads.
 First, we consider the simplest setup wherein each peer is equipped with a timer that is started
 when the peer joins the system and that triggers
  after a  possibly random period of time, with mean $1/\alpha$.  If the timer triggers before the download concludes,
  the peer abandons the system.  In this case, the system is stable, and the one club is bounded.
  The missing piece syndrome will not occur, as the mean number of peers in the system is bounded
  by $\lambda/\alpha$.

   Now, consider the case where peers can decide to abandon the system
immediately after completing the download of a block
(any block).  
In addition, to
simplify notation we assume that gifted peers do not prematurely abandon the
system. Indeed, such peers  have less incentives to prematurely abandon the
system than their non-gifted counterparts.
Because the gifted peers possess the rarest block, their
download progresses smoothly
unlike the non-gifted peers 
that may be suffer from very long delays waiting for the rarest block. 

After the download of each block, abandonments can be characterized either from a population-wide perspective (wherein the aggregate rate at which the population abandons the system is given) or from an individual peer perspective (wherein the individual probability that a peer abandons the system is given).      In what follows, we consider the latter. 
  Let $q_i$ be the probability that a peer abandons the system after obtaining its  $(i+1)$-th first block.    Then, $\gamma_p(i)$ is given by~\eqref{eq:rate_pop_j}, multiplying the right hand side by $q_i$,  
\begin{equation}
\label{eq:rate_pop_j_q_i}
\gamma_p(i) = \mathbb{E}[n_i] \mu' q_i.
\end{equation}
All the remaining equations in Sections~\ref{sec:queueing-model} and~\ref{sec:truncation} and Proposition~\ref{prop:propo1} still hold.

Recall that in Proposition~\ref{prop:propo1} the saturation throughput is the maximum throughput achievable by the open system.   
Let $\tilde{q}$ be the probability that a peer prematurely abandons the system.  If $(\lambda - U) (1-\tilde{q}) >    \left(\frac{(K-2)\mu}{\mu'}+1\right) U$ the arrival rate to the one-club is larger than the corresponding departure rate (Figure~\ref{fig:diagrama-1}),  the one-club still grows unboundedly and the arguments presented in Section~\ref{sec:qnetmode} remain valid to derive~\eqref{eq:open-lim}.

\section{Additional Policies}

\label{app:additionalpolic}

Most of the throughput analysis in this paper focused on a publisher adopting the most deprived peer/rarest-first block policy.
Next, we consider two additional policies.  We still assume that peers adopt the random peer/random block policy (RP/RB).  

\subsection{MDP/RB} 

The analysis of   most deprived peer/random block policy by publishers is similar to the 
one presented under the flow dynamics  in Section~\ref{sec:models}.  The key difference consists of the rate of newcomers that
become gifted, which must be replaced from $pU$ by $pU/K$.   Similarly, in the queueing network model the rates $\gamma_r(i)$ must 
be multiplied by $1/K$.  It can be verified that the maximum throughput capacity, when $\mu=\mu'$, equals $U$, i.e., the benefits
of MDP do not hold if the publisher adopts  RB in place of RFB.

\subsection{RP/RB with seeds}

Next, we consider peers that remain in the system after completing their downloads.   
 The flow dynamics analysis in Section~\ref{sec:models} can be extended to allow for peers that remain in the system as seeds
 after completing their downloads.  We consider a publisher that adopts RP/RB.

\begin{figure}[htb]
\center
\includegraphics[width=1\columnwidth]{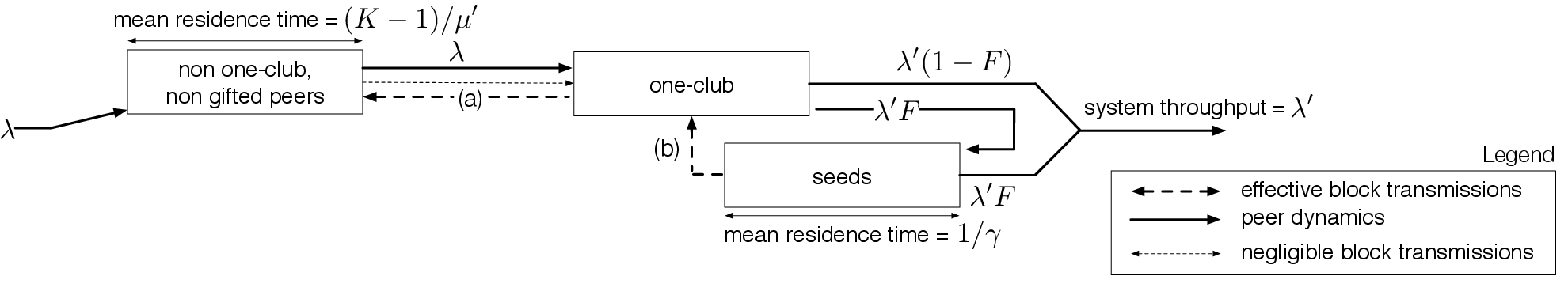}
\caption{Flow dynamics with seeds}
\label{fig:diagrama-1linger}
\end{figure}

The flow dynamics is shown in Figure~\ref{fig:diagrama-1linger}.  Let $\lambda'$ be the rate at which peers leave the one-club.
It follows from Little's result that there 
are on average $\lambda'F/\gamma$ seeds in the system.  Each seed serves at rate $\mu'$.
Then,  in steady state we have
 $\Gamma=\lambda'= \min(\lambda, U+\lambda'F \mu'/\gamma)$.    Equivalently, 
 \begin{equation}
 \lambda'= \min\left(\lambda, \frac{U}{1-\mu' F/\gamma}\right) 
 \end{equation}
 Note that the equation above holds if  $ F \mu'/\gamma \leq  1$. In particular, 
 if $1/\gamma = 1/(F\mu')$ we have $\lambda = \lambda'$ and peers will experience bounded delays.  
 If $1/\gamma > 1/(F\mu')$, delays are further reduced as peers count with the assistance of additional seeds.

 Note that a slightly more restricted version of this result has already been proved in~\cite{zhu2012stability}.  Here, we illustrate the
 simplicity of the flow balance dynamics to  derive this and  similar results.


\end{document}